   \PassOptionsToPackage{table}{xcolor}
   \PassOptionsToPackage{numbers,sort&compress}{natbib}
   \documentclass[sigconf]{acmart}
\usepackage[colorinlistoftodos]{todonotes} % provides the \todo{} command, \notice{} etc
\usepackage{algorithmic}
\usepackage{algorithm}

\usepackage{xspace}

\usepackage{fontawesome5}  % to allow funky checkmarks
\definecolor{carnationpink}{rgb}{1.0, 0.65, 0.79}
\definecolor{cinnabar}{rgb}{0.89, 0.26, 0.2}
\definecolor{darkviolet}{rgb}{0.58, 0.0, 0.83}

\newcommand{\name}{BotScan\xspace}
\newcommand{\nameMod}{BotScan-Mod\xspace}
\newcommand{\nameHi}{BotScan-Hi\xspace}
\newcommand{\smartProbing}{behavior-adaptive\xspace}
\newcommand{\SmartProbing}{Behavior-adaptive\xspace}

\newcommand{\dynamic }{dynamic\xspace} % playbook 
\newcommand{\threeclass}{three-class\xspace} 

\newcommand{\playbook}{replay-based\xspace} 
\newcommand{\Playbook}{Replay-based\xspace} 
\newcommand{\entries}{packets\xspace}

\newcommand{\hierCnC}{hierarchical\xspace} 
 
\newcommand{\ptopCnC}{P2P\xspace}

\newcommand{\host}{C2 server\xspace}
\newcommand{\Host}{C2 server\xspace}

\newcommand{\hosts}{C2 servers\xspace}

\newcommand{\module}{module\xspace} 
\newcommand{\Module}{Module\xspace} 
\newcommand{\modules}{modules\xspace} 
 
\newcommand{\moduleA}{A\xspace}
\newcommand{\moduleB}{B\xspace}
\newcommand{\moduleC}{C\xspace}

\newcommand{\fixed}{fixed\xspace} 
\newcommand{\Fixed}{Fixed\xspace} 
\newcommand{\tricky}{custom\xspace} 
\newcommand{\Tricky}{Custom\xspace} 
\newcommand{\obfu}{obfuscated\xspace} 
\newcommand{\Obfu}{Obfuscated\xspace} 

\newcommand{\Initial}{Initiation\xspace} % packets that can start a C2 interaction
\newcommand{\Response}{Response-pair\xspace} % packets that respond
\newcommand{\Responses}{Response-pairs\xspace} % packets that respond

\newcommand{\localDensity}{Local Density\xspace} %% the metric percentage of hosts in a subnet

\newcommand{\segment}{segment\xspace} 
\newcommand{\Segment}{Segment\xspace} 
\newcommand{\segments}{segments\xspace} 
 
\newcommand{\intraSeg}{intra-segment\xspace} 
 \newcommand{\interSeg}{inter-segment\xspace} 
\newcommand{\IntraSeg}{Intra-segment\xspace} 
 \newcommand{\InterSeg}{Inter-segment\xspace} 

\newcommand{\maxNoPorts}{MaxPorts\xspace} %Maximum number of ports to probe per IP address

\newcommand{\segk}{\eta\xspace}  % the prefix size of a \segment NOT the optimal value though

 \newcommand{\Phase}{Step\xspace} % Majore parts of our exploration in Methodlogy section
 
 \newcommand{\phases}{steps\xspace}

\newcommand{\ruleset}{communication pattern\xspace} 
 
\newcommand{\rulesets}{communication patterns\xspace} 
\newcommand{\Rulesets}{Communication patterns\xspace}

\newcommand{\replay}{replay\xspace}
\newcommand{\Replay}{Replay\xspace}

\newcommand{\replayable}{replayable\xspace}

\newcommand{\unreportedALL}{112\xspace}
\newcommand{\totalALL}{896\xspace}

\newcommand{\tifs}{threat intelligence feeds\xspace}
\newcommand{\Tifs}{Threat intelligence feeds\xspace}
\newcommand{\tif}{threat intelligence feed\xspace}

\definecolor{dkgreen}{rgb}{0.1,0.5,0.4}
\definecolor{gray}{rgb}{0.5,0.5,0.5}
\definecolor{mauve}{rgb}{0.58,0,0.82}

\newcommand{\vnote}[1]{{\color{blue}[Vivek: \color{red}{\textbf{#1}}]}}

\newcommand{\krish}[1]{ {\textcolor{red}{[SK:}} \textcolor{brown}{#1]} }

\newcommand{\miii}[1]{{{\textcolor{blue}{[MF:} \textcolor{red}{#1]}}}}
\newcommand{\miok}[1]{ {\textcolor{blue}{MF:}} {\textcolor{dkgreen}{#1}}} %%% This command is to highlight text that we want to leave in the final paper!

\newcommand{\zt}[1]{{{\textcolor{dkgreen}{[ZT:} \textcolor{teal}{#1]}}}}
\newcommand{\zok}[1]{ {\textcolor{dkgreen}{ZT:}} {\textcolor{mauve}{#1}}} %%% This command is to highlight text that we want to leave in the final paper!

\newcommand{\mim}[1]{{\textcolor{red}{[Mim:} \textcolor{blue}{#1]}}}  %%% Comment to say ask a question

\newcommand{\mimok}[1]{{\textcolor{red}{[Mim:} \textcolor{mauve}{#1]}}}  %% to highligh text that you changed but you want to leave eventually

\newcommand{\vj}[1]{{\textcolor{red}{[VJ:} \textcolor{blue}{#1]}}}

\newcommand{\revupdates}[1]{{\color{blue} #1}}

\newcommand{\eatreminders}[1]{
   \renewcommand{\miii}[1]{}
   \renewcommand{\miok}[1]{##1}
   
   \renewcommand{\zt}[1]{}
   \renewcommand{\zok}[1]{##1}
   
   \renewcommand{\mim}[1]{}
   \renewcommand{\mimok}[1]{##1}
   \renewcommand{\krish}[1]{}
   \renewcommand{\vnote}[1]{}
   \renewcommand{\revupdates}[1]{##1}
   \renewcommand{\vj}[1]{##1}
}

\eatreminders      %%% UNCOMMENT THIS TO REMOVE REMINDERS! and allow the "ok" reminders to appear as text
\newcounter{takeaway}

\newcommand{\takeaway}[1]{%
  \refstepcounter{takeaway}%
  \noindent {\colorbox[RGB]{230,190,250}{\parbox{0.975\columnwidth}{{\textit{\textbf{Observation \arabic{takeaway}}. #1}}}}}%
}

\definecolor{LightCyan}{rgb}{0.88,1,1}

\usepackage[most]{tcolorbox}

\newcommand{\rcone}{\textbf{{RC-1}\xspace}}
\newcommand{\rctwo}{\textbf{{RC-2}\xspace}}

\newcommand{\rInt}{$R_{int}$}
\newcommand{\rExt}{$R_{ext}$}
\usepackage{pdfrender}

\definecolor{ForestGreen}{rgb}{0.13, 0.55, 0.13}

\newcommand*\halfcirc[1][1ex]{%
  \begin{tikzpicture}
  \draw[fill] (0,0)-- (90:#1) arc (90:270:#1) -- cycle ;
  \draw[thick] (0,0) circle (#1);
  \end{tikzpicture}}
\newcommand*\fullcirc[1][1ex]{\tikz\fill (0,0) circle (#1);}

\definecolor{Red}{RGB}{220,0,0}
\definecolor{Orange}{RGB}{255,165,0}
\definecolor{ForestGreen}{RGB}{34,139,34}

\newcommand{\ourGreen}{\textcolor{ForestGreen} \fullcirc}
\newcommand{\ourYelllow}{\textcolor{Orange} \halfcirc}
\newcommand{\ourRed}{\textcolor{Red} \faTimes}
\usepackage[T1]{fontenc}
\usepackage{graphicx}

\usepackage{amsmath,amssymb}
\usepackage{booktabs}
\usepackage{tabularx}
\usepackage{multirow}
\usepackage{enumitem}
\usepackage{subcaption}
\usepackage{float}
\usepackage{tikz}
\usepackage{pifont}
\usepackage{url}

\newcommand{\xmark}{\text{\ding{55}}}

\usetikzlibrary{arrows.meta,positioning,fit,calc}
\AtBeginDocument{%
  }

\setcopyright{acmlicensed}
\copyrightyear{2026}
\acmYear{2026}
\acmDOI{XXXXXXX.XXXXXXX}
\acmConference[ASIACCS '27]{ACM Asia Conference on Computer and Communications Security}{July 12--16, 2027}{Macau, China}

\begin{document}

\title{\name: An adaptive  active probing approach for identifying live IoT Botnet \hosts at scale}

\begin{abstract}

% C2 server == malicious host or host
% C2 server communication protocol = botnet communication protocol or malicious host communication protocol
% C2 IP address = (malicious) host or botnet IP address

How can we actively search and identify live  \hosts 
of botnets at scale? 
%Detecting the botnet infrastructure is a critical capability to contain botnet-based cybercrime.
%We use the term active to refer to a capability that can probe a target IP space to find live \hosts.
%, which we define as devices that constitute the botnet's infrastructure.
The scalability requirement introduces the need to utilize resources efficiently in terms of computation and number of probing packets.
%\krish{I like the prior commented text. This does not tell us why the problem is important. It implicitly seems to assume that it is. Abrupt transition from quesiton.}
%use of resources in order to enable large-scale and frequent probing explorations.}
%\miok{Naturally, we want to achieve this goal with an efficient use of resources in order to enable large-scale and frequent probing explorations.
%}
% \miii{@ZT: does this do the job? We have numbers later in the intro}
%\zt{I think we should also use 1-2 sentences to claim the budget is pretty stringent in real world (to better motivate the problem)}
% especially C2 servers
%including C2 servers.
% by expanding the concept of C2 servers
% We use the term active to refer to a capability that can probe a target IP space to find live \hosts in malicious infrastructure  especially C2 servers.
%as opposed to 
%relying on a passive analysis of network traffic.
%We see a need for a widely-used reference active probing capability that can operate at scale.
We propose \name, an approach for actively probing a large IP space to find the highest possible number of live \hosts.
%given a probing budget. 
The novelty of \name revolves around two insights, which we establish empirically. 
First, contrary to popular PC-centric observations,
%contrary to popular belief, many modern IoT botnet communication protocols use packets with minimal customization.
many modern IoT botnet communication protocols use packets with minimal customization, which we 
observe %measure and taxonomize 
across six major families.
%unlike PC based botnets.
%that are more sophisticated.
%\zt{From my understanding based on our previous meeting, the key point here is to highlight something like ``contrary to conventional wisdom where the replay is not feasible given encryption of botnet communication, it is still feasible (for IoT family only?) to ...'' 
%}
Second, \hosts exhibit exploitable behavioral patterns, such as strong spatial locality.
%exhibit some behavioral patterns, such as strong spatial locality.
% \miii{@ZT: Did this address your comment - which is commented out for clarity?}
We substantiate the first insight by developing a streamlined approach where, given a malware binaries, we measure and taxonomize the "replayability" of its C2 communication protocol.
%: we activate it, extract its C2 communication trace, and check replay the trace elicits a response from a server. 
% we first develop an approach for generating a {\it \dynamic} and lightweight 
% %\miii{We don't make the case in the paper for lightweightness, so it may look as unsupported claim. If end up doing it, we can add it.} \krish{My understanding is - playbook based probing compared to sandbox based probing is lightweight and informed probing as opposed to random probing of the IP space is lightweight (high hit rate).}
% probing playbook by extracting information from the behavior of malware binaries. 
% }
%\zt{If playbook is a term we created, I suggest using some plain explanation here}
Then, we introduce a \smartProbing\ probing strategy that:
(a) exploits the spatial locality of \hosts
using a two-level \segment-centric approach, and (b) adapts dynamically to the success of its  probes. 
%to prioritize and explore target IP addresses efficiently. 
%\zt{I suggest making this part more direct to capture the insights. Instead of just saying ``such as...", we should explicitly say ``we exploit xxx, xx and xx to ...''}
%\miii{@Team: use "spatial placement" or "non-random spatial placement" instead?}
%%
%\krish{I suggest you make this more clear. Who knows what "spatial" locality is wrt C2 server behavior. For example, say that C2 servers often have common IP prefixes}
%\miii{Good point,let's revisit later}
%We conduct experiments to evaluate and substantiate our approach. 
We validate the effectiveness of our method using 1,842 recently collected IoT binaries, 
%\miii{Let's report only the ones that activate and connect, that's the baseline}
%which successfully activate and engaging with their C2 servers, 
% Colelcted 1,842 succesfully engaging
and we explore a target space of 2.5M IP addresses.
%Our experiments  validate the \miok{applicability and} effectiveness of \name: 
First, a replay-based method is applicable for at least 72\% of the malware binaries.
%\miii{A replay-based method is applicable for 80xx\% of the XXXX "C2-engaging" binaries} 
Second, our method outperforms baseline methods by finding approximately double the live C2 servers for the same number of probes. We also conduct two case-studies where we identify 896 live servers including 112 unreported C2 servers.

\end{abstract}


%\ccsdesc[300]{Networks~Network measurement}

\author{S M Maksudul Alam}
\email{salam031@ucr.edu}
\affiliation{%
  \institution{University of California, Riverside}
  \country{USA}
}

\author{Vivek Jain}
\email{vjain014@ucr.edu}
\affiliation{%
  \institution{University of California, Riverside}
  \country{USA}
}

\author{Zhaowei Tan}
\email{ztan@ucr.edu}
\affiliation{%
  \institution{University of California, Riverside}
  \country{USA}
}

\author{Srikanth V. Krishnamurthy}
\email{krish@cs.ucr.edu}
\affiliation{%
  \institution{University of California, Riverside}
  \country{USA}
}

\author{Michalis Faloutsos}
\email{michalis@cs.ucr.edu}
\affiliation{%
  \institution{University of California, Riverside}
  \country{USA}
}

\maketitle

\section{Introduction}
\label{sec:introduction}
%%% SCOPE and motivation
\label{sec:intro}
%\krish{Identifying botnet infrastructure sounds vague. Coming as the first phrase, I would suggest changing it} 
Identifying the  Command and Control (C2) servers of botnets and their infrastructure is essential for detecting and mitigating botnet-driven cybercrime. 
Botnet activities have increased in recent years at an alarming rate~\cite{crowdstrike2024}. 
C2 servers\footnote{
Recently, the term C2 server has expanded to include both hierarchical and peer-to-peer botnets~\cite{nappa2014cyberprobe,xu2014autoprobe} and it refers to devices that can act as ``communication gateways" to the botnet: they  respond and help  bots join a botnet (see section \S\ref{sec:motivation}).
%\krish{Why revisit?}.
} 
are the operational ``headquarters'' of botnets, and, therefore, they are a good target for security countermeasures.
According to Sekoia~\cite{sekoia_2022}, the number of identified C2 servers in 2022 was 65K, 
and seems to be growing by 20-50\% every year.
%which is a 50\% increase compared to 2021. The trend continued in 2023, with the number rising to 85K servers.
Once we identify these servers, we can implement measures such as monitoring or blocking traffic aimed at them. %This strategy is particularly valuable for IoT devices~\cite{C2Miner-Davanian-2024, alrawi2021circle}, since they often lack the computational capability needed for more sophisticated on-device defenses (e.g., antivirus software).

%In the IoT space, 
%C2-centric 
%The exploitation of IoT devices by hackers introduces novel challenges for cybersecurity.
The emergence of IoT devices as a threat vector introduces novel challenges for cybersecurity.
First,  these devices have lower computing and defense capabilities and there is a lack of mature end-point protections~\cite{victor2023iot, C2Miner-Davanian-2024, alrawi2021circle, tanabe2022disposable}. 
%their resource‑constrained hardware and lack of mature endpoint protections mean that proactive monitoring and blocking of botnet communication  is often the only practical defense~\cite{C2Miner-Davanian-2024, li2019privacy, canavese2024security}.
%As a result, securing IoT devices is urgently needed. 
Second, the underlying device population is exploding into tens of billions of  devices, dramatically increasing the pool of potential bots~\cite{gelgi2024systematic, iotanalytics2024state}.
Third, IoT devices are disproportionately exposed and often misconfigured, compared to other well‑managed IT assets~\cite{forescout2025devicevuln, sasi2024comprehensive, alzaylaee2025systematic}.

%We use the term {\bf malicious \host} to expand the definition of the traditional C2 server to also include
%compromised devices that listen and respond to botnet communication requests.
%Our work focuses on identifying \hosts.

%This is a commonly-used definition~\cite{nappa2014cyberprobe,xu2014autoprobe},
%and it is motivated by the need to include \ptopCnC botnets, which do not follow a traditional \hierCnC  structure. \mimok{A P2P bot, unlike traditional C2 servers, can be responsible as a server and a bot in the decentralized botnet.}
%\zt{I'm a bit confused here: 
%For p2p, since there's no server, I get it that we are finding devices because there's essentially no center server. 
%But for traditional structure, are we also interested in finding compromised devices? If not, I will have to re-write this part.
%}
%\miii{This is a good point: we can finesse this, or we can add a forward reference and discuss this in Section 2 to avoid confusing the reader up front.}

%%% Problem
\textbf{Problem:} 
\textit{How can we actively probe for live  \hosts  in an efficient way?}
% This is the question that we address in our work. 
We use the term efficiency to include the computational resources, and the number of probing packets.
%We can state the problem as follows.
Specifically, the input to the problem is a large target IP space of interest. %(e.g. a remote and possibly hostile network).
%, and (b) a probing budget (e.g.  total number of probe packets).
The desired output is the largest possible number of {\it live} \hosts within the target IP space in a cost efficient way.
In other words, we frame the problem as a
 {\em  Return on Investment (RoI) maximization}: we want the highest  number of identified live \hosts over the number of IP:port pairs that we will probe 
 (which we will refer to as {\em probing budget} for ease of reference). 
 %\mim{A proposed statement to avoid any reviewer concern: "We want a higher number of live \host detection over efforts (in terms of target exploration) in comparison to the existing approaches."}
 %\miii{Mimon: Current phrasing is good} \mim{okay, sure!}
 %For ease of reference, we will use the term refer to the number of probing packets 
 %which we can refer to as the {\em probing budget}.
%A key goal is to be  {\bf maximize the Return on Investment (RoI)}: the number of live \hosts detected over the number of probe packets.
%The requirement for efficiency is motivated by the need to explore large IP spaces, or establish an ongoing monitoring capability as we discuss in the next paragraph.
%\zt{I suggest better motivate the problem.
%ROI is important for sure, but an easy counter-argument is, detecting malicious hosts is so important that coverage is more important. So the argument needs to be amplified. The current arguments ``explore large IP spaces'' and ``an ongoing monitoring capability'' are too vague. I would love to see something like: ``Even for a xxx subnet, there is XXX combinations and enumerating all of them would take xx years on a server.'', or ``XXX family changes frequently and needs to be probed daily to understand its dynamics''
%These are just examples, but some arguments like these will make the work much stronger. (and can be used for your future work, as well)
%} \mim{Added a few paragraphs before the problem definition.}
From a practical point of view, 
we allow the solution to consult archival information on  \hosts in the target space, if such information exists.
%In addition, the user could provide a preferred set of malware families, but we do not discuss this option further. 
%\zt{what does this sentence mean?}
We distill the problem into two  research challenges \textbf{(RC)}:

\noindent {\bf \rcone: What to probe with?} 
%Creating the packet that we need to send to a target IP address is not straightforward: 
%a wrongly crafted packet will not elicit a response from  a malicious \host.
%Crafting a packet that will elicit a response from a potential botnet \host is not straightforward: these packets should trigger a response only from such \hosts and not from benign IP addresses. 
Crafting a  probing packet\footnote{
The two main options are: (a)  
generating or replaying packets, and (b) getting packets from an activated binary, as we discuss in  section \S\ref{sec:motivation}.
 In contrast to popular PC-centric beliefs~\cite{securelist2025malware}, 
 recent IoT malware does not seem to use sophisticated encryption, as we show in this study, which makes \playbook probing feasible.
 %We discuss prior work in section \S\ref{sec:related}.
} 
is not easy, as it should  elicit  ``incriminating''  responses  only from  \hosts. 
%but not  benign devices. 

%Recent studies~\cite{securelist2025malware,C2Miner-Davanian-2024} argue that modern malware uses sophisticated encryption, which makes the problem harder, but they seem to refer to PC-based malware, %botnets, 
%and not IoT malware, which are less sophisticated.

%and they focus on computationally expensive and brittle methods that sometimes require deep protocol reverse-engineering.
%(a) using activated malware, or (b) full reverse-protocol engineering \cite{}. 

%\zt{The key is to make it looks like a challenge. Given the existing work of probing, they must also have a strategy to generate probes. What's the gap between their approach and our goal? 
%Are they being too simplistic, or they cannot target the IoT families like what we can do?
%That should be the RC1 here. } \mim{updated and ready for review.}

%and we should be able to separate malicious from benign devices.
%we want to maximize the likelihood of eliciting a response from a live malicious \host. 
%\krish{``host'' must be singular. I think still the probes being lightweight is important. Otherwise, it violates budget constraints. Not coming across. Suggest  "Creating a small set of packets so as to elicit a response from a potential botnet host is not straightforward; these packets should trigger a response only from such hosts (and not from benign IP addresses)."}

\noindent {\bf \rctwo: Where to probe?}  Given a large IP space, the prioritization of which IP addresses and port numbers to explore is essential for large scale exploration. 
% \miii{@Vivek: I commented out your addition due to space considerations.}
%\vj{Scanning all 65K ports across 3.7 billion IPv4 addresses — a nearly $2^{48}$ search space — would take 5.6 years with ZMap~\cite{durumeric2013zmap, izhikevich2022predicting} at 1 Gbps, making it seemingly intractable.}

%\miok{{\bf An uninformed exploration can be expensive.}
{\bf Motivation: the need for efficiency.}
We can illustrate the scope and operational constraints of the problem with some approximate calculations.
First, although we experiment with IPv4 in this work,
let us consider both the IPv6 and IPv4 spaces, which consist of $3.4\times10^{38}$  and 4.3 billion IP addresses respectively.
Second,  a full exploration would require probing  all 65K ports,
especially since
C2 servers often use diverse ports~\cite{C2Miner-Davanian-2024, MalNet-Davanian22, paloalto2024newmirai, C2Store-platform}.
%\miok{requiring a smart exploration of the 65K possible ports}.
%requiring 65K ports to be explored to ensure full coverage.
Third, several studies have shown that sending one probe is not enough to guarantee discovery: C2 servers do not always respond to a probe even within the same day~\cite{MalNet-Davanian22}.
%\miii{Assuming X repeated probes,}
As a result, we get to $2.21\times10^{44}$ for IPv6 and $2.80\times10^{15}$ for IPv4 for a single probing session with 10 repeated probes per IP-port combination. 
Note that we may want to repeat this process periodically in practice.
The need for a scalable approach is apparent: we want to reduce the search space both in terms of IP addresses and port numbers. 
We revisit  practical considerations in \S\ref{sec:motivation}. %\mim{I think we should also mention the diverse IoT C2 port usage to justify the 65k port consideration in the calculation. I can add a couple of references/evidence on that.}
% If we want to probe say once a week, this creates a non-trivial burden.
%IP v6 281,474,976,710,656

%\krish{Transition issue: We end with practical challenges and next sentence talks about research challenges. What is the difference?}

%%% Previous work
\textbf{Previous work:} 
To the best of our knowledge, a large-scale IoT-focused botnet probing capability is not currently available, despite several noteworthy efforts.
% \mim{Again, Cyberprobe and AUTOPROBE claim about their scalability}
Previous active probing studies  can be grouped into the following categories: (a) binary activation methods~\cite{davanian2021cnchunter, farinholt2017catch, C2Miner-Davanian-2024, neugschwandtner2011detecting};
and
(b) protocol reverse engineering, protocol emulation and packet replay efforts ~\cite{xu2014autoprobe,fu2021realtime, tripathi2023delays, nappa2014cyberprobe}.
%including some older studies that developed packet replay approaches
There are also passive detection methods using network traces~\cite{entelecheia13,KimCFBFL08, troutman2018detecting, hafeez2020iot, rezaei2019deep}. 
Overall, most prior efforts either: (a) are not designed for scalability, (b)  do not develop an adaptive target-address prioritization strategy, or (c) use a small or fixed number of ports
as we discuss in \S\ref{sec:related}. Further, prior works do not assess the replayability of IoT malware at scale, as we do here. %\mim{could we write (a) .. for scalability due to resource-intensive design, (b) .. an adaptive target address prioritization, (c) .. designed on the convention of PC-based malicious communication.} 
%\mimok{adaptive} IP address %and port 
%prioritization \krish{Not clear what this means. Comes out of the blue.},
Note that repositories and feeds~\cite{C2Store-Vivek-2023, threatfox, VirusTotal,FeodoTra51:online} 
provide  historical information, but not whether the \host is {\em currently} live.
 %\mim{PAM reviewer questioned. could we add: .. or provide outdated information on a few PC-based families~\cite{FeodoTra51:online}.}
%without the current liveliness status \krish{Not clear what "the current liveness status means.}.
% \zt{These shortcoming should echo the RC.} \mim{reflected shortcomings in RCs}
% \miii{For Mimon: let's create one way of grouping prior work and use this both here and in the next section. I think you do not mention anything about reverse engineering efforts, see Sina's papers he was mentioning some such studies.} \mim{(i) I assumed Cyberprobe and AUTOPROBE are examples of reverse engineering efforts. Not sure, isn't active probing also falling into reverse engineering? (ii) Which one will be (d) category here, as mentioned we have 4 main categories?, (iii) we will be describing the same categories in detail in the related work, right? (iv) won't the work be cited in related work section?, for C2-Miner and CnCHunter, will I re-cite them again in related work?}
%\miii{Revisit citations} \mim{I will take care of it!}
%\miii{This paragraph should be either removed or merged with the above or merged within the next section}
%\krish{Add something to the effect: Importantly, these efforts all generate information using offline sources, and often do not contain up to date information relating to liveness of such servers.}
%%% Contributions

\begin{figure}[t]
    \centering
    \includegraphics[width=\columnwidth]{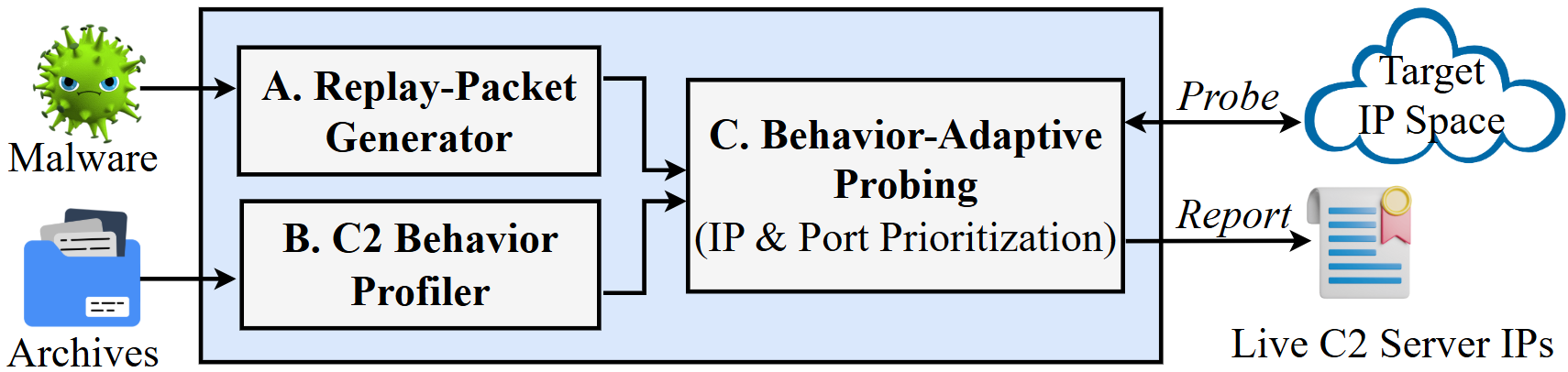}
    \vspace{-20pt}
    \caption{\small{The \name framework and its three modules: (A) Replay-Packet Generation, (B) Botnet C2 Behavior Profiling, and (C) \SmartProbing Probing.} %\miii{Fix module ids if changed}} \mim{changed. also added a C2 after botnet.}
    }
    \vspace{-10pt}
    \label{fig:overall_framework}
\end{figure}

% =========
% NEWEST Contribution

\textbf{Contribution:} We propose \name, a \smartProbing  resource-efficient approach to actively scan for \hosts at scale.
Our work makes the following contributions which are implemented by the modules shown in Fig.~\ref{fig:overall_framework}.
\miok{First, we present the first large-scale assessment of C2 replayability of modern IoT malware, introducing a \threeclass taxonomy. We show that \playbook probing is feasible for many recently-reported  IoT malware binaries across major families (\module \moduleA), which  substantially reduces the probing cost compared with malware-activation approaches (\S\ref{sec:computational})}.
%%
%\mimok{we show that \playbook probing is feasible for many recently-reported  IoT malware binaries and develop a \threeclass taxonomy of their communication behaviors (module A).}
%we present the first large-scale measurement of C2 replayability in modern IoT malware, introducing a \threeclass taxonomy to classify communication behaviors. Our analysis shows that \playbook probing is feasible for many recently-reported  IoT malware binaries (module A), achieving this at a fraction of the resource cost of conventional malware-activation approaches (\S\ref{sec:computational}). 
%%%
%\mimok{we present the first large-scale measurement of C2 replayability across modern IoT malware: we introduce a replayability taxonomy (\Fixed, \Tricky, \Obfu)}. We show that \playbook probing is feasible for many recently-reported  IoT malware binaries and for several major families (module A) \mimok{and it costs only a fraction of the probing resource compared to the malware-activation based approaches, as shown in \S\ref{sec:computational}}. 
%This observation challenges popular belief\cite{} of extensive use of sophisticated encryption which are often focusing .
Second, we propose an efficient probing strategy (module C), which is the first to combine:
(a) a two-level \segment-centric approach to exploit the spatial locality properties of \hosts (observed in module B),
and (b) 
an adaptive \segment re-prioritization in "real-time" based on prior probe responses. 
%\mimok{To our knowledge, our approach is the first probing strategy that re-prioritizes targets  during execution.
%rather than fixing them once from seeds as prior work does.
%}
 Finally, as an engineering contribution, we develop an automated pipeline that continuously collects recently-reported malware binaries, determines their replayability, and extracts their replay-based packets (module A). 
%Our work consists of three intertwined capabilities as shown in Fig.~\ref{fig:overall_framework}.
The focus on efficiency permeates every step of \name: we want to find live \hosts with the least possible cost.

We evaluate our method using 1,842 recently-collected IoT binaries, which we activate and study. We search for these malware in an IP space of 2.5M addresses across 8 Autonomous Systems (ASes).  We highlight our key results.
%Our key results are summarized below.

%with two experimental studies.
% First, %in our \playbook generation, 
% using our streamlined approach,
% we extract the C2 communication traces from 1842 IoT bot malware binaries recently reported by \tifs.
% Second,  we explore a total of 68K IP addresses over a combined span of \caseStudyDuration. 
%We summarize our results in the following takeaways.

%\miok{Note that we focus on IoT malware, as it is an emerging, worrisome, and less studied threat, compared to PC-based malware.
%}
%We use ... \miii{If there are some general numbers we can mention: number of IPs probed (total), number of probes send total -- ballpark or "more than 400K probes total" is ok.} 
%\mim{To report the probe count, I need to discuss on it first}
% \miii{I think a reasonable ballpark is fine, it could include}

% 1842 collected binaries, 1634 (89\%) successfully activated, and 1536 (94\% of activated binaries) established connections with external servers. 

%{\bf a. Creating an effective playbook-based approach.} 

{\bf a. Feasibility: A \playbook approach works for six major IoT malware families.
%\mimok{and their variants}.
} 
% We show that a \playbook approach is compatible with many modern malware families. 
Starting with our 1842 binaries, we activate them and study their C2 communication packets.
%Starting with 1842 recently-collected IoT binaries,
%we activate and observe their communication patterns, which we
% distill into a \playbook with 182 distinct \entries and related packet modification guidelines.
We find that a \playbook approach  works
for {\bf at least 72\% of  our  malware binaries}, %\krish{Not clear what "our binaries" means}
and covers six popular families, including Mirai, Gafgyt, Mozi, and Hajime (see \S\ref{sec:results})
and several of their variants (see \S\ref{sec:res-binary-activation}).
%\miii{Let's add a number and explain somewhere or drop.}
\miok{
For the remaining binaries, only 3\% of our them are not amenable to \playbook engagement, while 
for 25\% of them, results are {\it inconclusive}: either they don't activate or fail to connect with or manage response from their C2 server.
}

{\bf b. Efficiency: higher RoI in finding live \hosts.} 
%We show that our approach manages to outperform baseline approaches by detecting {\bf  70\% more \hosts}\miii{Revisit} in our evaluation.
Our target space has 2.5M IP address, which corresponds to 163.9B %\mim{2.5*1000000*65535 = 163.9B over 162.5B} 
IP:port pairs, and a seed set of 11 historically-known \hosts. Using \name,  we manage to find \mimok{33} new live \hosts  by strategically probing only 1M IP:port pairs \miok{(or 0.0006\% of the IP:port space)}. 
%\miii{Do we want to say: 0.0006\% of the space}
Specifically,  we find {\bf approximately \miok{three times}} more live \hosts 
%\mim{currently, seed 11, we find 34, thus new 23} 
compared to the seed set and {\bf \miok{approximately twice as many}} 
compared to the closest baseline probing approach.
%\mim{20 vs. 34, ignoring 4 seeds, 16 vs 30. So, "at most double new C2s" sounds more suitable to me} 

%\miii{Revisit numbers.}

%%   7 seeds only 3 live ---
%    11 seeds 7 live   --- 
% 
%

%and 122\% more in an emulated scenario using archived data. 
%\miii{The 70\% should also be updated, right? let's do a command again. I thought it was roughly 100\% for both or not? Maybe we can say more than 100\% on two scenarios.}
%We also show that the relative advantage increases as the IP space and probing budgets increase. 
%\mim{We should discuss on the out-performance of our approach. It is not fixed as I stated earlier.} 
%\miok{
%We show that our approach  achieves significant benefits by combining: a) spatial locality, b) historical information, and c) adaptive target prioritization.
%Our approach detects C2 servers with  a \textbf{F1-score of 85\%}.
%\miii{I removed last line of F1 score: Not sure it is dilates the message.}
%} 

{\bf c. Practical impact: Identifying live \hosts  and interesting behaviors.}
We conduct two case-studies where we probe a target space over a longer period.
%to demonstrate the kind of studies that \name can enable. 
%that \name can
%discover live and even unreported 
%C2 servers 
%\hosts.
%shed light on their operational patterns. 
First, we find \totalALL live \hosts
including \unreportedALL not reported by  three widely-used \tifs  %across both case studies.
(64 and 48 new \hosts in our two case-studies). 
%\mim{The 68K is the quantification of IP space, but if we consider 10 ports/IP, it becomes 680K exploration space. A higher number could reflect the effort and scalability.}
%In our longitudinal study,
Second, we  observe that we keep finding new \hosts as we repeat our probing over time: the average response rate is 1 out of 10 probes during a 20 day interval.
%\mim{Could we remove the 1/10 as a flakiness estimation of measurement? I am not sure about the estimation of 1/10 responsiveness. I found C2s responsive in most cases.}
This flake responsiveness suggests that 
% We also find that \hosts responses to probes are flaky: with roughly only 1 in \mimok{3} probes 
% being responded in alignment with earlier studies~\cite{C2Miner-Davanian-2024}.
% \miii{Here we should report OUR result or nothing at all! If not, we should delete also the line below too . FYI: C2Store: "We find that 95\% of these servers responded to fewer than 12\% of the probes (less than 20) over the 7-day period."}
%This observation suggests that 
a thorough study should probe each IP:port pair repeatedly, which further enhances the need for cost-effective probing.

%\mimok{We observe that over 60\% of the C2 servers go down within 2 days of being online. Another thing is that, a C2 server can use different C2 protocols at different times.}
%\miii{Let's find the 2-3 best observations.}

% {\bf Pre-empting some concerns.} We would like to pose the following questions pre-emptively: 
% (a) {\it "Can we detect the C2 servers for all families and their variants?"}, and (b) {\it "What happens if the malware  changes its behavior in the future?"}
% It is our belief to consider useful any capability that: (a) detects new malicious entities, or (b) forces the malware to alter its behavior.
% We revisit these questions in section~\ref{sec:discussion}. \krish{I would remove this paragraph here. It comes across as too defensive.}
% \miii{Valid point: let's think it through. Maybe replace it with a Our Work in Perspective paragraph below?}

{\em Our work in perspective.}
We consider this approach as a step towards ``offensive'' security: we can put botnet operators on the defense.
%by proactively discovering their infrastructure.
First, \name is a critical capability as it enables us to actively and on-demand, probe an IP space from the ``outside'' and without requiring any privileged access to devices or the traffic of the target network.
Second, the efficiency of \name  makes it suitable for large-scale and frequent explorations. 
We discuss uses and impact in \S\ref{sec:discussion}.
% \miok{
% Our approach can support important investigations including:
% (a) providing fresh and on-demand information  to \tifs,
% (b) finding the live \hosts for an emerging malware family, and
% (c) evaluating the health of a network as part of a security audit.
%  We discuss potential applications in \S\ref{sec:discussion}.
%  }
% We anticipate that \name could be used by:
% (a) security researchers, (b) security practitioners,
% (c) operators of \tifs, and (d) network operators.
% The tool could be used for a plurality of studies including:
% (a) finding the live \hosts for an emerging malware family,
% (b) evaluating the health of a network space,
% (c) monitoring a botnet by listening into the issued commands in a longitudinal study.

%\miii{For Michalis: How can we add: why we need active probing, and who will use it and how and connect it with the Discussion section}

{\em Artifacts and sharing.} We will make our platform and datasets available to the  community. For now, related artifacts can be browsed anonymously~\cite{BotScan-sharing}.

\section{Motivation and Related Work}
\label{sec:motivation}

% We present context, motivation and assess related work. 

% \miii{Do we want to revisit the two challenges here?}

% {\bf Part A. Definitions and terminology.}
% We provide a quick overview of concepts that we will use in this work.

   \textbf{Malware binary and family, Bot and C2 server}: A malware binary refers to a program that infects a device and operates as a bot within a botnet. These binaries are grouped into malware families, defined by common functionalities, communication protocols, and propagation methods.
   %\krish{In the intro you use the term binary. I think we need clarity there like what is provided here (or at least make clear it is malware on a Bot - and not a C2.}
   %\miii{Added "bot malware" in the introduction}

   %\miii{Let's delete: Grouping malware into families helps in understanding their behavior and designing effective mitigation strategies.}

A bot is a compromised device infected by malware which makes it a part of a botnet. 
In a botnet, a C2 server is a device  with an elevated role to act both as the contact point for new bots and  command center for the botnet. 
In a \hierCnC botnet
%\krish{botnet?} \mim{corrected bot -> botnet}, 
the role of a C2 server is  clearly defined.
As we said earlier in a footnote, the term \host is being used to include nodes of a 
\ptopCnC botnet~\cite{nappa2014cyberprobe,xu2014autoprobe}, as long as these nodes can act as gateways to the botnet: they listen at an open port and "welcome" new bots.
We use the term {\bf \host}  in this more broader definition.
%to refer to  devices that function as a communication gateway to a botnet, thus referring to both \hierCnC and \ptopCnC bots listening for new bot members at an open port.  
%\krish{The concept of a "gateway to a botnet" is unclear.}
% In our results, we often report these two types of botnet \hosts separately.
% \miii{We don't need this last line, do we? We can always provide more detail if we choose to.}
 % \mim{could not resolve "C2" terminology.}
 % \miii{I think it is good}
% \miii{Let's make sure the above makes sense...}
%control bots that "report" to it. It acts as the operational hub by sending commands to its bots.
   We use the term {\bf botnet communication protocol} to describe the 
   %communication \mimok{policy} of engagement
packet interaction between \hosts and bots. 
   %\krish{It is still unclear what is a C2 server in a P2P botnet.} \mim{@Prof. Michalis, I also think we need to tweak one/two lines here to introduce the C2ness of P2P botnet more clearly.}

 % \krish{I think the following comes out of the blue. This is your contribution. It should not be part of motivation and related work -even before you motivate. Move it to a section that provides an overview of your approach.}

  %\textbf{Network probing}: We use the term probing to refer to the exploration of the devices and components of a network. We use the term to refer both to: (a) general purpose probing, such as finding open ports, and (b) probes that want to detect whether a C2 server is hosted by a device.

%\miii{@Michalis: let's revisit where to place the below}
\smallskip
\noindent {\bf Related approaches.}
\label{sec:related}
 % There have been relatively few efforts that address the problem of interest.
 Prior efforts can be grouped into the following  categories.
% Comparing them with \name is shown
 % in Table~\ref{table:comparison_of_works}.

\begin{comment}
    \begin{table}[ht]
\centering
\def\arraystretch{1.5}
\resizebox{\columnwidth}{!}{
\begin{tabular}{|l|c|c|c|}
  \hline
  \rowcolor[HTML]{DADAF8} \textbf{Approach} & \textbf{Robustness} & \textbf{Scalability} & \textbf{Intelligent targeting}\\ 
  \hline
  C2Miner & \cellcolor[HTML]{98FB98}\checkmark & \cellcolor[HTML]{FF7F7F} \xmark & \cellcolor[HTML]{FF7F7F} \xmark\\
  \hline 
  CyberProbe & \cellcolor[HTML]{FF7F7F}\xmark & \cellcolor[HTML]{98FB98}\checkmark & \cellcolor[HTML]{FF7F7F}\xmark\\
  \hline
  AUTOPROBE & \cellcolor[HTML]{98FB98}\checkmark & \cellcolor[HTML]{FFBA08}$\bullet$ & \cellcolor[HTML]{FF7F7F}\xmark\\
  \hline \hline
  {\bf C2Scanner} & \cellcolor[HTML]{98FB98}\checkmark & \cellcolor[HTML]{98FB98}\checkmark & \cellcolor[HTML]{98FB98}\checkmark\\
  \hline
\end{tabular}
}
\caption{Comparison of our approach to indicative solutions from different families of approaches: \checkmark  \space
indicates high quality or performance, while \xmark \space denotes limitations or challenges in the respective category. \miii{Are we going to add Port targeting too as feature?}
}
\label{table:comparison_of_works}
\end{table}
\end{comment}

  \textbf{a. Active probing with binary activation.} 
  Many active probing methods rely on activating malware binaries and using them to contact servers with two different philosophies.
  %The methodswith two different flavors.
  First, several methods simply observe the activated malware hoping that it will connect with a live C2 server~\cite{antonakakis2017understanding, farinholt2017catch, nadji2011understanding, neugschwandtner2011detecting}, 
  without the option  to actively scan a space.
  %\krish{Why?}
  Second, some recent methods~\cite{C2Miner-Davanian-2024, davanian2021cnchunter}, activate the malware, intercept, and re-direct the  communication packets to a target of their choice.
  We use the most recent work, C2Miner~\cite{C2Miner-Davanian-2024} as a representative, 
  \miok{which {\bf differs from our work substantially}: (a) it relies on  malware activation (whose resource-cost is quantified in \S\ref{sec:computational}); 
  %\mimok{for probing each IP:port}; 
  (b) \miok{it  does not claim or attempt to develop an adaptive IP:port prioritization strategy.}} %it does not claim or attempt to develop an adaptive target-prioritization strategy as this work.}
In addition, some platforms claim to actively collect botnet information without:
(a) revealing their probing approach, so we cannot compare, and  (b)  continuous updating~\cite{FeodoTra51:online}.

\textbf{b. Active probing with protocol emulation.} 
Some efforts reverse engineer the communication protocol by analyzing the traffic~\cite{su2018detecting, fu2021realtime} or binary code~\cite{xu2014autoprobe, borzacchiello2019reconstructing}. 
The task is not easy~\cite{De2017Botnetprotocol, palo_alto_2},  
especially if we consider that hackers can vary their compilation tools and parameters.
% complicating the analysis of the binary.  
AUTOPROBE~\cite{xu2014autoprobe} is a representative method from this group which uses symbolic execution on the malware binaries.
%\krish{No need to put (a) and (b) for these types of claims. Also, I don't understand what you mean by "potentially sensitive to changes. Please make it precise. Is it "C2 server interactions are dynamic and are not captured?"}. 
%Some other efforts follow a more light-weight approach, where they lean more towards pure "replaying" of packets without trying to understand nuances and context. 
Other efforts~\cite{gu2009active, nappa2014cyberprobe} focus on replaying of packets without any customization effort and often in a small set of specific ports.
%Albeit simpler \krish{It is not clear it is simpler compared to what; moreover symbolic execution is complex - not simple?}, 
For example, some packets may contain custom information in their payload, such as the client’s IP address, port number, and timestamp, as we show in Fig.~\ref{fig:information-sent-by-malware-binary}. If the values are not properly set,  
a C2 server could detect the inconsistency and refuse to respond~\cite{xu2014autoprobe, crowdstrike2024}. 
As we will see later, a non-trivial percentage of our 
%15\% to be exact
binaries generate packets that contain such variable pieces of information.
%%% Remove
%% MF: deleted may
%require customization. 
%\mim{Could we write: .. that contain the varibale information ? ``..that require customization'' seems to be a strong claim that the information needs to be changed to elicit a response. In the experiment, we can see, the C2 servers still respond even without changing this information. The references given in the earlier line were basically for PC-based C2s.}
%These approaches could suffer from low success rates, if the packets require some context-sensitive customization~\cite{xu2014autoprobe}.
%\krish{Do we show that they result in low success rate? Otherwise this claim seems to have no basis.}
%\miii{We don't: but I propose we leave it for now, if we have time we can rethink it. It is a plausible argument}
% \miii{Can we cite one more works that support this? A Tech Report or blog?} \mim{I did not find any which discuss in this context.}.
%\miii{@Mimon: do we have any indication that sending "stale" packets leads to failure? If so, let's mention it!}
%\mim{In many cases, stale packets succeeded but it is stated in AUTOPROBE: "If the
% values of these fields do not match with the sender’s, the C\&C
% server can detect such inconsistency and refuse to respond."}
CyberProbe~\cite{nappa2014cyberprobe} is a seminal work that uses such simple packet replaying to detect PC-based malware servers focusing on ports 80 and 8080. 
%\mimok{Apart from this, CyberProbe relies on trace-driven stateless request \& response pair replay with signature-generation constraints focusing PC-based malware, limiting its applicability in the IoT landscape.}
%\miii{I am not sure I understand: RRP is not defined and rigid signatures does not really make sense: I know what you want to say, but a reviewer won't. Let's revisit this carefully!}\mim{Does it sound good now?}
%\mim{Do you think, to establish the difference, we also need to mention CyberProbe's issues (a) RRP replay: sending a request and hoping it will elicit a response, but Mirai needs multiple packets in correct order (stateful) to be sent to elicit a response, (b) it's assumption of signature token to be at least 5 bytes -- many IoT C2 communications contain C2 responses with less than 5 bytes in length.}
Note that both CyberProbe and AUTOPROBE were focused on PC-based malware as IoT malware had hardly appeared back in 2014. 
\miok{Overall, none of the prior works has systematically measured and taxonomized the replayability of IoT malware for probing \hosts.} Further, they do not develop an adaptive IP prioritization strategy.% as we do here.
%\mim{Should we mention ``PC-based'' here to reinforce the difference and also their limitation of port prioritization realization?}  
% \miok{However, these works~\cite{nappa2014cyberprobe, xu2014autoprobe} %\krish{Which prior works?}\mim{added citation } 
% rely primarily on a static probing target selection grounded in historical data, without optimizing the probing  strategy or dynamically adapting to the success or failure of the probes.}
%\miii{@Mimon: check the above, as I touched it}
% \mim{should we glorify them as ground-breaking? Cyberprobe citation count: 74, AUTOPROBE citation count: 64.}
%which could be readily used nowadays.

% \krish{(older comment) Do we not also do some form of replay? Highlight difference between Cyberprobe. Symoblic execution does not scale. I don't know about brittleness/inaccuracy - if so need to cite reference.
% }

Several efforts focus on determining the liveness and operational status of an IP address~\cite{durumeric2013zmap,censys,shodan}. 
%ZMap Censys, and Shodan
% Some of these efforts are commercial and do not fully reveal how they operate or collect the information.
Obtaining a liveness status of an \textit{individual} IP address is a significantly different problem from identifying live \hosts.

\begin{figure}
    \centering
    \begin{tcolorbox}[
        colback=gray!5!white,
        colframe=gray!75!black,
        boxsep=2pt,
        left=2pt,
        right=2pt
    ]
    \small
    \texttt{[31m\textcolor{ForestGreen}{Sokkyo} [37m| [31m\textcolor{blue}{device} [30m\textcolor{red}{10.0.2.15} [37m\ [31m\textcolor{blue}{arch} [30m\textcolor{red}{ARM4} [37m\ [30m\textcolor{red}{Linux} [37m\ [30m\textcolor{red}{Dropbear}}
    \end{tcolorbox}
    \vspace{-15pt}
    \caption{\small Example of a Gafgyt C2 packet payload (in ASCII) of the \Tricky category: it contains invariant fields (in \textcolor{blue}{blue}), and context-specific variable fields (in \textcolor{red}{red}).}
    \label{fig:information-sent-by-malware-binary}
    %\vspace{-20pt}
\end{figure}

%\miii{We could move (c) and (d) below to appendix!!!}

\textbf{c. Passive approaches and synergistic efforts.}
Passive methods analyze network-flow~\cite{fuller2021c3po, gu2008botsniffer, jacob2011jackstraws, troutman2018detecting, vandetecting, hafeez2020iot} or botnet behavior to detect botnet traffic~\cite{spitzner2002honeypots, wang2020iotcmal, sethia2019malware,anghel2023peering, trajanovski2021automated, tanabe2020disposable, almazarqi2021profiling, blaise2020botnet}, but require cooperative network permission and cannot query arbitrary IPs. We discuss complementary efforts in Appendix A.
We discuss these efforts in Appendix~\ref{sec:Appendix-Related_work}.

\noindent {\bf Design Goals, Motivation, and Challenges:}
% We discuss the challenges, specify the features of an ideal probing approach and compare with previous approaches. % qualitatively.
% {\bf Challenges and the breadth-vs-depth trade-off.}
Each of the two research challenges in the introduction are hard in their own right. Together, they make  designing an effective and low cost solution difficult.
Any budget-conscious approach needs to grapple with several inherent trade-offs between the breadth and the depth of the exploration.
First, we have \miok{to select} the number of distinct IP addresses that we want to explore.
Second, we have \miok{to select} the number of ports to explore for each address.
%\krish{associated with any given?}.
Third, we have \miok{to select}   the number of distinct botnet communication protocols that we can try on each port.
%\krish{How do we circumvent this? Comes out of the blue. Also not clear what we mean by protocols.}
%playbook \rulesets.
Fourth, we have \miok{to select} the number of repetitions  for the same probing (same packet, IP address and port) to increase our confidence in the results.
For example,
given a limited budget, we need to strike the balance between:
(a) exploring many distinct IP addresses, and (b) exploring  many ports for each IP address. % \krish{Not clear what we mean by "exploring each IP address exhaustively}.
Note that this tuning of the operation can depend on the purpose of the study. 
For example, a study on the longevity of \hosts could focus on fewer \hosts  but with more intense probing rate per \host and over a longer duration. 
  
  \textbf{\textbf{Feature \#1}. Computational efficiency}:
  Using as little as possible computational resources is a desirable property for a probing method.
  We argue that prior methods that rely on activating the binary in a sandbox to probe a target address are computationally expensive, which we quantify in \S\ref{sec:computational}. %\krish{Don't need the first sentence. We can just start with "Establishing ..."}
  Establishing a sandbox, injecting and activating the malware increases overhead: (a) substantial computational effort, and (b) non-trivial delays as we need to wait for the malware to  initiate contact with its \hosts by itself.

\begin{table}
\centering
%%\vspace{-20pt}
\def\arraystretch{1.3}
\resizebox{\columnwidth}{!}{
\begin{tabular}{|l|c|c|c||c|}
  \hline
  \rowcolor[RGB]{90,240,110} \textbf{Feature} & \textbf{C2Miner} & \textbf{CyberProbe} & \textbf{AUTOPROBE} & \textbf{\name}\\
  \hline
  \cellcolor[RGB]{253, 219, 121} Adpt. IP Priority & \ourYelllow &  \ourYelllow & \ourYelllow & \ourGreen\\
  \hline
   \cellcolor[RGB]{253, 219, 121} Port Priority & \ourYelllow & \ourRed & \ourRed & \ourGreen\\
  \hline
  \cellcolor[RGB]{253, 219, 121} Comput. Effic. & \ourRed & \ourGreen & \ourGreen & \ourGreen\\
  \hline
  \hline
  \cellcolor[RGB]{253, 219, 121} Overall & \ourRed & \ourYelllow
  & \ourYelllow & \ourGreen \\
  \hline
\end{tabular}
}
\caption{\small{\name is comprehensive \& efficient: comparison with representative active probing approaches. \mimok{\ourRed = Unsupported, \ourYelllow = Partially supported; \ourGreen = Fully supported.}}
}
\vspace{-25pt}
\label{table:comparison_of_works}
\end{table}

\textbf{Feature \#2. Target address prioritization:} 
An ideal probing approach will probe IP address strategically to maximize the likelihood of success.
Without an a priori determination of promising targets for probes, methods are bound to be wasteful. The challenge lies in identifying and focusing on the most relevant IP sub-spaces to maximize the RoI: the number of live {\hosts} per number of probes. We saw the expansiveness of the search space in the introduction.\looseness=-1
  %\mim{again, later we use the term budget as the number of IPs we can explore, not probes.} 
  %This feature relates to Challenge 2 in the introduction. \krish{Again no need to refer back I feel.}

%\miii{Maybe revisit here the impracticallity of brute force}

\textbf{Feature \#3. Target port prioritization:}
An effective probing approach will need a strategy to probe the ports that will maximize success.
For context, we mention that: (a) there are 65K ports, and (b) botnets seem to utilize this vast port space.
As result, a probing approach needs to strike a balance between full port exploration and scalability.
We argue that most previous methods have not developed a full port selection strategy.
For example, CyberProbe probed only 80, 8080 for TCP and 16471 for UDP, which made a lot of sense given its focus on PC-oriented \miok{botnets} in 2014. \miok{In contrast, we measure IoT C2 activity spread across 6,350 ports (\S\ref{sec:port_behavior}), 
which creates the need for a IP:port prioritization strategy.
%indicating a small and fixed port set would have low detection success.
}

%which renders a small, fixed port set ineffective in this setting.}
%\krish{Not clear what are malwarwe protocols!}\mim{corrected as malware-communication-protocols}.
% \mim{should we hide the name Cyberprobe? Won't mentioning Cyber/Autoprobe frequently seem like our work is highly focused on these works?}

%\miii{I removed Robustness/Agility}
%\textbf{d. Robustness:} Many existing approaches rely heavily on historical data or observed behaviors from more than a decade ago. These factors can limit their relevance for novel malware. 
  %ability to adapt to the ever-evolving and dynamic nature of the threat landscape. 
  %Additionally, methods that attempt to fully reverse engineer the communication protocols are often sensitive to changes. For example, the effectiveness of the methods that analyze the binary could be affected by the compilation techniques which could make the code harder to reverse engineer.
%\miii{Still we may need to reframe, maybe flexibility, generalizability..., or drop.} \krish{Is this robustness? I think it relates to "staleness and handling temporal dynamism}"

We compare indicative methods from the different families of approaches in Table~\ref{table:comparison_of_works}. 
\name is comprehensive with a focus on large-scale active probing.
The overall scalability and efficiency are supported as
\name consists of both \mimok{adaptive} IP prioritization and port prioritization strategies, while none of the other solutions has a fully developed strategy for optimizing target address and port selection.
Second, \name uses a \replay-based approach, reducing the cost to craft probing packets compared to requiring an activated binary like C2Miner~\cite{C2Miner-Davanian-2024}.

\section{Overview of Our Approach}
\label{sec:overview}
%\label{sec:overview}

%\krish{How can we use these patterns to effectively probe large IP spaces online, to identify live C2 servers.}
%\krish{I think we should make these questions more precise.}

% These three \modules together address the two challenges from the introduction:
% (a) ``how to probe?'', (b) ``where to probe?''.
% \krish{I would again focus on "small set of interactions" and "targeted probing to accomplish high hit rates." I would also refrain asking the reviewer to go back to intro.}

%\miii{Let's align the problem statement with the introduction}

At a high level, 
the input to the problem is an IP space that we want to explore for \hosts.
Our approach analyzes and leverages information from: (a)  malware binaries, and (b)  observed IP addresses of \hosts from publicly available sources.  
%\krish{At this point the reader is left wondering what information you are referring to. If information is already publicly available, then what is the goal beyond?} \mim{I think, explicitly mentioning the \textit{``historical data on \hosts''} can clarify the issue.}
The output is an efficient strategy for exploring the given IP space.
\name consists of three synergistic \modules, each addressing a question:

\noindent
 \; \; {\bf \Module \moduleA:} We want to identify the packets with which we can engage \hosts for a newly-captured binary. We need to: (a) determine if the malware binary is amenable to a \replay,
and if so, (b)  extract the packets that we can \replay to detect its \hosts.

%\krish{How to craft a minimal set of packets that can engage with C2 servers.}
%\miii{SK  suggested to add "minimal" }

\noindent
\; \; {\bf \Module \moduleB:} We want to ingest  historical information of \hosts and identify behavioral patterns that we can use to optimize our effort.

%\krish{Can we extract crucial C2 patterns that can help efficient probing yielding high hit rates? }

\noindent
\; \; {\bf \Module \moduleC:}
We want to develop a smart and adaptive probing strategy to maximize RoI in finding live \hosts. 
We combine and leverage the information from the first two \modules.

\vspace{-5pt}
\subsection{\Module \moduleA: Determining \replay \entries} 
\label{sec:Method-ModuleA-replay}
%\vspace{-5pt}
 In a nutshell, we collect newly reported malware binaries from \tifs. We activate them and extract their communication packets to their \hosts and generate \replay \entries used for probing. Note that not all binaries can generate \replayable \entries.
The processes hide several subtleties.
\mimok{First, we introduce a taxonomy for botnet protocols w.r.t. their amenability to \playbook probing, which we use in \S\ref{sec:results-validation-replay}.}
%%% amenability to replay probing
%First, we introduce the following taxonomy for botnet protocols.
 %categorize C2 communication protocols
%into the following groups:

\noindent
 \textbf{a. \Fixed} protocols use packets that do not require any change to be replayed for probing.
%\mimok{Fixed \rulesets follow consistent patterns and can be used in probing directly without modifying their payloads.}.
%\miii{Let's use full sentences. This is an important concept say let's discuss it properly. I wonder if we can even move this in Background, we are not inventing this, we are quantifying it.}

\noindent
 \textbf{b. \Tricky} protocols use packets that require some modifications in their payload to ensure that the \hosts do not perceive inconsistencies and decline to respond.
 %\miok{containing fields, such as the operating system or the current date.}
 %to ensure that the \hosts will not perceive inconsistencies and decline to respond.
%. For example, the payload may include the current date.

% \mimok{The payload typically contains variable fields that need to be adjusted according to the runtime environment for effective probing.}.

\noindent
\textbf{c. \Obfu} protocols \mimok{are all those that are not \Fixed or \Tricky}. 
The packets use obfuscation or encryption method and \miok{at the same time,} vary substantially across binaries executions.
%Therefore, to reverse engineer the packet is not straightforward.
\miok{These protocols are not amenable to a \playbook approach. 
}

With the taxonomy, we follow the steps below, which are {\bf fully automated}. 
 % Note that the {\bf results and evaluation of each step are presented in \S\ref{sec:results}}.
 
 %\mim{We should clearly state that the detailed data statistics of each step are shown in the evaluation and result section, and add a one-liner result summary at each step here.}

{\bf Step 1: Collecting malware binaries.} 
%\miok{
We collect recently-reported malware binaries from \tifs
Using recent malware binaries will increase the likelihood that they will manage to connect to an active \host once activated. 
Here, we use MalwareBazaar~\cite{MalwareBazaar}. 
%Naturally, our approach is not dependent on MalwareBazaar, as we can use any \tif with open and timely reporting of malware binaries.
%}
% \vj{
%Our work focuses on critical IoT binaries selected using industrial standards~\cite{alrawi2021circle, davanian2024c2miner, jain2023c2store}.
We determine key information about each binary, such as their family and architecture type, using existing tools~\cite{sebastian2020avclass2, alvarez2014yara, plohmann2017malpedia, alrawi2021circle, oberhumer2004upx}, %\miok{
and we follow the majority rule in case of disagreements. 
%}.  
%\krish{Again, the reader will be left wondering - if this information is available on Malware bazaar, why probe? The reasoning and objective is not coming across up front or here.} 
% \mim{\textit{Malwarebazaar} and the \textit{determined Key Information (such as architecture, family)} do not provide the C2 server IP: Port. The goal of probing is to detect C2 IPs in the wild/given space. Should we explicitly mention it here?}
% \miii{Agreed: we can leave this for Srikanth to see}

%\mim{In the case of different opinions of the tools, we agree with the majority's opinion.}
%miii{@mimon: can we add half a line of how we consult all these tools, do we use some kind of majority rule or typically only one source reports info about a binary?}
%and sometimes VirusTotal's comments for added confidence~\cite{C2Store-Jain-2023}.
% } 

% \vj{Between October 2023 and February 2024, we collected samples across various architectures covering major IoT malware families such as Mirai, Hajime, and Tsunami.}

{\bf Step 2: Activating the malware binaries.}
We build a binary activation capability on top of the
QEMU sandbox with best practices to improve the binary activation rate~\cite{bellard2005qemu, darki2020riotman}.
% We adhere to the following activation procedure.
%First, 
We let the malware run for three hours and capture the network traces 
%\mimok{produced by the malware,}
using \texttt{tshark}~\cite{combs2012tshark}.
%{\bf Multiple binary activations in time and space.}

\begin{comment}
   Second, we repeat the activation across {\bf 27 world-wide locations} using a Kubernetes cluster. 
%\miii{@Vivek: Do we repeat the activation in 27 locations, or when we do the playback we do it from different locaions?}
This allows us to classify the communication pattern as \Fixed, \Tricky or \Obfu. \mim{Reviewer Alert: We have not sync this 27 location activation in the next steps. Should we skip it as I found no evidence from Vivek's data collection? Or, incorporate this in the next steps?}  
\end{comment}

%\krish{Why does repetition help in classification?}
 %\krish{How do we observe? Is this a manual effort. Is there some code that you wrote that isolates comms with C2 server? It seems almost as if "we observe it believe us".  We need detail.}
%}
%\miii{Did we ever see any differences? If not, we should tone it down.}
%Our activation platform records the system calls and captures network traffic in each execution using \texttt{tshark}~\cite{combs2012tshark}. 
%\mim{should we say the activation from 27 different places? not sure about it.}

{\bf Step 3: Identifying the C2 communication packets.} 
% Identifying the bot-oriented communication packets  is not trivial
This task is not trivial as a bot  generates other types of traffic, such as reconnaissance and proliferation.
We adapt and finetune state of the art methods~\cite{darki2020riotman, C2Miner-Davanian-2024}, shown to produce accurate results (see  Appendix~\ref{appendix:C2Miner}). 
By identifying communication packets, we also identify the \hosts that the  binary tries to reach, which we use in the next step.
%Disambiguating C2-specific traffic amidst this activity presents a significant challenge, which we discuss next.
%At a high level, 

%\miok{We omit further details due to space limitations, and because we do not claim to make algorithmic contributions in this step.  
%}

\miii{If we have space we can add: For completeness, we provide a high-level summary of how the methods~\cite{darki2020riotman, C2Miner-Davanian-2024} operate. First,... Second,... Third... }

{\bf Step 4: Determining Replayability.} 
\miok{
% We assess the replayability of the C2 protocol of a malware binary in the following manner. 
By now, we have identified: (a) the packets, and (b)  target \hosts that the bot attempts to communicate with.
}

{\em a. Packet analysis.} We analyze the communication packets and distinguish two cases:
(a) the packet seems to have connection-specific information that we can easily detect with a simple parser, such as bot-side IP addresses or hardware information as shown in Figure~\ref{fig:information-sent-by-malware-binary},
and 
(b) the packet does not contain such information.
In the first case, we use \textit{Scanning function} (see  \S\ref{sec:scan-function}) to craft the appropriate packet.
In the second case, we simply replay the captured packet.

{\em b. Reconnecting with the \host.}
We attempt to connect to the same target \host from a different IP address by reusing  appropriately modified packets.
If the server responds and engages, we consider the malware binary as: (a) \Tricky, if we had to modify the packets, and (b) \Fixed, if we did not.
If the \host does not engage, we place the  binary in the \obfu category.

We compile the data from Module A into our 
{\bf replay packet generation dataset  \textit{(RP-DS)}}. The \textit{RP-DS} contains 1,842 malware binaries across six major IoT families, along with their key information; 1371 captured traces with their packets, and classified into three replayability categories.

\subsection{\Module \moduleB: \Host behavior profiling}
%%\vspace{-5pt}
The goal of this \module is to identify behavioral patterns of 
\hosts to form an efficient probing strategy.
This profiling requires continuous investigation to keep up with the evolving behaviors of the different families of malware.

{\bf a. Collecting \host data.}
We consider and consolidate all possible available datasets of detected \hosts including: (a) our own measurement studies,
and (b) third party sources, such as \tifs ~\cite{MalwareBazaar, threatfox} or aggregators, such as C2Store~\cite{C2Store-Vivek-2023}.
In its most general form, this information could include:
(a) IP address, (b) port, (c) time of observation, (d) malware family.
However, in practice, we may only have a subset of this information. 
We add this historical C2 data into our  \textbf{behavioral profiling dataset (\textit{BP-DS})}, which contains 18.2K unique C2 IP:port pairs across five popular IoT families over five years.

%\mim{There might be a potential question about the data specification. Should we show the statistics of the data in a table \& figures in resultt section to make it more presentable? It would be great if we could show some data in support of the three points mentioned in the following paragraph and the port prioritization.} 

%To analyze the IP and port-level behavior patterns of malicious \hosts, we gather real-world %C2 server and malicious 
%botnet \host  [IP:port] data from: (a) C2Store~\cite{jain2023c2store}, a super-aggregator of many sources of information, and (b)  our own experiments including malware activations and our probing results.
% \miii{Any more info on what sources, and what we do? } \mim{dynamic analysis of binary from i) playbook generation, ii) Sina's work}

{\bf b. Profiling behavioral preferences and locality.} 
\label{sec:methodology-spatial-locality}
% \zt{BTW One thing I noticed is that, there's huge space between the previous paragraph and this subsubsection. You might want to use a different macro that doesn't create additional spacing.} \mim{agree, I will take care of the spacing once comments are addressed.}
We study our \textit{BP-DS} dataset to identify patterns in the spatial distribution of \hosts.
%As in many practical situations,
We find that \hosts exhibit family-centric spatial preferences~\cite{C2Miner-Davanian-2024,C2Store-Vivek-2023}.
First, \hosts are distributed in the IP space in a fairly skewed way.
Second, different malware families exhibit different location preferences.
Third, there seems to be a locality in the placement of C2 servers.
All of the above are good news in the sense that they can help us optimize a probing exploration.

%as they tend
%o be "near" each other in the IP space 
%and \mimok{use specific ports frequently}
% Michalis ports have nothing to do with locality though...
%\zt{OK so this study module B is indeed inspired by some existing observations. If so need to briefly mention in intro}
%The goal is to use this locality feature to increase the success of our probing.

%\zt{So is your point: ``although people have been made such observations, but there is lack of quantification of the locality.''?}
%\miok{

We propose to exploit historical information as follows. 

{\bf 1. Exploiting address-space locality.}
The question we want to answer here is the following:
{\it What is the size of the neighborhood of a known \host that we should explore to increase the chances of finding more live 
\hosts?}
%\zt{Some quick reasoning here - basically mention that this will be useful for module C. Also is this size the same for each family or different across families?}

Our own measurements and earlier studies suggest that there is locality in the distribution of \hosts~\cite{C2Store-Vivek-2023}, but it has not been quantified systematically.
Here,
we use the  term  {\bf \segment of size $\segk$} to refer to the subnet of prefix /$\segk$ to refer to the neighborhood of an IP address.
Consider the following scenario in a probing study:
given a reported \host, what is the size of its neighborhood that we should expect to find more \hosts?
This information will inform our IP prioritization strategy as we will see later.  %how we will prioritize the addresses to explore in order to increase the chances of finding live \hosts. \krish{Please check last sentence. It does not make sense.}
%In our probing, and assuming that there is locality, we want to prioritize the exploration of the neighborhood, segment $\segk$, of an identified \host.

% We want to strike the following balance. On the one hand, if the \segment size is small, we may fail to find neighboring live \hosts.
% %\zt{Following my previous comment: If you don't mention how segment is useful for module C, people don't get why it is associated with likelyhood}
% On the other hand, if we explore a large \segment, our hit rate will most likely decrease and the exploration of one \segment will be costly in terms of packets.
%The latter is important if we consider a tight probing budget: we may 

%{\bf The \localDensity metric.} 
Following a \segment-centric approach,
we introduce the metric {\bf \localDensity}, which we define as follows.
For a given \segment, we count the percentage of IP addresses 
that have been reported as \hosts and
%This percentage serves as our locality metric. 
leverage historical information.
We examine \host IP addresses across different \segment prefix sizes, starting from /31 and moving to larger subnets (smaller prefix size). 
%We consider subnets that have at least
For each  prefix size, we calculate the \localDensity, 
which helps us determine the right \segment  prefix size {\it $\segk$} \ \miok{empirically, rather than assuming a fixed granularity.
}
%a priori, as prior work does}.
%% I am not sure we want to put down prior work, let alone prior work also suggests /24 anyway!
\miok{We describe our study in 
\S\ref{sec:results}.
}
%as we show \S\ref{sec:results}.
%\zt{This is really unclear - so you do have local density for each segment size. How do you design $\segk$? I thought that's your major contribution here to strike the balance}

% \mim{We need to revise this whole sub-section -- we are no longer quantifying the segment size. The plot highlighted /24 as the highest concentrated block seems to be incorrect.}

%}
% \mim{I think ... a bit more clearly.}
% \miii{@Mimon: Great comment. Does it read better? Note I introduced \segments in overview}\mim{yes, cool!}

% The locality density strikes a balance between two factors:
% (a)  increases the chances of finding more malicious \hosts,
% (b) 
% By considering factors such as locality tendencies and the impact of subnet size on malicious-priority computations, we determine the right subnet size {\it N}. %This optimal size ensures that the chosen subnet covers a significant number of C2 IPs, while keeping the computational overhead of subnet priority manageable.

{\bf 2. Exploiting  port persistence.}
\label{sec:methodology-port-selection}
\miok{
Identifying how many and which ports to explore is critical for efficient probing given that there are 65K ports.
We propose to leverage historical information, when available. 
We order ports based on the frequency with which a port is used by \hosts as reported in the available database and traces.
%by counting 
%To analyze the most popular C2 ports, we focus on port usage frequency.
%the number of incidents in which a \host  used that port.
%We rank the ports in descending order based on their usage counts.
We keep track of the malware family, and the date of the reported incident, if such information is available. 
% \miii{Michalis idea to add: on time-sensitive reputation - if we have space}
%For longterm archival data, we can consider the age of a reported incident, and give higher weight to port numbers of recently reported incidents.
%Here, we combine our expectation f
Some port numbers are used repeatedly, which we can exploit in our probing as we discuss later.
}

\subsection{Module \moduleC: \SmartProbing probing }
\label{sec:methodProbing}
\label{sec:intelligent-probing}
%\vspace{-5pt}

% \miii{Mimon: the paragraph below is out of place here:}
% \mimok{One limitation of prior work~\cite{nappa2014cyberprobe, xu2014autoprobe} is that it relies solely on historical data to guide locality-based probing. Such approaches do not incorporate the outcomes of ongoing active explorations to reprioritize the search space. Given the rapid churn of malicious host addresses, a purely static, history-driven strategy is incompatible with timely detection. In contrast, our approach adapts target prioritization dynamically, updating locality estimates and rankings at runtime based on the latest probing outcomes.} 

% In a nutshell, our approach combines the following aspects:

% {\bf 1. What:} We leverage an extensive library of replayable traces. This library will be continuously update with newer traces as we collect and analyze new malware binaries.

% {\bf 2. Where:} We exploit historical behavioral patterns of \hosts in terms of their spatial placement and their use of port numbers.

% {\bf 3. How:} Our approach specifies how we can adapt our target prioritization dynamically, updating  ranking and port preferences at runtime as we discover new live \hosts.
% % It adapts in "real-time" to the discovery of new live \hosts during our probing. As we discover,  new live servers,  we re-prioritize our targets, as we will see below.

How can we translate the information of the first two \modules into an actionable probing strategy? To answer this question, we combine three key concepts. 

%We want to transform information and historical observations into an algorithmic framework.
% We highlight the key concepts and novelty of our  strategy with the following points below. 

\indent 
{\bf a. Two-level \segment-centric exploration:} We decompose the IP space into \segments, and use the term {\bf reputation} to refer to the desirability of a \segment for probing.
 We exploit locality at two levels of abstraction: (a) the activity within the \segment ({\bf \intraSeg reputation}) %\mim{could we add any hypothesis why we are following this two-level approach?}
and (b) the reputation of nearby \segments ({\bf \interSeg reputation}).

\indent 
{\bf b. Defining reputation:} In calculating the \intraSeg reputation, we consider  the number of: (a) confirmed live \hosts, and (b) historically reported but currently not responsive \hosts. %\mim{could we justify why?} 
% We use the term {\bf online IP address} to refer to an address that has a responsive device. We use the term {\em online} on purpose to differentiate it from the term {\em live} that we use of a \host. 
%The rationale here is that, all other things being equal, we prefer to explore a space that seems to have more responsive IP addresses. \mim{IP or C2 IP?}
The \interSeg reputation captures a \segment's proximity to neighboring \segments with high \intraSeg reputation. 

\indent 
{\bf c. Adaptive re-prioritization:} We develop a strategy to adjust target prioritization in "real-time" based on prior probe responses.  Namely, finding live \hosts and responsive benign IPs triggers a recalculation of \segment reputation,  and thus, the choice of the next target \segment. 

We define all the above metrics and our algorithm rigorously below.
As shown in \S\ref{sec:motivation}, we are not aware of any prior approach that proposes an adaptive re-prioritization capability, as we do here.
%a prioritization strategy, let alone one with an adaptive re-prioritization capability.

%\zt{Probably this is due to reorganization, but the current structure is a bit weird to have the overall step and then some details.
%My suggestions:
% 1. First introduce the major challenges in utilizing the locality insight. E.g., how to design ranking function, etc. Need to put them in plain language.
% 2. some other challenges, if any. 
% 3. Now put everything together, basically your steps. I suggest putting everything into an algorithm with some simple explanation.
% } 
%\mim{I have added a few lines at the top of this sub-section.}

%\miii{I added the overview above great idea ZT! If we have space: we can introduce problem and challenges: we have info, but how do we translate it into actionable strategy?}

%either: (a) consider an exhaustive scanning and thus did not propose a prioritization,

%as follows.
%At a high level, our approach consists of the following main components.

%\subsubsection{\textbf{Our exploration framework.}}
%We present some key aspects and design choices of our approach.

% {\bf a. Establishing a \segment-level granularity in probing.} 
% We divide the target address space into \segments of prefix-size $\segk$
% and we use  \segments  as the minimum unit for prioritizing our exploration.
% We already discussed the importance of selecting the right value for the prefix  (we select $\segk=24$ in the next section).

% %, as we defined earlier. \miii{\S ?}

% {\bf b. Reputation-based \segment prioritization.} 

% {\bf c. Port exploration and prioritization.}
%\vspace{-10pt}
\subsubsection{{\bf Part 1: The \phases of our probing approach.}} %\mim{should we ignore "Part 1", i can't see other part anywhere in the paper.}
Our probing  algorithm (detail in Appendix~\ref{appendix:probing-algorithm}) takes as input a target space, a set $Sus$ of historically reported IP addresses, which could be empty, a probing budget, %\mim{should we remove it?} \miii{We can't remove it: when do we stop then? -- IF we have not defined it yet, let's do so (defined as the maximum number of probing packets)} \mim{I think, the given target space itself can be a limit. The reviewer was questioning that the budget in this era of high scanning capability seems unrealistic.}
 and several parameters governing the prioritization strategy, including the \segment prefix size ($\segk$), neighbors ($K$)  and decay factor ($\lambda$), which we explain below. The output consists of a list of live \hosts.
The probing consists of the following \phases.

{\bf \Phase 1.  Establishing a \segment-level granularity in probing.}
We divide the target space into \segments of prefix-size $\segk$, which are used as minimum unit for prioritizing exploration.
We have already discussed the importance of selecting the right prefix value, as we revisit in the next section. 
%(we select $\segk=24$ in the next section).
%\zt{This 24 is out of nowhere. I'd rather don't bring it up at all.
%}
%\mim{I agree, can we refer to Autoprobe behind our selection of /24?}

{\bf \Phase 2. Initial scanning. }
We start by probing the historically-reported \hosts addresses within our target space, if they exist.
%\mimok{to calculate the intra-segment reputation for the segment}. 
%% HERE we don't calculate anything, we probe the historical reported!
Otherwise, we proceed to the next step directly.

%\krish{Not clear what "online IP addresses" means and how you find them.}
% \miii{Maybe explain what happens if no such information exists: we resort to exploring segments in the order defined by finding responsive IPs ie allocated address space as captured in our ranking function below.} \mim{It's tricky, but can we say in this way that as follows? ``If the IP space is not large enough to hold any candidate/historical C2 IPs, we focus on the port associations of the live IPs to prioritize. We check the liveness of the top C2 ports popular in the malware family of interest. Please note that checking the openness of a port for an IP is not an expensive task -- a SYN-scan can achieve the goal. Based on the findings, we can rank the IPs to generate our candidate/historical C2 IPs list for initial probing.''}

{\bf  \Phase 3. \Segment ranking.} We calculate the reputation of all \segments within the target space and rank them using the  \Segment Ranking
function. The details are explained later.
% We use the term {\bf reputation} for a \segment to refer to the likelihood that the \segment may have significant numbers of \hosts. 
% First, we consider both: (a) the historically reported activity of \hosts, which may not be live anymore, and (b) currently live \hosts detected by our probes.
% %We probe segments using the reputation.
% Second, we calculate the reputation by considering both the \intraSeg and \interSeg reputation as we discuss below.
%\krish{Dangling sentence.}

%\miii{We need to define \interSeg and \intraSeg reputation in this section.}
%\S\ref{sec:segment-ranking}.
%\zt{This is weird. This is exactly section 4.3}

{\bf \Phase 4. \Segment selection. } 
We select the next \segment to probe to be either: (a)  the top ranked \segment, or (b) a randomly selected \segment with a small probability (\textit{r-explore}). The value of \textit{r-explore} can change the selection from strictly by ranking to purely random, and it is provided as an option to the researcher conducting the study. We provide parameter values that yield good results in our evaluation in ~\S\ref{sub:eval1}, which can be used as default choices.
% \krish{Again, it is unclear who will conduct this study and why.  It was never brought up in the earlier parts of the paper.}
% \miii{We added some explanations in introduction and discussion: not sure how else to address it.}
%Note that if we have an empty $Sus$ set, the
%\miii{I am getting sec}

{\bf \Phase 5. \Segment exploration.}
For the selected \segment, we probe all its IP addresses, starting with the addresses closer to known \hosts if exist.
% For each IP address, we select which ports to probe strategically as below.

{\bf \Phase 6. IP address exploration with port prioritization.}
 Once we select an address to probe, we still need to identify which port numbers to probe. We elaborate on this when we discuss our Scanning function below.  %discussed in ~\S\ref{sec:scan-function}.

{\bf \Phase 7. Repeat or stop.} We stop if our budget is depleted (or the desired time is reached, etc). Otherwise, we repeat \Phase 2 and recalculate the ranking with potentially new information on live \hosts from \Phase 3.

\noindent \subsubsection{\textbf{Part 2. \Segment Reputation and Ranking Functions.}}\label{sec:segment-ranking}
%We  dive deep into how we prioritize \segments.
The reputation of  \segment $i$ is based on two primary factors: 
\textit{\intraSeg  reputation (\rInt)}, and \textit{\interSeg reputation (\rExt)}:

\vspace{-10pt}
\[
Rank(Seg_i) = R_{int}(Seg_i) +  R_{ext}(Seg_i, K, \lambda)
\]
%\vspace{-15pt}
%\alpha
% For ease, we assume that we have contiguous \segments and the index $i$ varies from one to the maximum \segment.

\textbf{a. \IntraSeg reputation score (\rInt):} This score quantifies activity within a \segment.
To be comprehensive,
we combine  the following factors in decreasing priority: 
(a) count of confirmed live \hosts ($N_{live}(Seg_i)$), 
(b) the number of suspicious addresses, which are not confirmed as live in our probing ($Sus_{silent}(Seg_i)$), and 
(c) the number of suspicious addresses ($Sus_{down}(Seg_i)$), which have been unresponsive or offline.
%\miii{Nothing to justify: if they don't respond, they don't... What is to explain?} \mim{Sure, I was talking about the reasons for combining these factors: live, silent, and down in decreasing order. The reviewer was saying we have taken many decisions without any justification.}
%\mim{again, we should have some justification for these components}.
All these numbers contribute to the reputation of a \segment, but we wanted to differentiate their importance. We apply weighted coefficients ($\omega_x=10^x$), which is often used in such scenarios, with a higher priority assigned to $N_{live}(Seg_i)$. 
The equation for \rInt{} is as follows:

\vspace{-20pt}
\begin{multline*}
R_{int}\left(Seg_i\right) = \omega_2 \cdot N_{live}(Seg_i) + \\
\omega_1 \cdot Sus_{silent}(Seg_i) + \omega_0 \cdot Sus_{down}(Seg_i)
\end{multline*}
\vspace{-15pt}

%\miii{@Mimon: ***  For the FUTURE: I always thought that we also consider the number of online benign IPs... But I guess we don't according to the equation. This leaves a gap: what if we have zero initial servers? We resort to random selection then as all segments have equal value R-int*** -- I GUESS we can leave it as is for now}

\textbf{b. \InterSeg reputation score (\rExt)}: This score considers the reputation of surrounding \segments.
In other words,  we want to prefer \segments that are near \segments with significant C2 server activity (given by $R_{int}()$). 
%\krish{What is malicious reputation and how do we measure?} \mim{I think, the malicious reputation is already mentioned as ``reputation of a segment i'', $Rank(Seg_i)$}
Naturally, we want to consider the distance, so that nearby \segments have a stronger influence.
The contribution of each neighboring \segment (upto $K$ neighbors on each side) is weighted inversely to its distance from the  \segment with a decay factor  $\lambda$ as commonly done.
The formula below calculates the average reputation of $K$ \segments on each side of the \segment:
% which we normalize by the factor $2 \cdot K$ in order to avoid having the external reputation $R_{ext}$ overwhelm the internal reputation $R_{int}$:
% NeighborhoodReputation evaluates the influence of surrounding subnets by considering their individual LocalReputation scores. Contributions from neighboring \segments are weighted inversely by their distance from the target \segment, scaled using the factor $\epsilon$. The formula is:
%\vspace{-10pt}
{\small
\[
R_{ext}(Seg_i, K, \lambda) = \frac{1}{2 \cdot K}
\left( \sum_{j=1}^{K}  \frac{R_{int}(Seg_{i-j})}{\lambda ^ j} + \sum_{j=1}^{K}  \frac{R_{int}(Seg_{i+j})}{\lambda ^ j}\right)
\]
}
%\vspace{-10pt}
% \vnote{something wrong with the equation?} \mim{resolved!}
% \vj{This approach ensures that closer neighbors have a stronger influence, reflecting the observation that malicious activity often propagates across spatially proximal subnets. This is based on our observation that malicious infrastructure tends to cluster within similar IP address ranges.}
% %\vspace{-3pt}

%%\vspace{-10pt}
The \interSeg effect makes intuitive sense within an Autonomous System (AS), since such a network is managed by a single authority
and \segments are under %are likely to be affected similarly by 
similar malware operational policies.
By contrast, \segments across AS boundaries may be less likely to exhibit similar behavior affecting the effect, due to manager and policy mismatch. 
%and the effect on each other's reputation should be less. This consideration only affects segments at the "borders" of the IP space of an AS.

%% which will a small percentage of all the segments in practice
%%
%\miok{In our experiments, this issue did not have significant impact.%,  as we discuss in Section~\ref{sec:results}.}

%\miii{Last phrase should be deleted if we don't revisit the discussion in the Results section.! I don't think we do as of yet. Maybe easier to drop the line.}
%\miii{Let's add a blurb in that section: we can say for practical purposes this would not have a material effect, as only a few segments will be affect at the borders of a network: maybe we can say only X\% of segments seem to have been: either close to AS borders - within say 3 segments -} \mim{added this task as things to do in the doc.}

%\vspace{-5pt}
\subsubsection{\textbf{Target IP exploration: \textit{Scanning} function.}}\label{sec:scan-function} 
We now discuss how we scan a \segment efficiently, corresponding to the \textit{Scan()} function in Algorithm~\ref{algo:ip-address-priortization}.
% This is a fairly involved operation with multiple possible approaches. We present the methods that we used, which ended up working well in practice as we see in the next section.

{\bf a. Identifying  "online" addresses.}
% \miii{I defined "online" earlier let's reconcile: keep it there if needed, or move it here.}
We use the term {\bf online} to refer to an IP address of a device that responds to network level probing, and keep the term {\it live} to describe live \hosts.
We check the status of each address in the selected \segment.
% which can be done in many different ways.
%We directly focus on TCP endpoints as most \mimok{malicious \hosts} operate over TCP~\cite{alrawi2021circle, davanian2022malnet}.
%\mim{should we check the context again for this line?} 
Despite its simplicity,
we do not use ICMP, as it is often blocked by firewalls and can falsely indicate a host as unreachable. Instead, we use Masscan~\cite{masscan}, an asynchronous SYN-TCP port scanner, to scan the whole port-range to discover the open ports. 
% \miii{@VJ: This needs explanation: what prioritized ports and what open ports? I thought this does pure port scanning, no?} \mim{yes, we do a whole port range scanning to discover open ports.}
We consider an IP to be live if it has at least one open port. The result is a list of detected open ports or the determination that the address is "dead" for probing purposes at least.

{\bf b. Port prioritization.}
At this point, we have identified the open ports for each online IP address in the \segment.
%\mimok{and want to prioritize the open ports. Port prioritization is important because a C2 protocol often requires maintaining a stateful packet exchange over a C2 port.}
% \miii{Don't see the need to add anything. I don't understand what you are trying to address here: I think we confuse things more. There are open ports and a bound on how many we can explore if the number is large! Did any reviewer complain?} \mim{Yes, one of the reviewers said that we did not clarify the need for port-prioritization. S/he also hinted at the need of port-prioritization for stateful probing.}
%\mim{Reviewer Alert: We need to establish the justification of port prioritization here. Port prioritization is important for stateful IoT C2 probings, which require (a) multiple packet replay in sequence, (b) more reliability than asynchronous stateless probing.}
%\miii{We find the open ports for all the online IPs in a a segment before exploring? I thought we prioritize be IP, no?} \mim{I think, at this level, we are talking about segment scanning. In a segment (256 IPs/segment), there is no prioritization.}
We set an upper limit, $\maxNoPorts$, on the number of ports per IP address, in order to not deplete the budget quickly, as some devices may have a large numbers of open ports (up to 3000 in our study).
%\mim{We need to rephrase if we want to remove the budget.} 
We rank the open ports in order of 
{\it %\mimok{ family-wise}
historical popularity}: we consider the number of times that a port has been used by \hosts 
%\mimok{of a specific family}
in our knowledge base %\mim{family-wise historical popularity -- to have a slight difference with C2Miner's general popular port}.
as we mentioned in \S\ref{sec:methodology-port-selection}. 
%Note that we combine information from C2Store and from our own probe studies.
We stop probing an address, when we: (a) explore all open ports, (b) get a \host response,
or (c) reach the  $\maxNoPorts$ limit.
%Note that in our study, we found \hosts with as many as 3000 open ports, so having a limit was important to ensure efficient use of resources.
%\miii{I wonder if the last statement creates more problems than not: if an IP has 3K ports, isn't that weird? Is it a sign of malice or benigness?} \mim{Yes, it can be a sign of a weird, potentially malicious/suspicious server. A benign server typically doesn't hold a large number of ports.}
%\miii{Can we comment } 

% Otherwise, we
%  We rank historical \mimok{malicious} ports~\cite{c2Store} based on their frequency of hosting \mimok{malicious} services. In addition, we give preference to the recently reported/observed \mimok{malicious \hosts}.

% \subsubsection{\textbf{{Port Prioritization}}} 
% \label{sec:port-prioritization} 
% \vj{For a given IP address, the next step is to prioritize ports to avoid brute-force scanning of all 65,536 ports. We take a simpler, yet effective, ranking approach, based on our  behavior profiling. We rank historical \mimok{malicious} ports~\cite{c2Store} based on their frequency of hosting \mimok{malicious} services. In addition, we give preference to the recently reported/observed \mimok{malicious \hosts}.}

%Next, we probe each port in descending order of priority for deeper analysis.
%\miii{@Mimon Find and revisit all mentions of entry, rule, etc} \mim{Done}

%\miii{@Mimon Revisit all "malicious \hosts" and remove "mailicious"} \mim{Done}

{\bf c. Port probing.}
 We begin with a TCP handshake for each target port. We then wait for the endpoint to initiate communication. 
 Some \hosts initiate communication, while others expect the bot to do so.
 In our experiments, the vast majority of \hosts that initiated contact will do so within 5 seconds after the TCP handshake, and 
 we use this time interval as our waiting period.
 If the  \host does not initiate contact, we start probing with the packets from the traces, maintaining the sequence and inter-packet delay. %between consecutive
 %we use our \Initial \playbook rules and send the first packet.
 %\miii{We keep}
 We probe each IP with all applicable traces, or until we get a positive response. %all applicable \Initial entries in our \playbook, 
 %until we get a response.
 %or we run out of \entries.
Similarly, for UDP connections, we  replay the packets  in the corresponding traces.

% \miii{Add a one line for UDP: "For UDP connections, we send directly the packets in our \playbook." Should we mention that we did not find any UPD rules or it will take away from the work?}

% After sending the initial payload, we allow up to few seconds (i.e., 15 seconds) for a response. 

% In the meantime, If we receive a valid payload, the \emph{ProbeGenerator} produces follow-up messages guided by our C2-Driven-Dialogues and defined timing intervals. On the other hand, if we detect no response or an unexpected one, we reset the TCP connection and move to the next top-ranked \emph{Initials} payload.

% To handle \tricky packets, we match incoming payloads against our C2-Driven-Dialogues. We prioritize exact matches, but we also accept partial matches for certain fields (e.g., dynamic timestamps, hashes, IPs, ports). We adapt to the dynamic changes in a received payload, and we also look for distinct, known-malicious payload signatures (e.g., “ELF file stream”, “UNIX shebang (\#!/) ”, “wget ”) to avoid missing any malicious servers.

{\bf d. When do we claim successful detection?}
We con-
sider a device to \host if it engages beyond the TCP
handshake by responding with one or more C2 communication packets. A packet is deemed to be part of a C2 communication if it matches a packet in our database, which
we derive in \Module  \moduleA.
Matching is performed at the payload level, allowing controlled flexibility for dynamic fields, such as timestamps and command identifiers, with strict checks for known malicious indicators, including ELF binaries and UNIX shell scripts. To reduce false positives, we also consider a "whitelist" of  payload indicators that also appear in packets in benign communications (which we collect during our active probing). %Although we plan to improve this packet matching in the future, 
We see that our detection strategy achieves high accuracy, as shown in our evaluation (\S\ref{sec:mod-c-bap}).

\section{Results and Validation}
\label{sec:results}
\label{result}

%\miii{TODO: add some observations about port analysis in Eval section}
% We present our experiments that substantiate our three modules. 
% These results provide information that we use to guide the exploration, or demonstrate the benefits of our approach.

%\subsection{Module \moduleA: \Playbook Generation:  traces and feasibility}

% We present the evaluation and results for each of our modules.

%\vspace{-0.2cm}
We show the evaluation and results for each of our modules.
%\miii{It is ugly to start with a subsection.}

\vspace{-5pt}
\subsection{Module \moduleA: Replay probe generation}
\label{sec:results-validation-replay}
%\vspace{-0.2cm}

\subsubsection{{\bf Binary Collection and Activation.}}
\label{sec:res-binary-activation}
To ensure information freshness, we
collect our own botnet communication protocol interactions using recently reported malware.
Naturally, we can also use any available information from other research efforts~\cite{trajanovski2021automated, C2Miner-Davanian-2024, davanian2021cnchunter}.
Although many studies focus on activating binaries, fewer studies focus on establishing and collecting malicious interactions~\cite{darki2020riotman, muhovic2020behavioural}.

{\bf a. Binary activation: 89\% success.}
%\mim{84\% or 89\%?}} 
%\miii{@Mimon, 89\% right?}
%\miok{
%We collect 1842 IoT malware binaries by scraping   MalwareBazaar~\cite{MalwareBazaar} hourly  from Oct.
%2023 to Feb. 2024 which we refer to as %}. 
We use
 {\textit{RP-DS}} dataset with 1842 IoT malware binaries in \Module \moduleA as we discussed in \S\ref{sec:Method-ModuleA-replay}. %\mim{the RP-DS does not only contain the collected binaries, but it also contains the network traces, the categorized network traces, and the packages, etc., as we defined and quantified in the definition at the end of \S 3.1} (see Appendix~\ref{}).
 The dataset includes malware from six major families: 
%\mimok{(determined mainly by employing AVClass2 and Yara rules)} 
{\bf 1347 Mirai}, {\bf 368 Gafgyt}, {\bf 19 Kaiji}, {\bf 12 Mozi}, {\bf 31 Hajime}  and {\bf  43 Tsunami}.
The dataset also includes  {\bf 22 Unknown} binaries, whose families we could not identify with the process described earlier. \miok{Note that this distribution of the families is what naturally emerges from the MalwareBazaar feed.
%   \miii{I deleted: across six IoT-supporting architectures} 
For ease of presentation,  we opt to report "major" families, which may encompass multiple variants. For instance, Mirai includes Miori, Satori, Moobot, Daddyl33t, Hakai, etc., while Gafgyt includes Bashlite, Yakuza, Qbot~\cite{avast_mirai_variants_2018, famera2025analyzing, unit42_moobot_dlink_2024, malpedia_mirai, malpedia_bashlite, palo_alto_2}. The presence of 46 distinct bot-C2 protocols in \textit{RP-DS}} 
 makes it a rich and fairly diverse dataset.

% \miii{Why not revisit everywhere where we report families and make it clear: there are "major families" and variant families. We cover X major families, and Y variant families. It would make our coverage of families look stronger. But we should introduce both in intro and in background.} \mim{the Y is not surely known by us. But, I am revisiting the places of families and adding variants along with it.}

Each binary was executed in a controlled sandbox environment for at least 3 hours, simulating real-world conditions, utilizing Google Cloud VMs in different geographic locations.
We were able to activate  89\%  (1634) of the collected binaries. 
%A summary of activation rates across different families is shown in Table~\ref{table:collection_activeation}.  
The summary of activation rate across different families is shown in Table~\ref{table:collection_activation} in Appendix~\ref{sec:ModuleA-Activation-Results}.
%~\ref{sec:ModuleA-Activation-Results}. 

%\miii{***}
%\miii{Discuss activation issues here if any} \mim{Vivek can describe it in a better way.}

%\miii{Explain how we know the family, discuss that we cover many families, so our initial dataset is decent} \mim{added the family info. It seems like we are repeating the methodology part here too!}

{\bf b. Establishing \host connection: 84\% success.}
The vast majority 
(84\% = 1536/1842)  of the  malware  establishes connections (e.g. a TCP handshake) with their \hosts. 
Interestingly, we get binaries from nearly every 
    % \miii{every? if Kaiji connects, but does not go beyond TCP? that can create confusion.
    % } 
family to activate and connect, as we show
in  Table~\ref{table:collection_activation} in  Appendix 
~\ref{sec:ModuleA-Activation-Results}. 
For example, we get 87\% of Mirai binaries to connect. Similarly, Gafgyt has a 84\% connection rate, with
%and Kaiji 58\%.
%\miii{DEFINITELY drop Kaiji here, report a different one}\mim{Removed Kaiji and only focused on Mirai and Gafgyt.}
 Tsunami has the lowest connection rate with 33\%.
\miok{Note that a successful activation of a binary does not guarantee an effective engagement with its \host. The reasons could be: (a) no connection attempt by the malware while executing, or (b) an unresponsive C2 server. For example, %as shown in Table~\ref{table:collection_activation} of Appendix~\ref{sec:ModuleA-Activation-Results}, 
14\% of the activated Mirai and 15\% of the Gafgyt binaries don't engage with their servers despite multiple activation attempts. \mim{please ignore any overlaps among the lines, it happened due to vspace, will be resolved when we eat reminder.}
}

\begin{table}[t]
    \centering
    \resizebox{\linewidth}{!}{%
      \begin{tabular}{|c||c|c|c||c||c|} 
        \hline
        \rowcolor[RGB]{90,240,110} {\bf Major Family}  & \multicolumn{5}{c|}{\textbf{Category}} \\ 
        \hline
        \rowcolor[RGB]{250, 200, 250} &  \multicolumn{3}{c|} {\textbf{Replayable}} & {\bf Non-Replayable} &  {\bf Undecided}\\ 
        \hline
        \rowcolor[RGB]{253, 219, 76}  & \;  Total \;   &  \cellcolor[RGB]{242,236,196}\Fixed &  \cellcolor[RGB]{242,236,196} \Tricky & \Obfu & Non-Revealing\\ 
        \hline
        %\cellcolor[RGB]{95, 200, 83} {242,236,196}
        Mirai  &  \cellcolor[RGB]{242,236,196} {\bf 77\%} & 73\% & 4\% & \cellcolor[RGB]{242,236,196} 3\% & \cellcolor[RGB]{242,236,196}20\%\\
        \hline
        Gafgyt & \cellcolor[RGB]{242,236,196} {\bf 72\%} &  37\% & 35\% & \cellcolor[RGB]{242,236,196} 2\% & \cellcolor[RGB]{242,236,196} 26\%\\
        \hline
        Mozi &  \cellcolor[RGB]{242,236,196} {\bf 50\%} & 0 & 50\%  & \cellcolor[RGB]{242,236,196} 0 & \cellcolor[RGB]{242,236,196} 50\%\\
        \hline
        Hajime & \cellcolor[RGB]{242,236,196} {\bf 58\%} & 0\% & 58\%  & \cellcolor[RGB]{242,236,196} 0\% & \cellcolor[RGB]{242,236,196} 42\%\\
        \hline
        Tsunami & \cellcolor[RGB]{242,236,196} {\bf 21\%} & 0 & 21\% & \cellcolor[RGB]{242,236,196} 12\% & \cellcolor[RGB]{242,236,196} 67\%\\
        \hline
        Kaiji &  \cellcolor[RGB]{242,236,196} {\bf 0\%} & 0 & 0 & \cellcolor[RGB]{242,236,196} 16\% & \cellcolor[RGB]{242,236,196} 84\%\\
        \hline
        Unknown & \cellcolor[RGB]{242,236,196} {\bf 11\%} & 0 & 11\% & \cellcolor[RGB]{242,236,196} 16\% & \cellcolor[RGB]{242,236,196} 73\%\\
        \hline
        \rowcolor[RGB]{250, 200, 250} \textbf{Overall:} & \cellcolor[RGB]{253, 219, 121}%{242,236,196} 
        {\bf 72\%} &  \cellcolor[RGB]{242,236,196} 61\% & \cellcolor[RGB]{242,236,196} 11\% & \cellcolor[RGB]{253, 219, 76} 3\% & \cellcolor[RGB]{253, 219, 76} 25\%\\
        \hline
      \end{tabular}%
    }
    \caption{\small{Replayability of binaries across malware families:
    For at least 72\% of the cases, the \playbook approach is viable}. Every line adds to 100\%, excluding \Fixed and \Tricky columns (accounted in Total). %Every malware family except Kaiji has a non trivial number of binaries with replayable C2 communication packets. 
    The Undecided are binaries that do not reveal their communication (fail to either activate or to engage with their \host).
    }
    %Only for Kaiji, 100\% of its binaries are in the \Obfu category.
    %\vspace{-40pt}
    \label{table:taxonomy}
\end{table}

%===========================
%\vspace{-10pt}
\subsubsection{{\bf\Replay-based probing is feasible for modern IoT malware (for now).}}
\label{replay_feasibility}

Our study shows that \playbook probing remains feasible and effective for recent-day IoT malware for many malware families 
%and their variants
found recently in the wild.
This comes to challenge the status quo. 
The community seemed to worry about the applicability of packet-replay  for IoT malware~\cite{davanian2021cnchunter, C2Miner-Davanian-2024},
possibly extrapolating from the trends of the more sophisticated PC-based malware~\cite{securelist2025malware, xu2014autoprobe}.
%and IoT-centric work has anticipated similar concerns as a potential challenge~\cite{davanian2021cnchunter, C2Miner-Davanian-2024}, 

%\miii{@Mimon or Vivek: Let's be a bit more clear}
%\miii{Also, if we want to stick to 72\% are the key number, let's create a natural evolution and justification: we activate 89\%, we connect for 89*94, we verify for 89*94*XX, we end up with 72\%.
%Maybe even a two line table? or in text?}
%Here we challenge  

\begin{comment}
As we mentioned earlier, some recent studies claim that modern malware is %too 
sophisticated to be detected by a \playbook approach
~\cite{davanian2021cnchunter,C2Miner-Davanian-2024}.% \mim{could we remove 'too' to make things soften? Also, can remove ~\cite{xu2014autoprobe} to make things specific to the IoT scope.}
We challenge this statement with the following  study.
Due to space limitations, we will present this part in a concise way, although it hides significant effort. \mim{C2Miner does not claim; it shows concern about the future.}
\end{comment}

\vspace{0.1cm}
\takeaway{\Replay-based probing  works for at least 72\% of the 1842  binaries in the \Fixed and \Tricky categories in our \textit{RP-DS} dataset.}
 \vspace{0.1cm}

We start with 1842 collected malware binaries and successfully activated 1634 binaries (89\% of 1842) in our sandbox. We then find that 1536 binaries (84\% of 1842) establish  a TCP handshake with their \host as 
 described in \S\ref{sec:Method-ModuleA-replay}.
 A more detailed discussion can be found in Appendix~\ref{sec:ModuleA-Activation-Results}.

We report the applicability of a \playbook approach to the  malware families in our dataset.
%major IoT malware families \mimok{and their variants}.
We plot the results in Table~\ref{table:taxonomy}.
We find that a \playbook approach can work for at least 72\%  of the initial 1842 binaries.
Furthermore, this behavior applies for most major families, whose variants use \Fixed and \Tricky communication protocols, except Kaiji. 
We see that among the binaries that  only 3\% were \Obfu.
 
This 72\% percentage is a conservative estimate and should be seen as a {\bf lower bound} of the possible \playbook-compatible binaries. 
As the table shows, 25\% of the binaries do not reveal their communication with the \host, and thus, we cannot determine their
\playbook category.
There can be several reasons for this:
 (a) binaries not activating in our sandbox,
 and (b) binaries not managing to connect to their \hosts, which could have died or simply declined to respond, which is not uncommon as shown in \S\ref{sec:case-study}.
 %As a result, we could not assess if they are amenable to a \playbook approach.

%{\bf Use of encryption in our binaries.}
{\bf Encryption: limited presence among our binaries.} 
We observed that only 15\% of the collected malware binaries across all families employ encrypted C2 communications. However, we find that 9.25\% of our binaries use "static" encryption: the packets are the same across different binary activations and thus, are replayable. 
By contrast, 2.25\% of the binaries are not replayable 
and 3.5\% of the binaries are in the Undecided category.
We revisit this limitation in \S\ref{sec:discussion}.

The outcome of this study is: (a) we establish the feasibility of a \playbook approach, and (b) we extract packet traces that we use to probe a \host.

\vspace{-5pt}
\subsection{\Module B: \host behavior profiling}
\label{sec:behavior-profiling}
%\vspace{-5pt}
\subsubsection{{\bf Spatial locality:}}
\label{sec:locality}
We quantify the spatial locality of \hosts using 
 the \textit{BP-DS} dataset obtained in \Module \moduleB (see  \S\ref{sec:methodology-port-selection}  
%\mimok{C2Store~\cite{C2Store-Vivek-2023}, and Threatfox~\cite{threatfox}} between 2020-2025} 
and Appendix~\ref{sec:appendix_dataset}).
 Recall that the dataset contains 18.2K bot-C2 communications across five  IoT malware families.

\vspace{0.1cm}
\takeaway{\host locality is strong within a /24 subnet but drops off quickly after that.}
\vspace{0.1cm}

%Note that each family is concentrated in different IP spaces and ASes.

%\miok{
% We study the distribution of each malware family separately.
% Although all malware families exhibit skewed spatial distributions, they have different areas of concentration.
% We study each family separately to avoid a superimposition of their spatial distributions.
% }

%We focused on Mirai, which is one of the most popular IoT malware families~\cite{mirai-popular-malware}. 
%\zt{I think the focus on IoT/Mirai shouldn't come this late. If the experiments is about IoT/Mirai specifically, in intro we have to make it clear this is the focus (and some explanation why this hasn't been well explored)} \mim{@Prof. Michalis, can you please check the comment?}

We study the spatial distribution of \hosts using the
\localDensity metric as we defined earlier. 
%We calculate the average \localDensity, as defined earlier, as a function of the prefix size of a subnet \mimok{for the}  IoT families in our database. 
%\miii{@Mimon: I thought we were focusing on Mirai!!!!} \mim{last plot was for only Mirai. The statistics here are calculated in percentage (\%) for different families.}
For each subnet of prefix size $k$, we calculate the \localDensity metric for each family. Note that we study each malware family separately to avoid cross-family interference in dataset.
\miok{We observe the effect of $k$ on the \localDensity and see a notable increase for prefix size of $k=24$ for all families.} 
%\zt{is it average over all C2 servers in the dataset?}
%This percentage serves as the locality metric. 
%, and we show the result in  Fig.~\ref{fig:locality}. 

For all our malware families, the local density drops by an order of magnitude once we go from /24 to /20 and two orders of magnitude if we go to /16. 
For example, Mirai and Gafgyt achieve a local density  of 0.75\%  and 0.53\% at /24, but the density drops to 0.06\% and 0.05\% for /20, and to 0.005\% and 0.005\% respectively for /16 prefix size. 
%\miii{@Mimon: please fix the numbers above: your table did not have Gafgyt} \mim{Done. Bashlite is another name of the Gafgyt family}
%\mim{the density here is in \% -- 0.73 means, on average, 1.87 == 2 C2 IPs per a malicious /24 subnet (a /24 subnet holds 256 IPs in total)}

%Therefore, we select a segment size of /24 as a target for the neighborhood a \host, which gives good results in our experimental evaluation.}

% \miii{I rewrote the above, and also in Overview the definition of \localDensity metric.}

We codify this results into the following guideline for probing:
the probability of finding additional \hosts is relatively high within a /24 subnet that already contains a \host. 
Thus, we use /24 segments ($\segk=24$) in our probing. This size strikes the balance between capturing locality and keeping the subnet size small, as the evaluation in Appendix~\ref{sec:appendix_dataset} shows.
Note that we can 
repeat this study periodically with newly available measurement data, when malware behaviors change and update this parameter, following similar analysis.

 %\mim{there could be a table for 4.2 that covers both family-wise quantification of data used, spatial locality, and port-level behavior, as we claim this profiling as our contribution.}
%\miii{Let's define N if we are going to use it: boldface etc.}\mim{used ($N=24$).}

% The findings show a strong concentration of malicious IP addresses within subnets ranging from /24 to /31. This trend remains consistent in this range, indicating that these subnet sizes are more likely to contain a significant proportion of malicious IPs. This insight highlights the effectiveness of focusing on /24 to /31 subnets for uncovering malicious activity.

% Based on this analysis, we selected /24 as the optimal granularity size (\textit{N}) for segmenting the search space. This decision was informed by two key observations: (a) the /24 subnet size preserves spatial locality effectively, and (b) it offers the widest range of IPs per segment, reducing the total number of segments and minimizing the overhead of segment prioritization in future steps.

\subsubsection{{\bf Port-level behavior.}}
\label{sec:port_behavior}
At which port should we contact a potentially \host? To answer this question,
we study the popularity and re-use of the ports empirically. 
%the observed behavior: the usage patterns of port numbers in our database.
We use the  \textit{BP-DS} dataset which has 18.2K recorded C2 communications across all our malware families.
Among them, we  identify 6,350 
unique port numbers used by \hosts. 
%, including Mirai, Gafgyt, Kaiji, and others.
%All of the records were reported over the last five years~\cite{C2Store-Vivek-2023}. 
%\miii{If we use BP-DS, we use whatever the db has, unless we explicitly focus only on the last five years, which again is confusing: last years in the db or 2021-2026?} \mim{yes, I think we can delete the last line. We have mentioned about the data source in the second paragraph of this subsection.}
%\miok{For each malware family}, 
\miok{Note that we create an ordered list of ports for each family based on their frequency of appearance  in our dataset. We use this list     when probing for a specific family.}
%and we refer to this ordered list as {\it ranked\_ports}. 

\mim{Could we state that our port level analysis was also family-focused?}
\miii{Interesting idea: we could, if we have space and it will not create more questions that answers. We could say the following: "Note that we create an ordered list of ports for each malware family, to make the probing efficient when we probe for a specific family."
}
 
 %\miii{Let's provide some quick statistics: I bet the top 10 or 20 most popular ports are vastly more popular: maybe the top 20 ports account for 80\% of reported ports?} 

 \begin{comment}
     We find that the top single port accounts for 22.99\% of reported C2 server records. Expanding our analysis, the top 50 ports cover 52.34\% of all entries, while the top 100 ports encompass 58\%. As we include more ports, the coverage increases to 65.81\% with the top 500 ports and 70.34\% with the top 1K ports. These results demonstrate that a relatively small number of ports handle a significant majority of C2 server communications.
 \end{comment}

{\bf There is substantial \host port reuse.}
We find that the \hosts of most families use some port numbers extensively as we show in
Appendix~\ref{sec:appendix_dataset}.
%We find that the most popular port (7173)  accounts for 18.3\% of the IP:port reported interactions, 
% \miok{
% We observe 6350 distinct ports \miii{Mimon had "records" are records different than ports} \mim{I assumed each IP:port as a single record.} used by \mimok{C2 servers} in  \miok{the \textit{BP-DS} dataset}
% \mim{this line is repetitive, already mentioned in the last paragraph.}.
Indicatively, we mention that the top 50 ports for Mirai cover over 52\% of observed connections, and  the coverage grows to 71\% for the top 500 popular ports. 
We leverage the port reuse in our port prioritization strategy in \Module \moduleC.

\subsubsection{{\bf Parameter Tuning for \name}} \label{sec:param-tuning}
\miok{We determine the parameters K and $\lambda$ using \textit{BP-DS} dataset. We vary these parameters and select the  values 
at the point of
%smallest ones that exceed
diminishing-returns  (specifically when the marginal improvement drops below 10\% of the initial gain). For instance, the elbow falls at K=5, $\lambda$=1.5 for Mirai and K=10, $\lambda$=2.0 for Gafgyt, retaining 97.5\% and 86.5\% of their maximum detection. For consistency, we adopt the larger values across families. The full results are shown in Appendix~\ref{sec:param-selection}. In our evaluation (§\ref{sec:mod-c-bap}), 
we show  the effect of these parameters empirically.
%we present two candidates of \name:  (a) \name-Hi: a tuned parameter version, and (b) \name-Mod: an untuned parameter version to show the effects of parameter choice.
%%%% COMMENT   READ THIS
%%%% There was too much detial here, plus talking about an "untuned" selection makes it look arbitrary: compare it with another tuned version... the reviewer will say.
}

%We find K = 5 and $\lambda$ = 1.5 for Mirai, achieving 97.5\% maximum detection with \name, while for Gafgyt, with K = 10 and $\lambda$ = 2.0, it achieves 86.5\% maximum detection. The full results are shown in Appendix~\ref{sec:param-selection}. In our evaluation section (\S\ref{sec:mod-c-bap}), we show two candidate versions of \name, with different parameter values, to show the effects of parameters.

%With this configuration, BotScan achieves 97.5\% of the maximum detection for Mirai and 86.5\% for Gafgyt. .} \mim{add the figure at the appendix and describe it more.}
%\mim{this statistics is applicable for Mirai. If we claim about the individual behavior profiling, should not we be specific here?, please look at the tables in Appendix \S\ref{sec:appendix_dataset}}

%Overall, a small set of ports handles most \mimok{malicious} communications. This information can be used as a guide in our study. In addition, a probing study can also adapt the relative priority of ports based on the responses that it receives.

% \miii{Given my revised text in 4.2 (b), maybe we can shirnk this text here a bit? At least connect it: As we saw in subsec xxx, etc} \mim{ I think 2 lines can be removed:\\
% i. last line of 1st para --- Starting 2nd para as: As mentioned in 4.2.2 we rank the port and find that... \\
% ii. Last line.
% }

 % To enhance the efficiency of Intelligent Probing, the \textit{ranked\_ports} list was later revised and refined, ensuring better targeting and prioritization during the probing process.
 %\miii{Revisit the above paragraph}

%\vspace{-10pt}
\subsection{\Module C: \SmartProbing probing}
\label{sec:mod-c-bap}
We evaluate our \smartProbing strategy using two evaluation studies and two metrics: (a) detection accuracy, and (b) probing efficiency.
%Our approach exhibits high accuracy with higher probing efficiency at the network level.

% {\bf a. Detection accuracy at the IP address level: 85\% F1-score.} Our approach exhibits high accuracy.
%We assess the accuracy of classifying an IP address as \host.
%This is our answer to the question:
%Given an IP address, how accurately can we determine if device is a \host or not?

%Using a small dataset of benign and malicious addresses, we show that our method achieves a true positive rate of 74.81\% with a false positive rate of only 0.88\%.

% {\bf b.  Higher probing efficiency at the network level:}
% %To measure resource efficiency, we compare our  strategy against two other active probing techniques used in prior studies. 
% We show that our \smartProbing method delivers 70\% more \hosts  for a given budget compared to two baselines.
%demonstrating its effectiveness in maximizing discoveries within a  probing budget. \mim{should remove budget?}
%\miii{Let's replace what used to be "optimization throughput" with resource efficiency}.

\textbf{A. Exhibiting high detection accuracy.} 
We first evaluate the accuracy of our detection approach. Given the absence of a benchmark, we create a 
\textbf{ground-truth dataset  {\textit{GB-DS}}} of  800 benign and 528 \hosts.  First, we collect 800  benign server IPs spanning eight service categories (100 IPs per category), including FTP, SSH, DNS, and others, along with Honeypot deployments. We verify that these IPs were responsive in Censys~\cite{censys}. %\mimok{Please note that we are aware of Sinkhole entities. Sinkhole entities capture C2 communications for analysis by listening to inbound C2 traffic without imitating or responding like actual C2 servers~\cite{bitsight_sinkhole_2021}.}
Second, for the malicious servers, we collect 528 live IP addresses across six popular IoT botnet families 
(Mirai, Gafgyt, Mozi, Hajime, Kaiji, Tsunami), comprising up to 200 active IPs per family, utilizing ThreatFox~\cite{threatfox} service. As a final check, we validate both benign and malicious servers with VirusTotal using a threshold of at least 3 threat engines' detection as per convention~\cite{peng2019opening, wang2022mfdroid}. %\mim{As we are mainly focused on the correctness of detection, we did not focus on the optimization part (IP optimization, Port optimization) here. Should we explicitly state that here?}

Our method can identify   \hosts accurately with a precision of 98.8\%, a recall of 74.8\%, and an  F1-score of 85.0\%. 
Arguably, the lower recall is the price we pay for the high  precision, which could be preferable depending on the intention of the study. Notably, the 800 benign servers include echo-prone services (FTP, SSH, DNS) that could respond to a replayed packet. The 98.8\% precision shows that our detection scheme (\S\ref{sec:scan-function}) rarely mistakes such responders for C2 servers, addressing a key false-positive concern.

\textbf{B. Probing efficiency at the network level.} 
\mimok{We compare \name against three baseline approaches, corresponding to the strategies used in prior works. \textbf{Baseline 1 [No locality]:} We use this method as a reference of not utilizing locality at all. Given the seed C2 servers, it probes them and then explores the remaining space with no prioritization until the budget is depleted. This corresponds to the C2Miner~\cite{C2Miner-Davanian-2024} approach, where the target IP:Port is given as direct input to the system. \textbf{Baseline 2 [\IntraSeg locality]:} This method uses only intra-segment locality. The method divides the IP space into \segments and probes in two phases: (a) it probes the seed C2 servers  and the \segments associated with them; and (b) it selects and probes the remaining \segments uniformly randomly. \mimok{This static intra-segment locality approach
\miok{corresponds to} the strategy of} AUTOPROBE~\cite{xu2014autoprobe}. The approach does not use: (a)  \interSeg  locality, or (b) adaptive target prioritization, as we do in \name. \textbf{Baseline 3 [ISP-centric locality]:} It exploits locality at the AS (ISP) level. Given a set of ASes, it ranks them by their %malicious reputation.
C2 server density,
namely,
the percentage of IP addresses of the AS that are known C2 servers (to account for the AS size). 
%Ok, to me percentage seems smoother, ratio typically needs to say ratio of X over Y
%(?) yes, percentage = ratio.
%Percentage = ratio * 100 , no? :-) yes, that why i explained, not direct percentage, rather ratio, as the C2 density type. We can go with percentage. Got it Professor. Yes, we treated known/seed C2 server density as malicious reputation to rank the AS for exploration order.
%{\small $R_{AS}(a) = Count_{a}(Seed \space C2s)/Count_{a}(Candidate \space IPs)$} 
 %\mim{R_as(a) = reputation score for an AS 'a', Count_a(SeedC2s) = Count of C2 seeds from the as 'a', Count_a(CandidareIPs) = Count of all IPv4 in the AS 'a'.}
The method probes the seed C2 IP first and then probes the IPs of the AS space %uniformly 
according to this AS ranking. This reflects CyberProbe's~\cite{nappa2014cyberprobe} provider-based localized probing.}

{\bf Two \name versions: high and moderate locality.}
As we stated in \S\ref{sec:behavior-profiling}, we evaluate our probing approach with different sets of parameters where we {\em consider locality with different intensities} in our prioritization: 
(a) {\bf \nameHi:} with high locality intensity ($K$=10, $\lambda$=2, \textit{r-explore}=0), \miok{selected based on our findings in \S\ref{sec:param-tuning}};
(b) {\bf \nameMod:} with moderate locality intensity  ($K$=15, $\lambda$=1,  \textit{r-explore}=0), \miok{downplaying the intensity of locality (wider neighborhood, smaller decay).
}
%for settings with limited locality knowledge}.
% \miii{Can we say: \nameHi values were selected by the parameter selection method as discussed in \S ZZZ -- Can you add?} \mim{Added, could you please check?}
For consistency, we set \textit{r-explore} to 0 in both cases. Note that a high value of \textit{r-explore} can affect the adaptability and may steer the algorithm towards baseline 2.

{\bf Probing RoI:}
Our evaluation metric is the count of live \host identified by a method as a function of probed IP:port pair count.
We set the maximum number of ports explored per IP, \maxNoPorts = 20, for all the approaches.
%which we can refer to as probing budget.

%\vspace{-15pt}
\subsubsection{{\bf Evaluation 1: Actively probing a large target space.}}\label{sub:eval1}
\miok{We evaluate the efficiency of \name by exploring  a target space of 2.5M IP addresses across 8 Autonomous Systems (ASes). Note that the exploration space is independent of \textit{BP-DS} dataset. A comparison over a "dead" IP space would not have been meaningful, so we selected the IP space to contain  recently-reported \hosts.
Using existing archives~\cite{C2Store-Vivek-2023}, our target space has 11 recently reported IoT \hosts, which belong to two major malware families and their variants  across 9 different /24 subnets.
We consider these \hosts as our seed set.
} \mim{Add the line if there is time: we list the ASes and prefixes in Appendix X.}
\miii{If there is space: it makes sense to report here which ASes, at least. Otherwise it is very "hidden". Ideally we would have a Dataset section that listed all these things, ASes, and IP prefixes, maybe point to an appendix?} \mim{Yes, a dataset section would be great if the space permits.}

\begin{comment}
   \mim{This part needs to be removed: We evaluate the performance of our approach by probing the IP space: 154.216.0.0/16, which consists of 65536 IPs \mimok{and multiple ASes}. We chose this IP space because of it has reported malicious activity: there are 7 C2 servers \mimok{of two different families} reported across 3 different /24 subnets within a week~\cite{C2Store-Vivek-2023}.}  
\end{comment}

%\mimok{In this phase of the evaluation, we discuss our target optimization capability rather than family-specific \host detection, which we show in \S\ref{cs-1}}
%This gives us the opportunity to assess the effect of prior knowledge. 
%\mim{We did not mention any family. The selected subnet hosted mainly Mirai and Gafgyt. Do we need to mention it? (i) Evaluation 2 also does not talk about any family, (ii) if explicitly said about Mirai and Gafgyt, will the reviewer question the scoping?, iii) if we don't say anything about family, will the reviewer question it?} \mim{We should also talk about the port prioritization -- we can say we focus top 20 ports, combining all top ports/family in our scope. In case of no such open ports, we explore at most 20 open ports per IP.}
 %We consider those \hosts as prior knowledge, which we use in our probing strategy. 

\begin{figure}[th]
    \centering
    \vspace{-5pt}
    \includegraphics[width=\columnwidth]{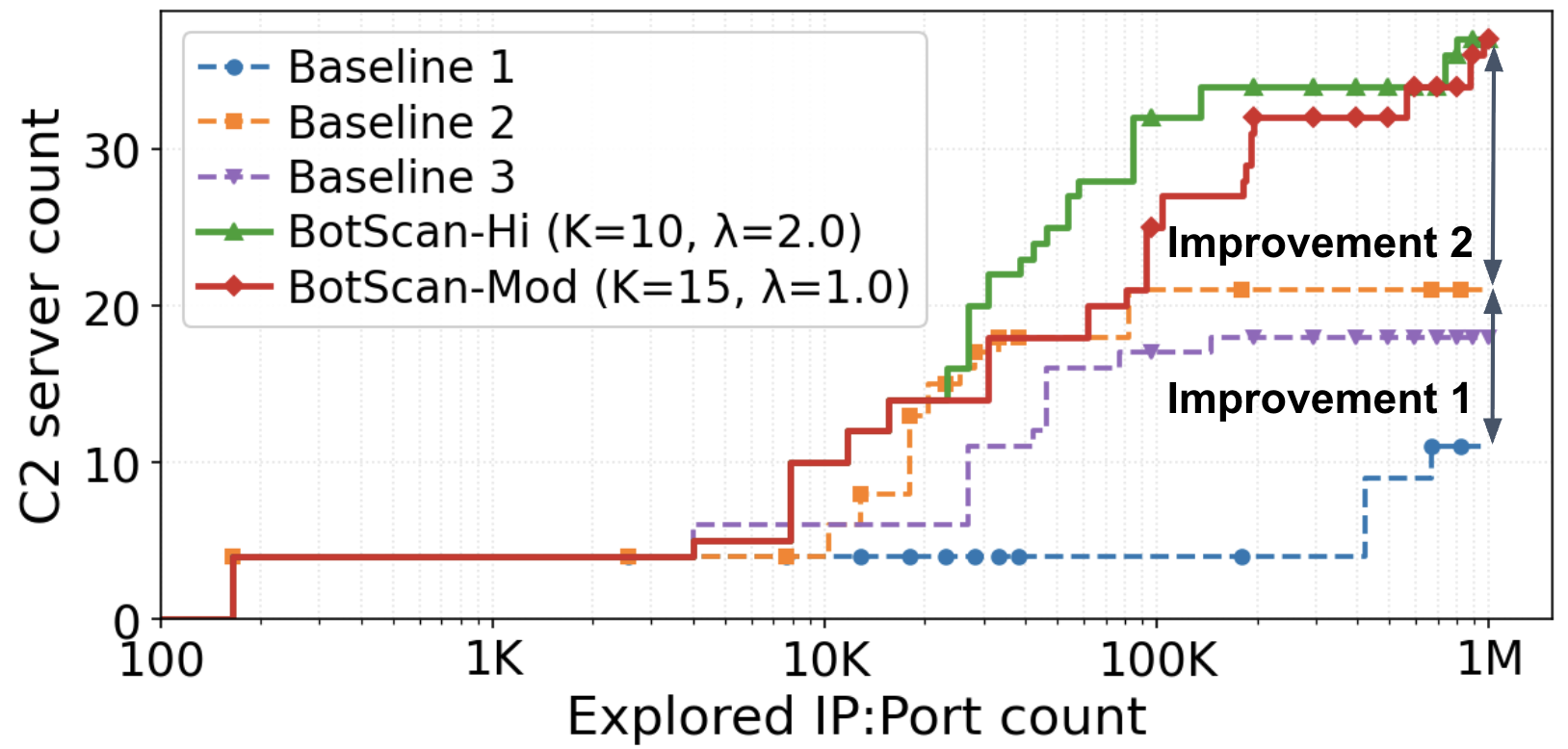}
    \vspace{-20pt}
    \caption{\small{Evaluation 1. Probing RoI: \name  outperforms the baseline methods. We plot the number of identified live \hosts versus the number of IP:port pairs explored. Improvement 1 comes from considering the \intraSeg locality. Improvement 2 is exclusive to \name due to: (a) \interSeg locality, and (b) adaptive probing. \name finds \mimok{37} live hosts compared to \mimok{21} for the closest Baseline 2 (including the 4 live \hosts of the 11 in the initial seed set).} %\mim{add justification of tuning and un-tuning.}
    %\mim{improved figure and checked the caption.}
    }
    %\caption{\small{Evaluation 1. Probing RoI: \name  outperforms the baseline methods. \miok{We show the number of identified live \hosts versus the number of IP:port pairs explored (which represents the probing cost). Improvement 1:  By considering \interSeg locality, our two approaches outperform Baseline 2 by finding 39 live \hosts vs. 22 for Baseline 2 including the 7 \hosts in the initial seed set.} \miii{Need to change the text in the paper too!} \mim{improve the figure.}}
    %}
    
    \label{fig:active-validation}
    \vspace{-10pt}
\end{figure}

%We 
%We limit all explorations to 3072  distinct IP addresses, which corresponds to 5\% of the space. \mim{Should we remove this line? --just to avoid the question why only 5\%?}

%\miii{If this is an option in our algorithm give it a name Randomization option} \mim{Revised!}.

%\miii{Rephrase and explain better this paragraph: I can't follow it.}

%Figure~\ref{fig:active-validation} illustrates the result of the experiment.

{\bf\name outperforms the baseline methods substantially.}
We plot the identified C2 server count as a function of probed IP:port pair count for each approach in Fig.~\ref{fig:active-validation}. 

\vspace{2pt}
\takeaway{ \name provides superior probing RoI: combining adaptivity, \intraSeg, and \interSeg locality awareness improves the \host detection rate substantially compared to the baseline methods.}
\vspace{2pt}

{\bf Improvement 1: \IntraSeg locality increases the detection rate significantly.}
We isolate the effect of \intraSeg locality by comparing Baseline-1 and Baseline-2, with the latter using the locality. 
Baseline-2 exhibits superior detection rate of live \hosts  in Fig.~\ref{fig:active-validation}. After exploring 1M IP:port pairs,
Baseline-2 finds \mimok{21} live servers compared to \mimok{11} found by Baseline-1. If we exclude the 4 live which were among the 11 initial seed servers, the difference becomes more pronounced: \mimok{17 vs 7} live \hosts. \miok{Baseline-3, which uses coarse AS-level locality, on the other hand, finds 19 live servers (15 excluding the 4 seed servers), outperforming Baseline-1. However, it falls short of Baseline-2, which suggests that the finer locality granularity is more effective than the coarser AS-level locality in Baseline-3}.

{\bf Improvement 2: Adaptivity and \interSeg locality increase the detection rate further.}
The two \name variants exhibit consistently higher detection rate compared to Baseline-2. We attribute this improvement to the use of: (a) adaptive address re-prioritization, and (b) the \interSeg locality, used in \name.
At the 1M IP:port mark, \name has identified \mimok{37} live \hosts compared to \mimok{21} of Baseline-2.
Excluding the 4 live seed hosts, the difference becomes \mimok{33} to \mimok{17}, i.e., {\bf almost double the number} of new live servers.

\miok{\textbf{\name is relatively robust to parameter selection.} \nameHi outperforms the \nameMod variant, demonstrating that our elbow-based selection is effective (\S\ref{sec:behavior-profiling}). At the same time, the relatively small  performance difference  suggests that our approach is relatively robust to the parameter selection.}

\subsubsection{{\bf Evaluation 2: \miok{Retrospective evaluation at scale with historical data.}}}
\label{sub:eval2}
To further validate scalability, we conducted a retrospective emulation over a 16M IP space (103.0.0.0/8) using historical data that does not overlap with \textit{BP-DS}. The space contains 207 historically reported \hosts accross 72 Autonomous Systems and multiple IoT C2 families. We uniformly randomly select 20 servers as our seed group and the goal is to identify the 187 (=207-20) \hosts, which we "pretend" we are not aware of. \miok{As detailed in Appendix~\ref{sec:Appendix-Eval-1} and Fig.~\ref{table:op_cost_appendix}, \miok{\nameHi}  ($K$=10, $\lambda$=2.0) \miii{this is the Hi right?}\mim{right, it is BotScan-Hi} consistently outperforms the closest baseline 2 by \textbf{a factor of two} with a budget of 550K probing IP:port pairs. 
This result corroborates the results of the evaluation presented above.
% Changes the last line and some words here and there
}.

\subsubsection{{\bf Computational overhead.}}
\label{sec:computational}
%We consider two aspects of the computational overhead.
We empirically compare the resource efficiency of the two main active probing approaches:
(a) \playbook used in \name,
and (b) activation-based used in C2Miner.
We conduct our experiment in
a 32 GB RAM Google Cloud VM running Ubuntu 22.04
LTS.

%\miii{This is correct, right?} \mim{yes, my VMs are generally on this configuration.}

%A \playbook approach is more computation resource efficient compared to  activation-based approaches, which activate a binary every time they want to probe and engage with a potential \host.
%%\vspace{-15pt}

We compare the two methods as follows. For the activation-based method,  we activate malware binaries in a sandbox and wait until they communicate with their associated \hosts. For our approach,  we probe the same \hosts by replaying the appropriate packets. For the comparison, \miok{we consider C2Miner as a representative of the binary-activation approaches.
%in the case where malware binaries successfully engage with the targets through packet redirection
}
We track the CPU utilization (as  a percentage), memory usage (in MB), and experiment duration (in seconds), until we receive a valid response from the \host. \mimok{We focus our comparison on the post port-discovery probing or redirection stage as the post discovery stage is same for all}.
\miii{I can't understand what you say: I think we want to say: Here, we focus on the resource use for probing a given IP:port pair. - right?} \mim{exactly, because, C2Miner, as it is now activates malware to probe each IP:port. Should we highlight it to differentiate?} \mim{another thing, should we also clarify, we also activate malware but for a single time to get the replay packets.}
We repeated the experiment three times with three different malware binaries.
%\mimok{that initiated the C2 communication to their C2 servers}. 
We show the results in Table~\ref{table:op_cost}. First, we see the substantial gains in computational resources with a \playbook approach, which is an order of magnitude for CPU and memory\footnote
{
Recall that a \playbook approach does not activate a binary every time that we probe. \name activates malware binaries in \module \moduleA, but only {\bf once} for each binary to extract communication trace.
During probing, we only replay the packets.
}, 
and two orders of magnitude for the probing duration.
The reason for the differences are: (a) the higher resources required by a sandbox that emulates a device to activate the malware (compared to simple packet replaying using roughly 250 lines of code), (b) the need to wait for the malware to initiate the contact, which adds delay in the execution.
%\miii{Maybe trim or delete last phrase if we need space}

\begin{comment}
\begin{table}[t]
\centering
%%\vspace{-6mm}
\resizebox{\columnwidth}{!}{
\begin{tabular}{|c|c|c|c|c|c|} 
\hline 
\rowcolor[HTML]{64e764}
\multirow{2}{*}{\textbf{Approaches}} & \multicolumn{3}{c|}{\textbf{Resources: probing one IP:port}} & \multicolumn{2}{c|}  {\cellcolor[HTML]{DADAF8} \textbf{Cost}}
\\ \cline{2-4}  \cline{5-6}
\rowcolor[HTML]{64e764}
\multirow{-2}{*}{\cellcolor[HTML]{64e764}\textbf{Approaches}} & CPU (utiliz. \%) & Memory (MB) & Duration (sec) & 
\cellcolor[HTML]{DADAF8} Duration (Days) & \cellcolor[HTML]{DADAF8} Cost (\$)\\
\hline 
Replay: \name  & 0.38 & 11.86 & 0.65 & \textbf{0.63} & \textbf{5}\\
\hline 
Activation: C2Miner  & 6.24 & 142.41 & 28.78 & \textbf{304} & \textbf{2330}\\
\hline 
\rowcolor[RGB]{182,232,205}
% warm yellow {253, 219, 76}
\textbf{Resource savings} & 16$\times$ & 11$\times$ & 45$\times$ & \cellcolor[RGB]{250,200,250}  482$\times$ 
& \cellcolor[RGB]{250,200,250}  466$\times$
\\ 
\hline
\end{tabular}
}
\caption{ \small{\Playbook methods are more resource efficient by an order of magnitude vs activation-based. This translates into higher parallelizability with the same hardware, which explains the significant gains in actual duration and cost for 10M IP:port pairs.}
}
%\vspace{-30pt}
\label{table:op_cost}
\end{table}
\end{comment}

\begin{table}[t]
\centering
\resizebox{\columnwidth}{!}{
\begin{tabular}{|c|c|c|c|}
\hline
\rowcolor[HTML]{64e764}
\multirow{2}{*}{\textbf{Approaches}} &
\multicolumn{3}{c|}{\textbf{Resources: probing one IP:port}} \\
\cline{2-4}
\rowcolor[HTML]{64e764}
\multirow{-2}{*}{\cellcolor[HTML]{64e764}\textbf{Approaches}} &
CPU (utiliz. \%) & Memory (MB) & Duration (sec) \\
\hline
Replay: \name & 0.38 & 11.86 & 0.65 \\
\hline
Activation: C2Miner & 6.24 & 142.41 & 28.78 \\
\hline
\rowcolor[RGB]{182,232,205}
\textbf{Resource savings} & 16$\times$ & 11$\times$ & 45$\times$ \\
\hline
\end{tabular}
}
\caption{\small
\name probing is substantially more resource efficient than
C2Miner, reducing CPU utilization, memory usage, and
probing duration by 16$\times$, 11$\times$, and 45$\times$, respectively.
}
\vspace{-28pt}
\label{table:op_cost}
\end{table}

\textbf{Cost: a back-of-the-envelope calculation.} 
\miok{Extrapolating to a 10M IP:port exploration under realistic parallelism assumptions (detailed in Appendix~\ref{appendix:cost_estimation}), \name completes the task in ~15 hours for an estimated \$5, compared to 304 days and \$2,330 for C2Miner. This analysis suggests that a 
\playbook 
probing can provide significant efficiency and cost gains for a large-scale or  frequently-repeated exploration.}

\miii{I commented out Mimon's details, if  space, we can add some more}

\section{Two Case Studies}
\label{sec:case-study}
%%\vspace{-5pt}
We conduct two case studies to show  \name  in action: (a) finding live \hosts cost effectively, and (b) studying 
temporal behavior of a "bad"
neighborhood.

\vspace{-10pt}
\subsection{Case-study 1: Finding live \hosts}
\label{cs-1}
Let us consider the following request: {\it Find as many as possible live \hosts anywhere in the Internet with a limit on the number of IP addresses that we can explore.} %\mim{Could we rewrite this and the following line, ignoring the limit?}
Here, we can recall the concept of \textit{RoI}: we want to find the maximum number of live servers by exploring a given number of addresses.

%This kind of study could be motivated by a research effort to identify and study botnet servers, listen-in to their communication,
%or by an actual counter-measure to take such servers down
%and even take over their botnets~\cite{YourBotnet-CCS09}. 

%\miii{Cite Yourbotnet is my botnet and anything related from there} \mim{cited}.
 
%\mim{We have not talked about the ``certain budget'' in detail here. It might confuse the reader.}
%\miii{We have defined budget in many places, and in a case study, we have some liberty to define it as we want, no?}
%the spatiotemporal behaviors of  \hosts across major IoT malware families.

{\bf Overview.}
We outline the steps that our case study will follow.
First, 
we identify and prioritize target subnets of interest by considering historical information based on the budget of the study.
 Second, we explore each target subnet.
We identify the open ports for each IP address we want to explore.
Here, we set a limit of $\maxNoPorts = 20$ ports that we will explore per target IP, to ensure that we do not deplete our budget.
For IP addresses with many port numbers, we prioritize ports based on their historical popularity which we obtain from \module \moduleB.
Finally, we actively probe each port for that address
with all the available traces.
Note that we stop the exploration for an IP address the moment
we get a \host  interaction on one of its ports.
Each of the steps above hides many subtleties, which we highlight  below.

%\miii{Mimon: please check my above revision} \mim{The write-up conveys the things that we have done. However, i think we need to define the budget of our experiment. In addition, in our evaluation, we have shown that our adaptive-probing outperforms AUTOPROBE (only historical /24 subnet centric). But here, it seems like we are only doing the AUTOPROBE stuff -- this is what we actually did. I wonder, if the reviewer questions it.}
%\miii{@Mimon: Great observations in comment. I rely on the fact that the reviewer will not figure out, but also we are NOT lying, we tell them what we do!} 
%\mim{Need to fix the ``budgeting'' stuff.}

{\bf The target space.}
%We assume the position of a practitioner:
%We establish a practically-interesting goal: we want to model the presence and longevity of C2 servers.
%{\it The space.} 
As the case-study problem setting constrains on target exploration count,  we target the spots that are more likely to have live \hosts.
We exploit the locality property and historical information of the space.
We  obtain 288 C2 IPs from historical data~\cite{C2Store-Vivek-2023}, referred to as {\bf Seed-Set}. 
These IPs belong to
 267  /24 subnets, for a total of 68,352 IP addresses.
 \miok{Even with the limit of $\maxNoPorts=20$ ports per IP,
 we may need to explore up to 1.4M IP:port pairs.}
\miii{Can we point to description of the space in Appendix? If we don't have it it is ok} \mim{the record can be found in the artifact, we have not orgranized the space of Case-Study 1 cleanly. We get /24 from urlhause, probe it exhaustively.}

{\bf Exploring the target space.}
Within our 267 /24 subnets,
 we find that 59\% (40,619) of the IP addresses seem to be "dead": they do not respond to any of our communications
 and seem to not have any open port.
%\miii{Just doublechcking: Never responded to any port we tried among 65k ports?} \mim{Found no open port while port scanning phase!}
We hypothesize that: (a) there is no device with that address, or (b) its communication is blocked.
We repeated our measurement three times 
 %over three different days 
 %one week apart 
to increase our confidence in our results.
%and overcome ephemeral behaviors and outages.
%\miii{Did we do this for port scanning}

%\miii{We may want to comment on how many and what types of ports were usually available?}
{\bf Port exploration: observations and strategy.} 
As a side investigation,
we wanted to model the use of ports by target addresses.
For this, we did a full port scan of each target address.
We found that many  \hosts often had a large number of non-standard ports open, ranging from 1 to over 10K open ports.
%Due to space limitations, we cannot expand on our results here.
%\miii{What was the median though? Maybe we can mention  that  is say 15 or 30 thus we justify our selection of 20 ports to explore: around the median} \mim{Unfortunately, the open port count was not tracked in those experiments. But the median is 8, based on the recent nmap-report data of the follow-up project.}
%\miii{This is an interesting observation we could provide more detail: distribution 60\% of \hosts had at least 20 open ports...etc} \mim{==> Our data is scattered in multiple VM in different folders holding different experiment result. My exploration, may not be 100\% accurate, shows 1864 IPs that have at least 20 ports. Please note that having more than 20 ports does not necessarily indicate maliciousness.}
% Our working hypothesis is that this high number of open ports can be either an evasion tactic to obfuscate a probing adversary, like us, or it could be an attempt to maximize its acquisition of potential bots. 
% \miii{Leaving it open seems problematic: could we at least say something with manual inspection: we found that the IPs top most open ports (>540 open ports) were indeed C2 servers. ELSE we could even remove this.}
As we stated earlier, we probe up to 20 open ports per address on their popularity which we discussed in \S\ref{result} to stay in exploration constraints.
We probe each such port using all the traces in our database. We establish a TCP connection and wait for the \host to contact us, and if this does not happen, we initiate a contact using the packets observed in the traces according to sequence and time. %our \Initial rules.
%\miii{Mimon, clarify what this means what you had said: "only found 5-10\% \rulesets are applicable". This could be a nice usage of rules study: we found that we only had to use 10\% } \mim{Yes, while probing, the \hosts responded to only a few \rulesets (5\% - 10\%) of our \playbook}

\begin{figure}[ht]
    \centering
    \includegraphics[width=\columnwidth]{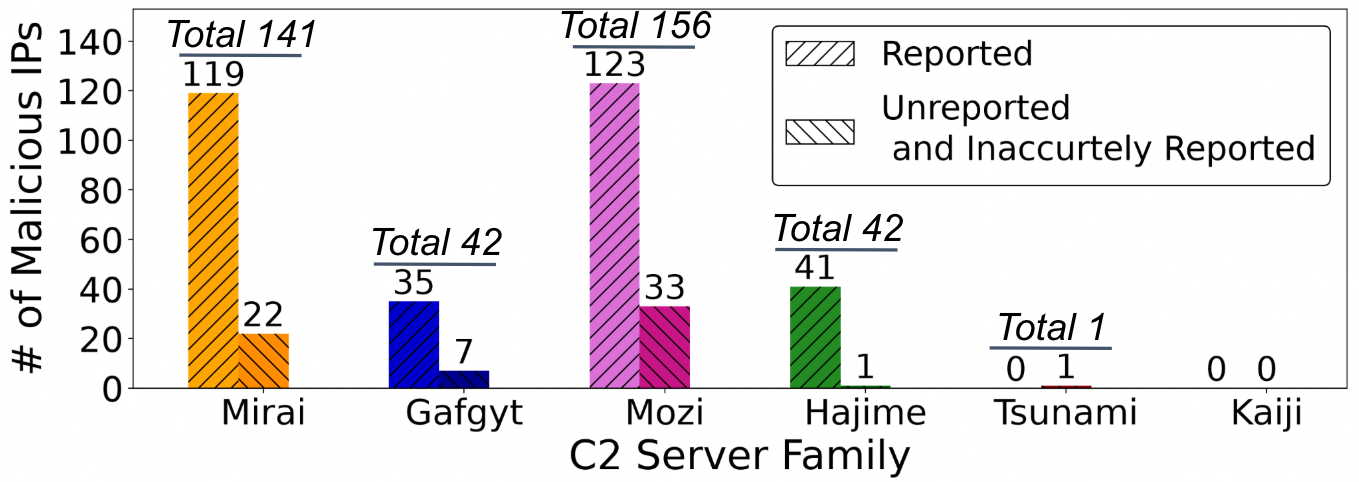}
    \vspace{-25pt}
    \caption{\small{Cross \mimok{major}-family coverage: The distribution of our live \hosts across malware families. Each second column lists the previously unreported \hosts.} }
    \label{fig:family_distribution}
    %\vspace{-20pt}
\end{figure}

\vspace{0.1cm}
\takeaway{Using historical information and prioritized probing can lead to cost-effective discovery of live \hosts.}
\vspace{0.1cm}
% We support this observation with the following results.

{\bf Result 1. We achieve high RoI with 382 live \hosts after searching 68K IP addresses.}
% \miii{Make this a highlighted Observation pink box -maybe the other ones too?} \mim{not sure if it fits as an observation.}
% \miii{Why not? It is a real scenario: request to find, we search a small space, and get good results! Let's discuss on Wed}
%\mim{yes, agree.}
%Recall that the required task was to find live servers without any guidance.
We find 382 live \hosts by exploring only 68K addresses in areas of the given IP space under the exploration constraints.
Among these live \hosts, we find \textbf{ 194 \hosts} that were not in the Seed-Set.
%    382 - 188 = 194
%\miii{Mimon: please clarify the previous numbers} \mim{is it better now?}
% 183 + 199 =   382
%%%%
%Among the live \hosts,  we have 183  C2 servers and 199 P2P bots. 
We claim that finding 382 live \hosts exploring 68K addresses is a relatively high return on investment.
%we were asked to find live \hosts anywhere in the Internet and %using hints from historical data,
%we explored  and found . 

 %\miii{Let's make clear definitions for the two commands I created hierCnC and ptopCnc in section two! I believe we should have one term that refer to both C2s and P2P.}
%\miii{I don't think  we want to highlight that most of these subnets were active: "we found  malicious activity in 198 of these subnets". }

 {\bf Result 2. We find 64 previously-unreported live \hosts}.
  %{\it Do we find unreported \hosts?}
 %The main goal for our study was to find live \hosts.
 Interestingly, among our 382 live \hosts, 
 we found 64 servers to be unreported by any of the three popular 
\tifs~\cite{VirusTotal,URLhaus,IBMXForce} at the time of the experiment as shown in Fig.~\ref{fig:family_distribution} and detailed in Appendix~\ref{sec:Appendix-TIF-misclassification-CaseStudy-1}. \miok{To rule out false positives, we manually inspected these newly discovered unreported \hosts. Their traffic consistently exhibited strong behavioral signature characteristics of their respective families, confirming their role as \hosts. To enable third party validation, we release the C2 IP:port pairs and the binary hashes via anonymous artifact~\cite{BotScan-sharing}}. 
\miii{"we will release" or "We release" in the related URL?} \mim{"We release" should be more suitable. Corrected.}

{\bf Result 3. Good malware family coverage: five major families.}
%\mimok{and their variants}.}
We evaluate if \name can provide results for different malware families \mimok{and their variants}.
% \miii{Let's add here the discussion about family breakdowns: let's again identify 1-3 main points and make those and refer to the figure for details. Btw, the color meanings in the figure is not easy to understand: let's simplify the message:
% a) we cover 5 important families, b) we find unique (non treat reported) \hosts for each family. We could drop Kaiji or leave it and mention zero activity, maybe not surprising as it is less widely used so far - if that's true.}
\name detects \hosts from five major families
\miok{(out of the six families that we probe for)}: 141 Mirai, 42 Gafgyt, 156 Mozi, 42 Hajime, and 1 Tsunami (Fig.~\ref{fig:family_distribution}). Note that a major family often includes several different variants\miii{ADD number: ", as we observe XXX (46?) distinct communication protocols"}. 

{\bf Result 4. Trade-off exploration: 100 ports cover 52\% of the \hosts.}
In a post-experiment analysis, we answer the hypothetical question: 
{\it How many live \hosts would we have found if we only explored the top 100 most popular ports?
}
In this limited exploration, we would have found  52\% of the live \hosts that we found now.
%\miii{@Mimon: the "now" refers to a smarter probing method regarding ports or}
%\miii{@Mimon: The difference is that  we explore up to 20 open ports, even if they are not in the top 100, correct?} \mim{yes, correct.}
% Although we scan all open ports for each address, our post-experimental analysis find that leveraging the {\it ranked\_ports} with strict scanning (only 100 ports per IP) can cover more than 52\% of our identified C2 \hosts. However, due to the wide port ranging behavior of P2P bot IPs, this optimization may not perform well.} \mim{I think, this part can be shifted to discussion.
This is indicative of the information that can help inform 
a probing study that wants to balance the trade-off of success and probing cost. 
\miii{I wonder if we can turn this into a trade-off study:
100 ports with a cost of X (possibly 100/65K) gives you 52\% detection of live servers. We could even report 50, 100, 200 ports, cost and success (IF we have space)
} \mim{great idea, Professor. We have data; we can describe it in detail in the appendix.}

%\mim{This para is repetitive with the last paragraph of \S\ref{sec:locality}}
\miii{I think: This paragraph focuses on a particular target space, as opposed to popularity of reported ports. We do active probing and its results are not included in the C2Store data. The fact that the numbers agree validates both measurements} \mim{that's a valid point.}

\vspace{-5pt}
\subsection{Case-study 2: Temporal behavior}
\label{sec:case-study-2}
%%\vspace{-5pt}
%\subsection{Case-study 2: The temporal behavior of a "bad" neighborhood}

%To study how malicious activity evolves over time, we extended our Spatial-Temporal Study by monitoring a selected subset of previously scanned subnets. 

In this study, we quantify the persistency of \hosts in terms of being responsive to probing over time.

%\vspace{0.1cm}
\takeaway{\host responsiveness is "flaky" in our study, necessitating repeated probing for reliable detection.}
%\vspace{0.1cm}

\miok{
We selected 38 subnets from Case Study 1 to measure C2 responsiveness over time: 12 Mirai, 5 Gafgyt, 15 Mozi, and 6 Hajime subnets, each containing 1–20 confirmed live \hosts. We probed all open ports every 2 days for 20 days using all available traces. The median response rate per server was 1 in 10, indicating that C2 servers are \textit{flaky} and repeated probing is necessary for reliable detection. Over the 20 days, we discovered 514 additional live C2 servers, 48 of which were unreported by the three threat intelligence feeds. We present this case study in Appendix~\ref{apppendix-Case_study_2}.
}

\noindent
\textbf{Combined impact.} Across both case studies, \name identified 896 live C2 servers in total, of which, 112 were previously unreported by the popular \tifs.

%Michalis major revision
%===============================

%\vspace{-15pt}
\section{Discussion}
\label{sec:discussion}
 We discuss  the practical impact and limitations of our work.

{\bf a. \name Impact: enabling important applications.}
%We see \name  as a missing weapon in the fight against botnet-based cybercrime. \name can have significant practical impact, especially as we intend to make our tool publicly available for deployment.
%code base.
%The user can deploy our tool as-is with the information, IP and port reputation and statistics that it incorporates.
We see \name as a practical, deployable capability against botnet-based cybercrime, which we intend to release publicly. We will also be updating the knowledge-base of \name periodically with external sources. %collecting data from external sources.
%, creating an updated \playbook and consider new reports on malicious addresses, port utilization patterns.}

{\em Intended user and usage.} 
\miok{
\name can serve security researchers, practitioners, and network and threat-intelligence operators to: (a) discover new live C2 for an emerging malware family, (b) validate the liveness of known C2 servers, (c) assess the health of a network space, or (d) monitor a botnet.
}

\begin{comment}
  We anticipate that \name could be used by:
(a) security researchers and practitioners,
(b) \tif operators, and (d) network operators.
The tool could be used for a plurality of studies:
(a) validating the liveness of a \hosts at the time of observation,
(b) finding the live \hosts for an emerging malware family,
(c) evaluating the health of a network space,
(d) monitoring a botnet by engaging the communication.  
\end{comment}

%\miii{Michalis can revise this: potential applications and users and have some of that in the introduction as well.}

%Detecting C2 servers is essential for mitigating cyber-crime and defending networks against malicious activities. Consequently, \name can be an important part of network defense systems, ensuring network security and contributing to a safer and more trustworthy Internet environment. 

%{\bf b. Does our approach work for all malware families?}
%We have seen that our approach works well for six popular families. 
%This success seems to suggest that our approach can extend to malware families in the \Fixed or \Tricky categories.

%with the following caveats.
%First, the communication protocol should belong to the \Fixed or \Tricky category.
%Second, we need to know its \ruleset, which we can get either by activating its binary, or from a captured network trace.

{\bf b. Are our datasets and results representative?} 
This is the typical concern for any measurement or deployment study. We argue that our study is reasonably representative since:
(a) we collect  recently-reported IoT malware binaries from  Malware Bazaar {\bf hourly}, %focusing on six major IoT-supported architectures,
(b) our method studies six major families and finds live \hosts for five of them,
and (c) we use arguably the most recent and comprehensive botnet information aggregator~\cite{C2Store-Vivek-2023}. 
The results in our evaluation seems to suggest that \name can apply to any malware family in the \Fixed or \Tricky categories.

%Furthermore, we argue that we made every effort to get binaries and information that is as representative and as unbiased as possible, as we recap below.
%This comprehensive approach we followed throughout the work ensures that both our dataset and results are representative of current C2 server behaviors and IoT malware trends. 
%First, our malware binaries are sourced from Malware Bazaar, focusing on recently reported IoT binaries. 
%containing a substantial number of recent real-life malware samples from various prominent families, ensuring comprehensive and relevant coverage. 
%Second, we activated the malware multiple times and from different geographic locations, when we extracted the  network traces, in order to ensure that our approach works from any location. 
%Finally, the authenticity and accuracy of our results are further validated through verification with leading threat intelligence platforms, including VirusTotal, UrlHaus, and IBM X-Force. 
% Third, we leverage  C2Store, arguably the most recent and comprehensive botnet information aggregator~\cite{C2Store-Vivek-2023}.
%a comprehensive reference database of malware activities, which critically synthesizes many different sources of C2 server reports.

{\bf c. Is the \name approach generalizable?}
Our approach will work to any malware,
 as long as the bot-C2 protocol is \Fixed or \Tricky. Note that with our streamlined capability, we are constantly collecting and profiling malware binaries (\Module \moduleA), and finding ways to manage them for replay. Similarly, our behavioral-adaptive algorithm is effective for any C2 family as long as there is locality, a well-established attribute of C2 servers~\cite{xu2014autoprobe, nappa2014cyberprobe, C2Miner-Davanian-2024}.
%the \host will respond to our probes.
%15\% are encrypted of which 9.25\% are replayable

{\bf d. Can hackers evade our approach, possibly using encryption?} 
Non-replayable bot-C2 communication protocols will pose a problem for the current implementation of \name. 
However, we argue the following.
First, as we saw, currently, for our binaries,
15\% are encrypted, and 9.25\% are replayable, although encrypted.
% 7\% 
% of IoT malware binaries employ any kind of  encryption. Further, the majority of these encryption-using binaries still generate replayable packets, which \name can handle as we saw in \S\ref{sec:results-validation-replay}. %Table~\ref{table:taxonomy}.
Second, generation of \Obfu category will likely cost hackers in terms of: (a) using more complex protocols, (b) requiring the distribution of secret keys, (c) requiring more computational capabilities and power from IoT devices.
 %(a) some IoT devices may not support encryption easily,
 %and (b) key distribution and management can add complications.
Third, reports predicting the use of encryption among IoT devices vary wildly~\cite{zipdo-cyber-threat-statistics, paloalto-iot-visibility-2023}.

%from  \miok{40\% to 60\%

Overall, we believe that  {\bf a security approach is successful, when hackers are forced to change their behavior}.

{\bf Ethical Considerations.}
 We adhere to best practices of ethical research. First, we do not manage any personally identifiable information as we only probe devices. Second, when analyzing malware, we use a fully enclosed sandbox environment, which filters out harmful scanning/propagation traffic and isolates execution to prevent unintended spread. We only allow communication with malware's own \host. Moreover, we do not execute any commands from the \hosts. Finally, our measurement was conducted in a way that had minimal effect on server load and network traffic by rate limiting our probing. 
 Additional discussion can be found in Appendix~\ref{sec:Extra-Ethical}.
\section{Limitations and Future Work}
\label{sec:limitation}
\miok{\name has a few limitations that can suggest future work. First, any \playbook approach will most likely not work for C2 protocols with sophisticated session-specific
encryption (\S\ref{replay_feasibility}). 
Such protocols require a more involved approach that could include reverse binary engineering of the communication protocol or resorting to malware binary activation.
%\miii{Check if what I say makes sense above. Your line was hard to parse.} \mim{looks great!}
%We plan to add a machine-learning classifier that fingerprints encrypted exchanges from side-channel features, with selective binary activation for this small tail. 
%\miii{I don't think the ML classifier was clear: it is fair to say: any packet replay approach will not work with dynamically session-specific encryption}
Second, detection depends on a C2
server's responsiveness: an inherent constraint of active probing is  the ``flaky'' ( 1-in-10 response rate \S\ref{sec:case-study-2}). We can mitigate this problem by repeated probing, while we can attempt to understand if there is a pattern or reason in the responsiveness.
%and plan to model responsiveness to schedule probes more efficiently.
}

\miok{We also have identified two  possible extensions. First, \name discovers live \hosts but does not yet close the loop: we want to introduce an automated pipeline that reports them to threat-intelligence
feeds and tracks them until they are taken down. Second, we want to extend our work to IPv6 by: (a) identifying if IoT malware exists in IPv6 addresses, and (b)  extending the segment-reputation model to the far sparser IPv6 space, where a /24-based granularity may need to be revisited.}

%\vspace{-5pt}
\section{Conclusion}
\label{sec:conclusions}
\label{sec:conclusion}
Our work introduces \name, a comprehensive and efficient approach to actively scan for IoT
\hosts at scale and answers our two research challenges: {\bf RC\#1: What to probe with?} We develop a \playbook probing approach that starts with a malware binary,   extracts the communication packets,  and
 determines if the packets are replayable;
 % (\Fixed or \Tricky category). 
 {\bf RC\#2: Where to probe?} We develop an efficient \smartProbing  probing approach that: (a) prioritizes the target IP addresses,  (b) determines which port  to explore, and (c) adapts dynamically as probes find live servers. 
 The approach is informed by the observed spatiotemporal behavior of \hosts, and can be updated as these behaviors evolve in future.
 %: (a) the \playbook, and (b) a measured-driven profiles of botnet behaviors.

As a proof of concept, we deploy \name to demonstrate the capabilities, observations and insights that it can provide.
We intend to expand, evolve, and share \name with the research community as an open-source project.

\bibliographystyle{ACM-Reference-Format}
\bibliography{AsiaCCS27_BotScan}

\appendix
\section{Open Science} We follow open-science principles by sharing our tools and data to help others reproduce our work and use it safely. We are releasing the \name tool and probing pipeline, the replay signatures, and the C2 IP:port pairs we found, along with the hashes of the malware we analyzed. However, we are not sharing the actual malware files to prevent any misuse. You can still identify these files by their hashes on MalwareBazaar. All of this information is available to the public anonymously~\cite{BotScan-sharing}.

\section{Related Work: Passive approaches and synergistic efforts.}
\label{sec:Appendix-Related_work}
\textbf{i. Passively-collected network-trace-based methods.} 
Many studies operate in a passive fashion:
   they analyze the network traffic to identify botnet communications~\cite{fuller2021c3po, gu2008botsniffer, jacob2011jackstraws, troutman2018detecting, vandetecting, hafeez2020iot}.
   The approach is passive in that it cannot ``query'' an IP address, but instead, it waits to observe suspicious behavior from that address, if and when this happens.
    These efforts fall within the greater family of network traffic and application classification~\cite{azab2024network, KimCFBFL08, rezaei2019deep, erman2006traffic}. 
   % \miii{cite 2-3 fundamental traffic classification papers }
   Unsurprisingly, many recent approaches develop AI-based classification algorithms ~\cite{Hugo, palo_alto, sampled-network-flow-and-novelty-detection, detection-with-hierarchical-attention, ongun2021portfiler}. 
   A different line of work relies on DNS traffic analysis to detect and block malicious servers ~\cite{botacin2020empirical, ghafir2015dns, wang2017dbod, zhao2015detecting}.
   % \krish{I again am not sure what is different. Are these offline? Highlight the differences and why those are not sufficient for live C2 server detection.}
   %However, the key limitation of these methods is the need to collect and access (nearly) all  the related network traffic. 
   Overall, these methods are passive, as they cannot actively examine an arbitrary IP space. Instead, they rely on having access to the related network traffic, which restricts their scope: they need a cooperative network owner who is capable of collecting and willing to share the traffic traces.
   \miok{For example, convincing an ISP to provide a substantial amount of their network data for a research study for an emerging malware is an uphill battle with highly uncertain outcome.
   Active probing is the most reliable and available option for a researcher.
   }

   {\bf ii. Synergistic related work.}
For completeness, we highlight some complementary and synergistic efforts.
First, there are some efforts that analyze network traffic to detect botnet traffic and behavior at large~\cite{antonakakis2017understanding,alrawi2021circle, tanabe2020disposable, almazarqi2021profiling, blaise2020botnet}.
Second, many efforts develop honeypots and archived datasets~\cite{spitzner2002honeypots, wang2020iotcmal, sethia2019malware,anghel2023peering, trajanovski2021automated}. 
% Third, there are some efforts for building scanner and service prediction~\cite{durumeric2013zmap, izhikevich2022predicting}.
Finally, many efforts develop effective sandbox platforms~\cite{darki2020riotman, kachare2022sandbox, jamalpur2018dynamic, alrawi2021circle, bellard2005qemu, cuckoo}.

\section{\Module \moduleA: Activation results}
\label{sec:ModuleA-Activation-Results}

%%%\vspace{-0.2cm}

%====== BEGIN Move to Appendix ==========

We start with 1842 collected malware binaries and successfully activated 1634 binaries (89\% of 1842) in our sandbox. We then find that 1536 binaries (84\% of 1842) that managed to  establish  a TCP handshake with their \host.
From those, only 75\% of them exchange more packets, and we say that they {\em connect} meaningfully to their server.
% We extract the \playbook entries for these binaries with our algorithm.
To test the replayability of a binary,
we
%to ensure the stability of the C2 server.
%in order to verify that the \playbook \entries will work.
 replay the same packet sequence,  \mimok{with minimal customization as discussed earlier, if required,} from a different IP address and attempt to connect to the same \hosts that the binary had just contacted in the previous step. 
 %\mimok{For the sophisticated cases, as discussed earlier, we check the replayability of distinguishable TCP service-based fingerprinting and utilize them to detect the C2 servers.}
This  step is successful for \mimok{at least} 1371 binaries (75\% of 1842). %\mim{at least 1371 binaries -- the other 9\% (connected 84\% - successfully attempted 75\%) failed due to their offline C2 server.}
However, among these 75\% binaries, we find that 3\% are %of them  
with \Obfu communication protocol. 
In conclusion, we claim  that %(72\% of 1842) %(1321) 
we can use our approach to probe successfully \mimok{at least} 72\%  non-\Obfu binaries from the initial 1842 binaries. %\mim{at least 72\%.}

\begin{table}[h]
  \centering
  \small
   \centering
        \resizebox{0.75\linewidth}{!}{%
      \begin{tabular}{ |c c c c| } 
        \hline
        \rowcolor[RGB]{90,240,110} \mimok{Major} Family & Collect  & Activate  & Connect  \\ 
        \hline
        Mirai & 1347 & 91\% & 87\%\\
        \hline
        Gafgyt & 368 & 87\% & 84\%\\
        \hline
        Kaiji & 19 & 79\% & 58\%\\
        \hline
        Mozi & 12 & 75\% & 50\%\\
        \hline
        Hajime & 31 & 84\% & 58\%\\
        \hline
        Tsunami & 43 & 63\% & 33\%\\
        \hline
        Unknown & 22 & 59\% & 32\%\\
        \hline
        \rowcolor[RGB]{250, 200, 250} \textbf{Total:} & 1842 & 89\% & 84\%\\
        \hline
      \end{tabular}%
    }
    \caption{\small{High activation and communication success: our  malware binaries by \mimok{major} family and their activation, and \hosts connection rates.
    Connection means establishing a TCP handshake.
    %Connection success rate is calculated for the activated malware.
    }
    %%\vspace{-20pt}
    }
    \label{table:collection_activation}
\end{table}

%====== END Move to Appendix ==========

\section{C2 Traffic Disambiguation}\label{appendix:C2Miner} 

We present a brief overview of the disambiguation approach~\cite{C2Miner-Davanian-2024}, which we adopt in our work.
To distinguish C2 traffic from scanning, exploitation, and other  traffic generated by IoT malware, we apply a target-centric traffic disambiguation algorithm. A {\em target} is defined as either an IP:port tuple or a DNS name. Since C2 traffic represents only a small fraction of the overall activity, the algorithm assigns each target a likelihood score indicating how likely it is to correspond to a C2 server.

The algorithm proceeds in two phases. First, it quantifies communication activity by analyzing TCP traffic only, filtering out unrelated protocols such as ICMP, ARP, DHCP, and NTP. For each observed packet, the algorithm tracks how often each target is contacted. For IP:port targets, this is the number of packets exchanged. For DNS-based targets, successful resolutions are treated as IP:port targets, while repeated failed DNS queries are counted directly, capturing attempts to reach blocked or unavailable C2 domains.

Second, a C2 likelihood score is computed for each target. DNS targets are filtered based on domain reputation, excluding highly popular domains using Alexa ranking; remaining DNS targets are scored by query frequency. For an IP:port targets, the score is computed as the number of packets sent to the target divided by the number of times the destination port is used across all IPs. This formulation favors endpoints with frequent communication while penalizing ports broadly reused for scanning or propagation. Targets are ranked by score, and the top-ranked target (or top N, configurable) is selected as the malware’s C2 server.

\section{\SmartProbing Probing Algorithm}\label{appendix:probing-algorithm}
%Our \SmartProbing probing is shown in Algorithm~\ref{algo:ip-address-priortization}.
% It takes as input a target space, a set $Sus$ of historically reported IP addresses, which could be empty, a probing budget,  and several parameters governing the prioritization strategy, including the \segment prefix size ($\segk$), neighbors ($K$)  and decay factor ($\lambda$), which we explain below. The output consists of a list of live \hosts. 

\begin{algorithm}[h]
\caption{BotScan's \SmartProbing probing}
\label{algo:ip-address-priortization}
\hspace*{\algorithmicindent} \textbf{Input:} \textit{IP-space}, \textit{Sus}, \textit{Budget}, \textit{r-explore}, \textit{K}, $\lambda$\\
\hspace*{\algorithmicindent} \textbf{Output:} Live \host address
\begin{algorithmic}[1]
\STATE \textit{Split(\textit{IP-space})}
\STATE \textbf{For each} \textit{ip} $\in$ \textit{Sus}:
\STATE \quad \textit{Scan(ip)} 
\STATE \textit{RankSegments(K, $\lambda$)}
\WHILE{\textit{Budget} $\neq$ 0}
    \IF{random() $<$ \textit{r-explore}}  
        \STATE \textit{r\_seg} $\gets$ Random(\textit{Seg(s)})
        \STATE \textit{Scan}(\textit{r\_seg}) 
        \STATE \textit{Seg(s)} $\gets$ \textit{Seg(s)} $\setminus$ \textit{r\_seg}
    \ELSE
        \STATE \textit{t\_seg} $\gets$ Top(\textit{Seg(s)})
        \STATE \textit{Scan}(\textit{t\_seg})
        \STATE \textit{Seg(s)} $\gets$ \textit{Seg(s)} $\setminus$ \textit{t\_seg} 
    \ENDIF
    \STATE \textit{RankSegments(K, $\lambda$)}
\ENDWHILE
\STATE \textbf{Return} results
\end{algorithmic}
\end{algorithm}

The Algorithm~\ref{algo:ip-address-priortization} illustrates the \SmartProbing probing in steps as follows:

{\bf a. Splitting the (\textit{IP-space}) into Segments}:
As described in Line 1, the process begins by splitting the \textit{IP-space} into segments of prefix size $\segk$. The value of $\segk$ is determined following the methodology outlined in \S\ref{sec:methodology-spatial-locality}.

{\bf b. Initial \textit{Sus} Scanning and Ranking (Lines 2–4)}: The algorithm begins by scanning each IP address in the set \textit{Sus}. Based on the outcomes of the scans, it ranks the segments using a scoring mechanism

{\bf c. Segment Exploration Loop (Lines 5–16):}
Given we have enough budget (checked in {\it Line 5}), during each iteration of the loop, we do as follows: \\
(i) As in {\it Line 7}, it selects segments probabilistically, favoring the top-ranked segments but occasionally randomly to balance exploration and exploitation, \\
(ii) The selected segment is probed \mimok{exhaustively to save space} in {\it Line 8 \& 12.} 
\miii{better say "probed exhaustively to save space" -- not sure what is } \mim{done. it means, no prioritization in a segment, all ips in a segment are probed sequentially.}
\\
(iii) An explored removed from the list of unexplored segments in {\it Line 9 \& 13.}\\
(iv) At the end of each iteration, the algorithm re-ranks the remaining unexplored segments on the updated information, as shown in {\it Line 15}. This ensures that the selection process in subsequent iterations is guided by the latest data.
\miii{We may want to revisit the explanation! Not sure it matches...} \mim{Re-ranking is already mentioned in Step 7 on page \#13}

{\bf d. Termination and Output}:
The loop continues until the budget is exhausted. Finally, the algorithm compiles and reports the findings of the exploration, highlighting the live \hosts discovered during the process.

\section{C2 Behavioral Profiling Dataset \textit{BP-DS} Statistics, Port Distribution, and Locality Analysis}
\label{sec:appendix_dataset}

This section provides a detailed description of the dataset \mimok{\textit{BP-DS}}, which we obtain from Module A, C2Store~\cite{C2Store-Vivek-2023}, and Threatfox~\cite{threatfox} between
2020-2025 and use in the historical data-driven C2 behavior profiling phase. We also present C2 behavioral patterns concerning spatial locality and port usage.

\textbf{a. Dataset \mimok{\textit{RP-DS}} in detail and Subnet Locality.}
\begin{table*}[t]
\centering
\small
\begin{tabular}{|c|c|c|c|c|c|c|}
\hline
\multirow{2}{*}{\textbf{Family}} & {\textbf{Total}} & {\textbf{Unique}} & {\textbf{Unique}} & \multicolumn{3}{c|}{\textbf{Locality Density}} \\
\cline{5-7}
 & \textbf{IP:port pair} & \textbf{IP count} & \textbf{Port count} & \textbf{/24} & \textbf{/20} & \textbf{/16} \\
\hline
Mirai  & 13,899 & 10,324 & 4,536 & 0.751 & 0.059 & 0.005 \\
Gafgyt & 2,566  & 2,271  & 572   & 0.526 & 0.045 & 0.005 \\
Mozi   & 715    & 711    & 668   & 0.397 & 0.029 & 0.003 \\
Hajime & 890    & 849    & 840   & 0.404 & 0.027 & 0.002 \\
Kaiji  & 144    & 134    & 41    & 0.463 & 0.032 & 0.002 \\
\hline
\textbf{Overall} & \textbf{18,214} & \textbf{14,120} & \textbf{6,350} & \textbf{0.670} & \textbf{0.055} & \textbf{0.005} \\
\hline
\end{tabular}
\caption{Data statistics of \mimok{\textit{BP-DS}} and subnet locality density across IoT malware families.}
\label{tab:subnet_locality}
%%\vspace{-20pt}
\end{table*}
Table~\ref{tab:subnet_locality} summarizes the size of our dataset and the subnet locality density across five major malware families. Our comprehensive \mimok{\textit{RP-DS}} contains a total of 18.2K unique IP:port pair, including 14.1K unique IPs and 6.4K unique ports. To address biases in the dataset, such as the dominance of Mirai compared to the lesser presence of Kaiji, we processed the data for each family to avoid cross-family inferences.

The table presents the locality density, calculated following the methods described in Section~\ref{sec:locality}, for prefixes: /24, /20, and /16. As mentioned earlier, the locality density significantly decreases when shifting the prefix from /24 to /20 and finally from /20 to /16.

Among all families, Mirai exhibits the highest subnet locality at 0.75\% at the /24 level. This indicates that its infrastructure tends to be more concentrated within shared or closely assigned IP ranges. Gafgyt shows moderate locality at 0.52\%, whereas Mozi, Hajime, and Kaiji demonstrate comparatively lower clustering, suggesting a broader distribution across subnet spaces.

\textbf{b. Port Distribution Analysis.} Table~\ref{tab:port_distribution} presents the distribution of C2 ports used across the the malware families. We highlight the most frequently utilized ports and provide cumulative coverage for the top 50 and top 500 ports. This analysis allows us to determine whether C2 communication is concentrated within a limited number of ports or if it is spread across a wider array.

\begin{table}[t]
\centering
%%\vspace{-5pt}
\small
\begin{tabular}{|c|c|c|c|c|}
\hline
\multirow{2}{*}{\textbf{Family}}
  & \multicolumn{2}{c|}{\textbf{Top Port}}
  & \textbf{Top-50 Ports} & \textbf{Top-500 Ports} \\
\cline{2-5}
 & \textbf{Port} & \textbf{Coverage} & \textbf{Coverage} & \textbf{Coverage} \\
\hline
Mirai  & 7173 & 18.3\% & 53.9\% & 70.8\% \\
Gafgyt & 23   & 20.2\% & 70.3\% & 97.2\% \\
Mozi   & 6881 & 1.4\%  & 13.6\% & 76.5\% \\
Hajime & 6881 & 0.8\%  & 11.1\% & 61.8\% \\
Kaiji  & 808  & 34.6\% & 100\%  & 100\% \\
\hline
\textbf{Overall} & 7173 & 14.0\% & 47.9\% & 66.0\% \\
\hline
\end{tabular}
\caption{Port distribution characteristics across IoT malware families.}
\label{tab:port_distribution}
%%\vspace{-20pt}
\end{table}

The Mirai and Gafgyt families demonstrate a moderate to strong concentration on a select number of ports, suggesting a reliance on common or reused configurations. Specifically, the top 50 most popular ports for Mirai account for 53.39\% of historical C2 IP addresses, while for Gafgyt, the coverage is 70.3\%. In contrast, Kaiji shows a high concentration of activity, with nearly all its communication focused on a very small set of ports. On the other hand, Mozi and Hajime exhibit a broader dispersion of ports, indicating more varied communication patterns.

In summary, port usage varies significantly among different malware families. Some display highly structured and concentrated configurations, whereas others distribute their communication more evenly across a wider range of ports.

\section{Selecting the Segment Parameters $K$ and $\lambda$}
\label{sec:param-selection}

This section details how we choose the neighborhood breadth $K$ and the
decay $\lambda$ of \name's inter-segment reputation (\S\ref{sec:behavior-profiling}).
We select a single fixed configuration on historical data, without using any
live evaluation data.

\textbf{Tuning data.}
We tune on BP-DS (\S\ref{sec:behavior-profiling}), our historical C2 dataset, which is
disjoint from the target spaces used in \S\ref{sub:eval1}, and
\S\ref{sub:eval2}. The procedure is family-agnostic; we apply it to
Mirai and Gafgyt, the two dominant families that together account for $93\%$
of our binaries. All tuning replays the algorithm on historical data; no live
probing is involved.

\begin{figure}[ht]
  \centering
  \includegraphics[width=\columnwidth]{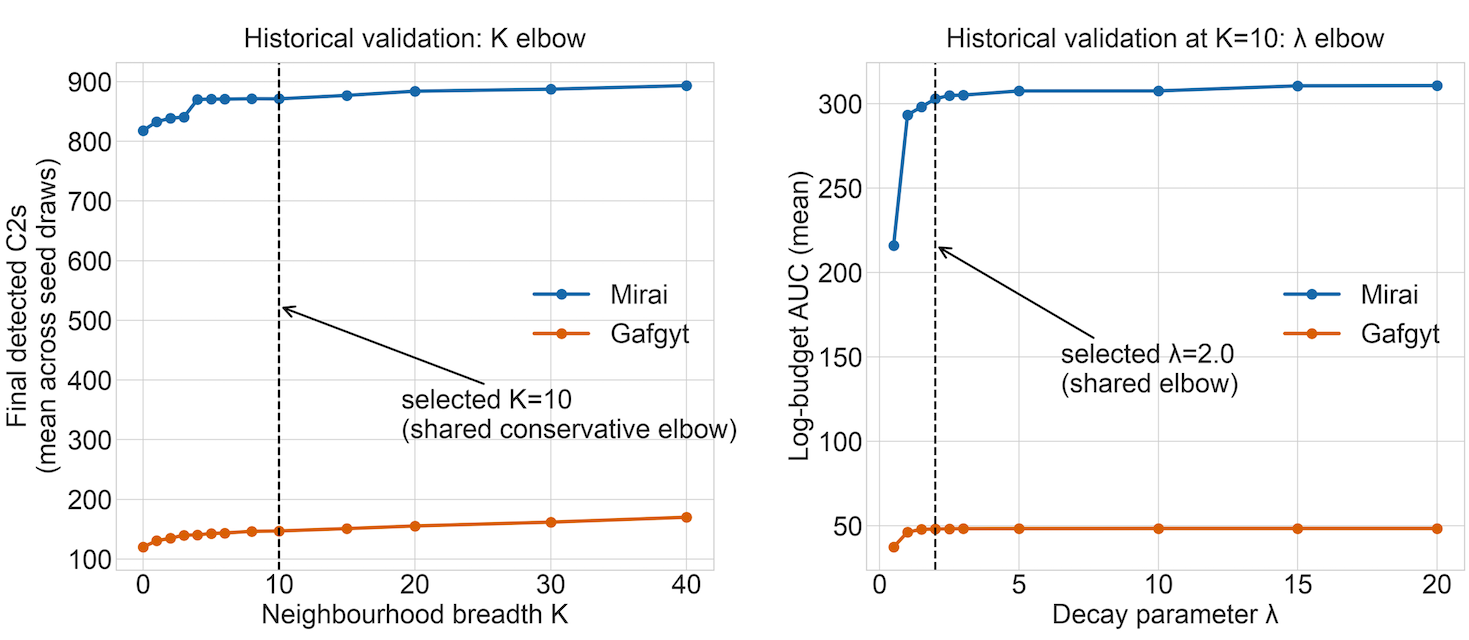}
  \caption{\textbf{Parameter selection on historical data (BP-DS).}
  Left: final detected C2 servers vs.\ neighborhood breadth $K$; the
  $10\%$-diminishing-returns elbow is at $K=10$.
  Right: budget efficiency (log-budget AUC) vs.\ decay $\lambda$ at $K=10$;
  the elbow is at $\lambda=2.0$.
  Curves are means over random seed draws for the two dominant families.}
  \label{fig:param-tuning}
\end{figure}

\textbf{Selection rule.}
For each parameter we sweep a grid and apply a diminishing-returns (elbow)
rule rather than taking the maximum. Let $\Delta$ be the improvement per grid
step and $\Delta_0$ the initial (steepest) improvement at the start of the
sweep. We select the smallest value at which $\Delta$ drops below $10\%$ of
$\Delta_0$. We use an elbow rather than an \emph{argmax} because performance
keeps rising slowly across the whole grid (Fig.~\ref{fig:param-tuning}): an
\emph{argmax} would pick the largest tested value for a negligible gain,
whereas the elbow captures most of the benefit at a smaller, more parsimonious
setting.

\textbf{Procedure.}
We sweep in two stages and average over multiple random seed draws to reduce
variance. First, we sweep $K$ and measure the final number of detected C2
servers to locate its elbow (Fig.~\ref{fig:param-tuning}, left). Second, we
fix $K$ at this value and sweep $\lambda$, measuring budget efficiency as the
area under the detection-versus-budget curve, to locate its elbow
(Fig.~\ref{fig:param-tuning}, right). We use two complementary metrics because
$K$ mainly affects how many servers are ultimately found, whereas $\lambda$
mainly affects how quickly probing concentrates on productive neighborhoods.

\textbf{Result.}
The elbow falls at $K=5, \lambda=1.5$ for Mirai and $K=10, \lambda=2.0$ for
Gafgyt. To obtain one configuration for both families, we adopt the setting
that covers both elbows: $K=10$ and $\lambda=2.0$. At this setting each family
retains most of its best observed detection---$97.5\%$ for Mirai and $86.5\%$
for Gafgyt, relative to the best value over the swept grid. We prefer a single
shared configuration over per-family tuning so that \name needs no per-target
calibration; the small per-family shortfall is the cost of this generality.

\section{{\bf Evaluation 2: Large-scale emulation with historical data.}}
\label{sec:Appendix-Eval-1}
We provide an additional evaluation 
on a larger space conducting a {\it thought experiment} \mim{\textit{retrospective analysis}}. We  "emulate" a probing study using historical data over a larger space.

\begin{figure}[th]
    \centering
    %%\vspace{-20pt}
\includegraphics[width=0.8\columnwidth]{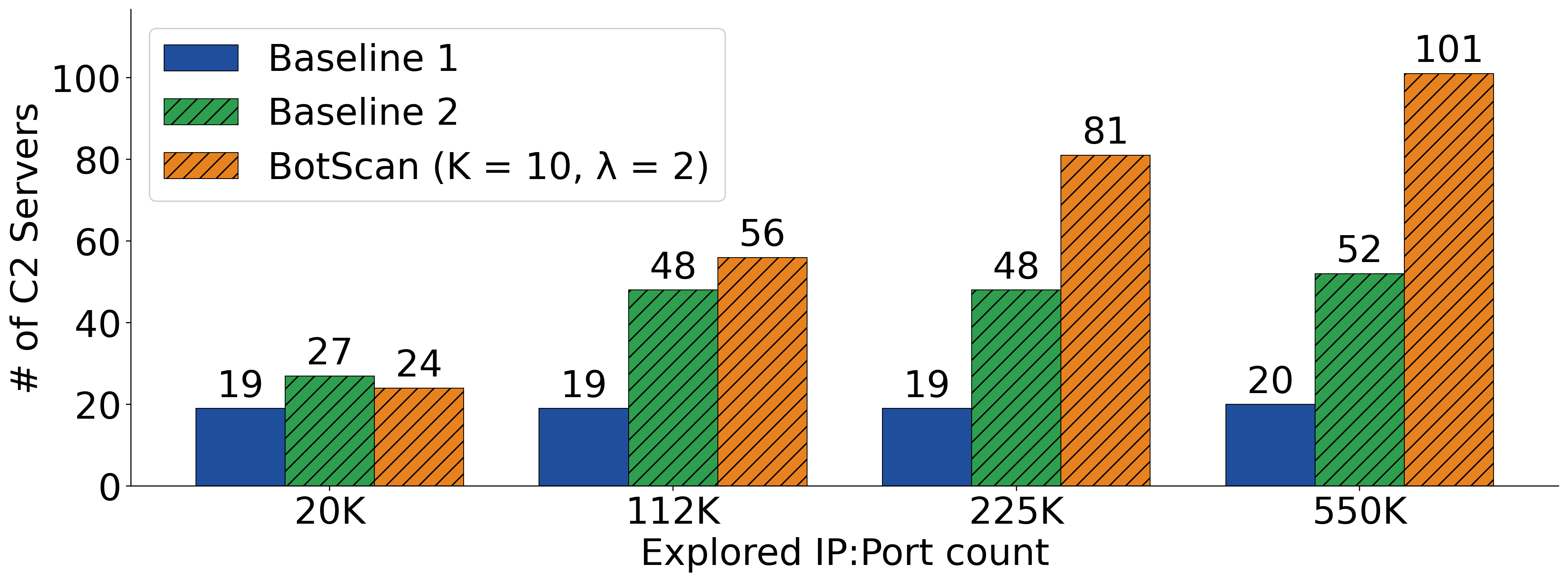}
    %%\vspace{-10pt}
    \caption{\small{Evaluation 2: The effect of the probing budget on the relative performance of the three methods. Our approach consistently outperforms the other two.} }
    %%\vspace{-20pt}
    \label{fig:budget_effect}
\end{figure}

% Address space: 103.136.40.140/8 --> ~16M total IPs
% Grountruth (historical servers): 207
% Seed information: 20
We select a large space 103.0.0.0/8, which includes 207 historically reported \hosts.
%% Explain how we selected the space: 
We chose this space due to its diverse coverage of IoT families
and its spanning 72 unique Autonomous Systems.
The exploration space is substantial with 16M IP addresses,
\miok{1.04 trillion IP:port pairs}.
We uniformly random select 20 servers as our seed group.
Therefore, the goal of a probing effort is to identify the 
187 (=207-20) \hosts, which we "pretend" we are not aware of.
% Our evaluation consists of the following steps.
% First, we consider the 207  reported \hosts as our groundtruth, which a probing method should identify.
% Second, in our probing, we pretend that we are not aware of these \hosts except for a small initial seed subset.

{\bf Our approach performs better for several probing budgets.} \mim{budgeting stuff.}
We  evaluate the impact the probing budget on the relative performance of our approach. 
We vary the probing budget and we repeat our evaluation for 20K, 112K, 225K, and 550K IP:port pair targets.
We consider that we explore on average 10 ports per IP address, which is close to the average number of open ports that we observed. Since this is the same for all three methods, it does not affect the comparison, but it helps us estimate the probing budget.
We plot the relative performance in Fig.~\ref{fig:budget_effect}. \mimok{Our approach consistently outperforms the baseline methods by approximately a factor 2 compared to Baseline-2.
Furthermore, both our approach and Baseline-2 outperform Baseline-1 significantly.
} %doing roughly twice as well as Baseline-2, and typically 10-20 times better than Baseline-1.
These results are a strong indication that both the \intraSeg and \interSeg locality are important properties to consider in an efficient probing strategy.

\section{Detailed Cost and Time Estimation on parallelism}
\label{appendix:cost_estimation}

To illustrate the practical implication at scale, we extrapolate the measured per-target costs to a 10M IP:port-pair workload under realistic resource and parallelism assumptions. Under these assumptions, BotScan completes the workload in approximately 15 hours at an estimated cost of \$5, compared with approximately 304 days and \$2,330 for an activation-based approach. This estimate captures the C2 engagement component and excludes the common port-discovery cost. The calculation is indicative and depends on deployment and parallelism assumptions.

This section outlines the detailed calculations used to estimate the time and cost of large-scale probing by considering parallelism, as we discussed in \Module \moduleC (see \S\ref{sec:mod-c-bap}) The cost comparison is shown in Table~\ref{table:op_cost_appendix}.

\begin{table}[ht]
\centering
\resizebox{0.85\columnwidth}{!}{
\begin{tabular}{|c|c|c|}
\hline
\rowcolor[HTML]{64e764}
\textbf{Approaches} &
\textbf{Duration (Days)} &
\textbf{Cost (\$)} \\
\hline
Replay: \name & \textbf{0.63} & \textbf{5} \\
\hline
Activation: C2Miner & \textbf{304} & \textbf{2330} \\
\hline
%\rowcolor[RGB]{182,232,205}
%\textbf{Reduction} & 482$\times$ & 466$\times$ \\
%\hline
\end{tabular}
}
\caption{\small
Estimated duration and cost for probing 10M IP:port pairs using
replay-based and activation-based approaches. The estimates are based
on the resource usage and parallelism assumptions.
}
\label{table:op_cost_appendix}
\end{table}

\textbf{a. System Assumptions}
We have a virtual machine with 32 GB of RAM running Ubuntu 22.04 LTS, which costs approximately \$230 per month. To account for practical deployment constraints, we limit CPU usage to 70\% and memory usage to 75\%, resulting in approximately 24 GB of available memory~\cite{gcloud_general_purpose, azure_cpu_linux, vmware_memory_limits}.

\textbf{b. Parallelism Estimation}\\
From Table~\ref{table:op_cost}, for \name, each IP:port probing task requires 0.38\% of CPU usage and 11.86\,MB of memory, with an average probing time of 0.65 seconds. The maximum CPU-bound parallelism is calculated to be 184 tasks \((70 \div 0.38 \approx 184.21)\). However, to account for practical inefficiencies such as scheduling overhead and network contention, we conservatively limit the effective parallelism to 120 tasks. The corresponding memory usage for 120 tasks is only about 1.42\,GB \(\left(\frac{120 \times 11.86}{1024}\right)\). This memory usage is significantly lower than the upper limit of 24\,GB, indicating that \name is CPU-bound.\\
Similarly, for C2Miner, each instance requires 6.24\% CPU and 142.41 MB of memory, with an average probing time of 28.78 seconds. The maximum CPU-bound parallelism is approximately 11 \((70 \div 6.24 \approx 11.22)\) tasks. Consequently, the corresponding memory usage is also relatively low, totaling 1.57 \(\left(\frac{11 \times 142.41}{1024}\right)\) GB. The reduced memory usage further indicates that C2Miner is also CPU-bound.

\textbf{c. Throughput Estimation.} We estimate an affordable probing rate for BotScan and C2Miner based on the assumed system specifications. 

We have already determined that BotScan can effectively manage 120 parallel tasks. As shown in Table~\ref{table:op_cost}, BotScan can produce approximately 185 probes per second at runtime, calculated as follows: \( \frac{120}{0.65} \).

Similarly, C2Miner can produce around 0.38 probes per second at runtime, calculated as \( \frac{11}{28.78} \).

\textbf{d. Probing Duration Estimation.} Assuming the use of a single probe packet for each unique IP:port, a total of 10M unique IP:port pairs would require 10M probes. To generate these 10M probes, BotScan will take around 15 hours, calculated as \(10M \div (185 \times 3600) \approx 15.02\) hours. This amounts to roughly \(0.63\) days, calculated as \(15.02 \div 24 \approx 0.625\).

In contrast, C2Miner will take nearly \(304\) days to complete the same task, calculated as \(10M \div (0.38 \times 3600 \times 24) \approx 304.58\) days.

\textbf{e. Cost Estimation.} The monthly cost of \$230 translates to a daily cost of approximately \$7.67, calculated as \(230 \div 30\). Therefore, to complete the job, BotScan will cost around \$5, calculated as \(0.625 \times 7.67 \approx 4.80\). On the other hand, C2Miner is estimated to cost about \$2,330, calculated as \(304 \times 7.67 \approx 2331.68\).

This analysis clearly demonstrates the advantages of BotScan over C2Miner, with an approximate \(482\times\) reduction in probing duration and a \(466\times\) reduction in cost while exploring a large target space.

\section{Case study: Threat Feed Misclassification}
\label{sec:Appendix-TIF-misclassification-CaseStudy-1}
  % , VirusTotal~\cite{VirusTotal}, URLhaus~\cite{URLhaus}, and IBM X-Force~\cite{IBMXForce}.
\begin{table}[ht]
%%\vspace{-20pt}
\centering
\resizebox{0.9\columnwidth}{!}{
\begin{tabular}{| c c c c c|}
  \hline
  \rowcolor[RGB]{90,240,110} 
  %\rowcolor[RGB]{95, 200, 83}
   \multicolumn{5}{|c|}{  Total Identified Malicious IPs: 382}  \\
  \hline
\rowcolor[RGB]{253, 219, 76}  \multicolumn{5}{|c|}{ Total \hierCnC C2 Servers: 183}\\
  \hline
  Category & \name & VirusTotal & URLhaus  & IBM X-Force\\ 
  \hline
  Malicious & 183 & 144 & 132 & 30 \\
  \hline
  Benign & & 2 & & 153\\
  \hline
  Unreported & & 37  & 51 & \\
  \hline
  \rowcolor[RGB]{253, 219, 76} \multicolumn{5}{|c|}{Total \ptopCnC IPs: 199}\\
  \hline
  Category & \name & VirusTotal & URLhaus  & IBM X-Force\\ 
  \hline
  Malicious & 199 & 156 & 135 & 50 \\
  \hline
  Benign & & 4 & & 149\\
  \hline
  Unreported & & 39  & 64 & \\
  \hline

\end{tabular}
}
\caption{\small{\tifs misclassify our live
\hosts. Reporting malicious, benign and unreported addresses among our malicious servers.} \mim{We also quantify how badly we should focus on active probing instead of depending on the \tifs.}}
\label{table:comparison_of_TI}
%%\vspace{-20pt}
\end{table}

We compared our identified live \hosts with well known \tifs.
  Specifically, we consider
 VirusTotal~\cite{VirusTotal}, URLhaus~\cite{URLhaus}, and IBM X-Force~\cite{IBMXForce}.
%  \footnote{
  For VirusTotal, an address is classified as malicious if flagged by at least one vendor, benign if unanimously deemed clean.
  %, and unreported if no data is available. 
  With URLhaus, we consider an IP "reported", if it is blacklisted by the platform. With IBM X-Force, an IP is deemed malicious, if its risk factor exceeds 1.0 and its behavior is classified as Malware, Spam, or Scanning; otherwise, it is labeled benign. 
%  } 

 We show the results in Table~\ref{table:comparison_of_TI}.
VirusTotal did not identify 82 of our \hosts with 6 of those being reported as benign. URLhaus did not repord 115 of our \hosts.
IBM X-Force did the worst as it misclassified 302 of our \hosts as benign.

\section{\Module \moduleA: Case-study 2: Temporal behavior}
\label{apppendix-Case_study_2}

In this second case study, we want to quantify the persistency of \hosts in terms of being visible in active probing.  We conduct a more intense  study focusing on fewer subnets.

\begin{figure}[t]
  \centering
  \begin{subfigure}[b]{0.48\columnwidth}
    \centering
    \includegraphics[width=\linewidth]{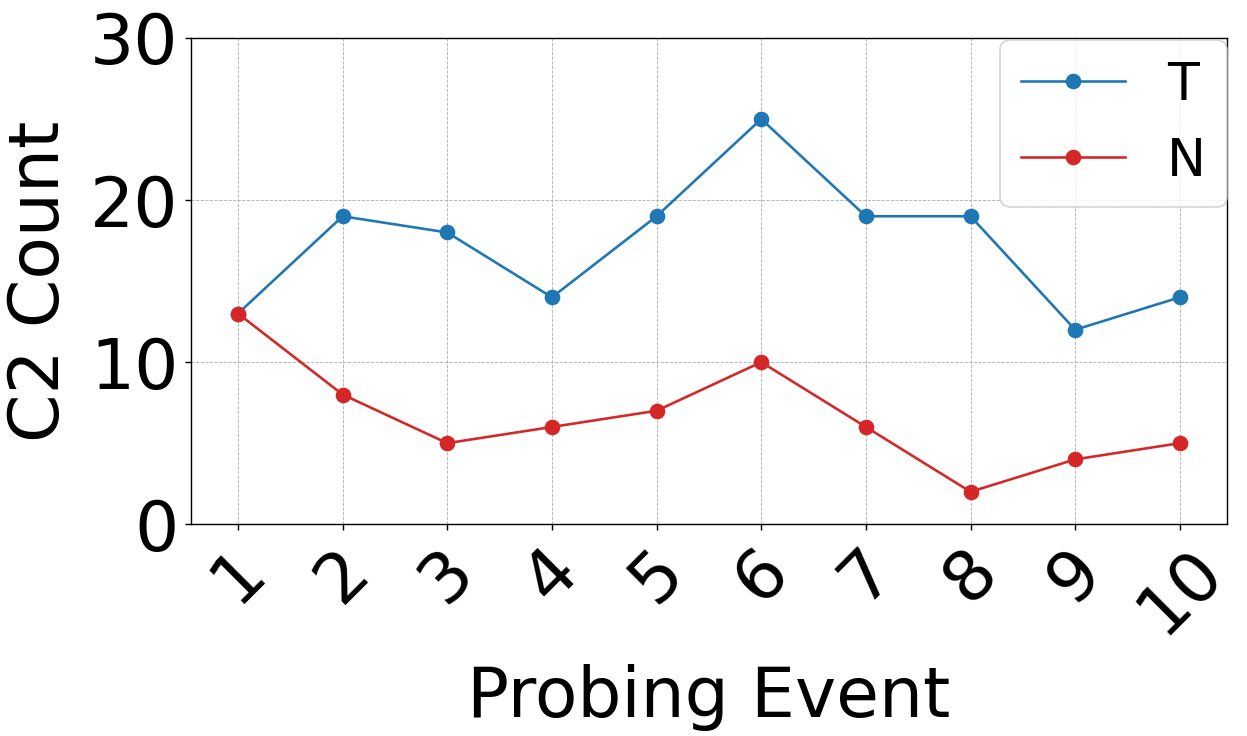}
    \caption{C2 servers}
    \label{fig:line_c2}
  \end{subfigure}
  \hfill
  \begin{subfigure}[b]{0.48\columnwidth}
    \centering
    \includegraphics[width=\linewidth]{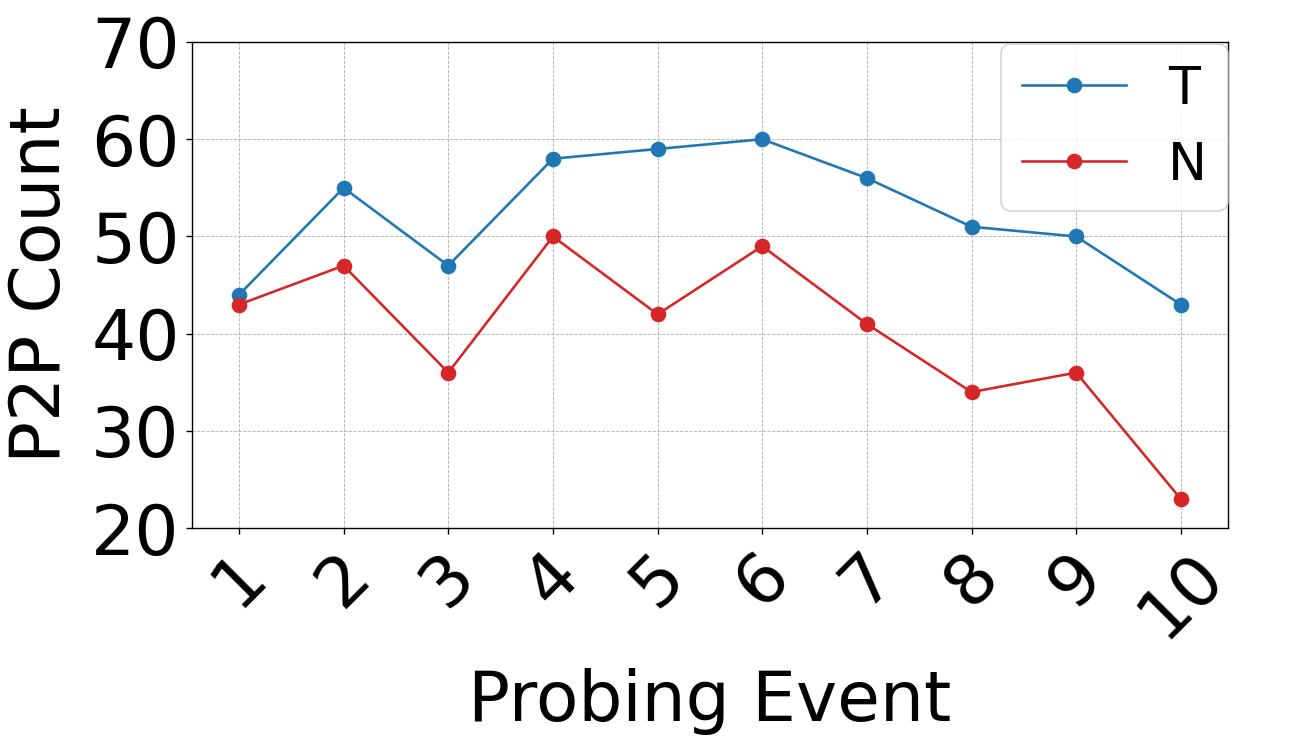}
    \caption{P2P \hosts}
    \label{fig:line_p2p}
  \end{subfigure}

  \caption{\small The continuous ``emergence'' of newly-identified
  \hosts and the unreliable responsiveness of already identified
  \hosts: Total (T) and New (N) live hosts on each probing event in
  case-study 2.}
  \label{fig:ip_count}
\end{figure}

{\bf The experimental setup.}
We selected 38 subnets among those used in our previous case-study. We selected subnets that contained \hosts from different malware families. 
%At the same time, they also covered a wide range of malicious activity. The malicious activity of a subnet refers to containing live \hosts in it.
%\miii{how did you pick those?  ***}
Specifically, the selected subnets cover the following families families: 
(a) 12 with Mirai, (b) 5 with Gafgyt, (c) 15 with Mozi, and (d) 6 with Hajime subnets. Each subnet contained 1–20  \hosts which were confirmed as live by our previous study. 
%(a) 12 dominated by Mirai with 1–10 C2 servers, (b) 5 dominated by Gafgyt (1–7 C2 servers), (c) 15 dominated by Mozi (1–20 P2P bot IPs), and (d) 6 dominated by Hajime (1–3 P2P bot IPs). 
We actively probe the IP addresses of these subnets every 2 days for 20 days: we explore all their active ports and use all our communication traces.
%\mimok{each one-time probe} 
 % for a total of 10 \mimok{probes}. 

% Table~\ref{table: Subnets_distribution} details the distribution of these monitored subnets.
% \miii{I am not sure this table provides an interesting message: I suggest we drop it, as we don't have space}

% \begin{table}[ht]
% \centering
% \resizebox{\columnwidth}{!}{
% \begin{tabular}{ c c c} 
%   \hline
%   Family & Subnet Count & Distribution (Observed\_IP\_Count/Subnet)\\ 
%   \hline
%   Mirai & 12 & 
%   8 Subs (1 IP/Sub) \\
%   & & 1 Sub (2 IPs/Sub), 1 Sub (6 IPs/Sub),\\
%   & & 1 Sub (7 IPs/Sub), 1 Sub (10 IPs/Sub)\\
%   \hline
%   Gafgyt & 5 & 
%   3 Subs (1 IP/Sub). \\
%   & & 1 Sub (3 IPs/Sub), 1 Sub (7 IPs/Sub)\\
%   \hline
%   Mozi & 15 & 
%   4 Subs (1 IP/Sub), \\
%   & & 3 Sub (2 IPs/Sub), 1 Sub (3 IPs/Sub),\\
%   & & 2 Sub (4 IPs/Sub), 1 Sub (5 IPs/Sub),\\
%   & & 1 Sub (7 IPs/Sub), 1 Sub (12 IPs/Sub),\\
%   & & 1 Sub (15 IPs/Sub), 1 Sub (20 IPs/Sub)\\
%   \hline
%   Hajime & 6 & 3 Sub (1 IPs/Sub), 3 Sub (3 IPs/Sub),\\
%   \hline
% \end{tabular}
% }
% \caption{Distribution of Subnets used in monitoring}
% \label{table: Subnets_distribution}
% \end{table}

%{\bf Key results.}

% used to say unrliable
%\takeaway{\host responsiveness is "flaky" with a median response rate of 1 out of 10 probes  in our study. As a result, probing an [IP:port pair] needs to be repeated multiple times for reliable results.}

The overarching observation is that \hosts are not responding consistently to all probes. 
We consider all the \hosts that responded positively to at least one of our probes.
If we assume that the servers were active for the 20 days, the median response rate per server is 1 in 10.
\miok{This is just an indication of the unresponsiveness and not an empirical rule, as the responsiveness varies among \hosts. 
However, the observation  strongly suggest that a thorough probing study should probe its targets repeatedly to ensure good recall.
}

%\mim{The line is critical. We saw that a C2 server responds again after 8 days from the last response. This is not aligned with C2Miner -- it says 31\% success rate assuming C2s were live while probing.}
%\miii{C2Miner probed a smaller number of reported C2s, no? It is fine to report our result as long as we are upfront with what we do and what we assume!}

%\mimok{This inconsistency stems from characteristic behaviors of malicious servers, such as short lifespans and frequent transitions between online and offline states.} 
This key observation is supported indirectly by the following results.
%This is a rather negative result as it suggests that a comprehensive probing study needs to send many probes to ensure that it does not miss a live server.  

{\bf a. Our 20 days of probing identifies 514 additional live \hosts.}
Probing the same space every other day over 20 days 
reveals an additional 514 \hosts  beyond those found in our first case study.

{\bf b. We find 48 previously unreported servers for a total of 112 such servers.}
In this case study, we find an additional 48 servers that have not been reported by well known \tifs on top of the 64 such servers that were identified in the first case study.

\miii{*if* we have space: we can  refer to coverage or even the distribution of new servers across families}
%These malicious addresses consist of 66 C2 servers and 448 \ptopnodes \hosts.
%\miii{let's find a term for this}. 

%\miii{Comparing with VT is not necessary here: let's keep the focus on evolution}
% Again, our probing provides information that does not exist in the three threat feeds. We find 9 new C2 servers and 39 new P2P bots that never appear in any of the three threat feeds, VirusTotal, URLhaus and IBM X-Force.

{\bf c. We keep finding new \hosts even on the 20th day of probing.}
%%9 \mim{9 or 18 days? 9 probing events = 18 days} probing days.}
Intrigued by this behavior, we investigate the evolution of the responses over time.
We track the total number of live \hosts across all subnets for each probing event. We also track the number of \hosts that are previously unobserved or new to us.
We plot the results in Fig.~\ref{fig:ip_count}.
% plotting both T (the total count of malicious IPs) and N (newly observed malicious IPs) for each probing cycle. We compare the results for subnets dominated by C2 servers vs. those dominated by P2P bots, showing how transient malicious IPs can significantly reshape the threat environment. 
%\miii{Please put Total instead of T and New instead of N in the plot. Do we probe also on Day 1? If we do we should show it or at least explain} \mim{putting "Total" and "New" make the legends overlap with the actual figure.}
Although the number of new \hosts drops over time,   we keep finding new ones even on the last day of probing.
%We share more detailed results in Appendix~\ref{sec:Appendix-consecutive-days-distribution}.
\miii{Check if we still have the APpendix!}

{\em Why are \host  responses flaky?} A working hypothesis is that the \hosts are responding selectively either: (a) on purpose, as a way to operate stealthily, or (b) due to inability,  because they are busy performing other tasks or  receiving too many incoming requests. We intend to investigate the root cause of this phenomenon in the future, and \name 
could enable a large study in this direction.

\section{Ethical considerations - extended}
\label{sec:Extra-Ethical}

We further discuss the ethical considerations of this research. We adhere to best practices for ethical malware analysis and Internet measurement and do not collect or process any personally identifiable information. Our study is limited to probing publicly accessible devices and services, and all malware samples are executed in a fully isolated sandbox environment. To prevent any unintended impact, malware behavior and network interactions are strictly controlled and monitored, as detailed in the sections below on traffic containment and probing safeguards.

%a. Detection of C2s: The algorithm we use is done in isolation. Thus, it does not interact with the Internet, as we “fake” connections to the malware in the sandbox. For sophisticated binaries that check for Internet connectivity, we deploy InetSim to
% simulate services like DNS and HTTP.

{\bf Traffic containment:}
During malware activation, we adopt the filtering method from C2Miner~\cite{C2Miner-Davanian-2024}. This filters out all traffic except the communication towards the malware's intended C2 server. The remaining traffic is just a standard call-home request. If the destination IP is not a real C2 server, the device simply ignores our probe or fails to complete the handshake. We never execute any commands received from the server.

{\bf Probing safeguards:} We only probe subnets that have a known history of malicious C2 activity. For each target IP, we first check if it has any open ports. We do not send probes to dead hosts. If a live port responds with a well-known benign banner (such as Apache or Nginx), we skip probing that port to avoid interfering with legitimate services. We also rate-limit all probes to keep network traffic minimal. We manually reviewed our sample traffic traces and found no cases of non-C2 communication.

\end{document}